\documentclass[]{aa}
\usepackage[dvipsnames]{xcolor}
\usepackage{soul}
\usepackage{ulem}
\usepackage{graphicx}
\usepackage[varg]{txfonts}
\usepackage[colorlinks,citecolor=blue,urlcolor=blue]{hyperref}
\usepackage[flushleft]{threeparttable}
\usepackage{natbib}
\usepackage{verbatim}
\usepackage{float}
\bibpunct{(}{)}{;}{a}{}{,}

\usepackage{siunitx}
\usepackage{dcolumn}
\newcolumntype{d}[1]{D{.}{.}{#1}}
\usepackage{verbatim}
\usepackage{float}
\def\mso{\,M_\odot}

\begin{document}

\title{Detailed models of interacting short-period massive binary stars 
}

\author{K. Sen\inst{1,2}
\thanks{email:ksen@astro.uni-bonn.de}
\and N. Langer\inst{1,2}
\and P. Marchant\inst{3}
\and A. Menon\inst{1,4}
\and S. E. de Mink\inst{5,4,6}
\and A. Schootemeijer\inst{1}
\and C. Sch\"urmann\inst{1,2}
\and L. Mahy\inst{7}
\and B. Hastings\inst{1,2}
\and K. Nathaniel\inst{1}
\and H. Sana\inst{3}
\and C. Wang\inst{1,2}
\and X.T. Xu\inst{1,2}
}

\institute{Argelander-Institut f\"ur Astronomie, Universit\"at 
Bonn, Auf dem H\"ugel 71, 53121 Bonn, Germany
\and Max-Planck-Institut f\"ur Radioastronomie, Auf dem H\"ugel 
69, 53121 Bonn, Germany
\and Institute of Astronomy, KU Leuven, Celestijnenlaan 200D, 
3001 Leuven, Belgium
\and Astronomical Institute “Anton Pannekoek”, University of 
Amsterdam, Science Park 904, 1098 XH Amsterdam, The Netherlands
\and Max Planck Institute for Astrophysics, Karl-Schwarzschild-Strasse 
1, 85748 Garching, Germany
\and Center for Astrophysics, Harvard-Smithsonian, 60 Garden 
Street, Cambridge, MA 02138, USA
\and Royal Observatory of Belgium, Avenue circulaire/Ringlaan 
3, B-1180 Brussels, Belgium
}

\date{Received November 3, 2021 / Accepted December 6, 2021}

\abstract 
{
The majority of massive stars are part of binary systems. 
In about a quarter of these, the companions are so close 
that mass transfer occurs while they undergo core hydrogen 
burning, first on the thermal and then on the nuclear timescale.
The nuclear timescale mass transfer leads to observational 
counterparts: the semi-detached so-called massive Algol 
binaries. These systems may provide urgently needed tests 
of the physics of mass transfer. However, comprehensive model 
predictions for these systems are sparse. }
{
We use a large grid of detailed evolutionary models of 
short-period massive binaries and follow-up population 
synthesis calculations to derive probability distributions 
of the observable properties of massive Algols and their 
descendants. }
{
Our results are based on $\sim$10,000 binary model sequences 
calculated with the stellar evolution code MESA, using a 
metallicity suitable for the Large Magellanic Cloud (LMC), 
covering initial donor masses between 10\,$M_{\odot}$ and 
40\,$M_{\odot}$ and initial orbital periods above 1.4\,d. 
These models include internal differential rotation and magnetic 
angular momentum transport, non-conservative mass and angular 
momentum transfer between the binary components, and 
time-dependent tidal coupling. }
{
Our models imply $\sim$30, or $\sim$3\% of the $\sim$1,000, 
core hydrogen burning O-star binaries in the LMC to be currently 
in the semi-detached phase. Our donor models are up to 25 times 
more luminous than single stars of an identical mass and effective 
temperature, which agrees with the observed Algols. A comparison 
of our models with the observed orbital periods and mass ratios 
implies rather conservative mass transfer in some systems, while 
a very inefficient one in others. This is generally well reproduced 
by our spin-dependent mass transfer algorithm, except for the 
lowest considered masses. The observations reflect the slow 
increase of the surface nitrogen enrichment of the donors during 
the semi-detached phase all the way to CNO equilibrium. We also 
investigate the properties of our models after core hydrogen 
depletion of the donor star, when these models correspond to 
Wolf-Rayet or helium+OB star binaries.}
{A dedicated spectroscopic survey of massive Algol systems
may allow to derive the dependence of the efficiency 
of thermal timescale mass transfer on the binary parameters, 
as well as the efficiency of semiconvective mixing in the 
stellar interior. This would be a crucial step towards 
reliable binary models up to the formation of supernovae and 
compact objects.
}


\maketitle

\section{Introduction}

In the last two decades, it has been well established 
that massive stars are preferentially born as members of 
binary or multiple systems \citep{vanbeveren1998,sana2012,sana2013,kobulnicky2014,moe2017,Banyard2021}.
Hence, the study of their evolution is complicated by 
the fact that they can interact with their companion, 
which can significantly alter their properties and lead 
to observable characteristics that differ largely from 
those obtained from single stars \citep{philip1992,selma2013,wang2020}.
This complexity makes the modelling of massive binary 
star evolution challenging, and many aspects of it are 
not yet well understood \citep{langer2012,crowther2019}.

Since stars tend to expand with age, mass transfer via a 
Roche-lobe overflow occurs naturally in many close binaries. 
In the tightest binaries, mass transfer may occur while 
both stars are still undergoing core hydrogen burning 
\citep{pols1994,vanbeveren1998,wellstein2001,selma2007}, 
initiating the so-called Case\,A mass transfer phase. 
Case\,A mass transfer is unique in the sense that it 
comprises a nuclear timescale mass transfer stage,
and it has the massive Algol systems  \citep[or contact 
systems, see][]{Menon2021} as observational 
counterparts, where the currently less massive, 
Roche-lobe filling star is transferring mass to a 
more massive star in a semi-detached configuration. 
Mass transfer in massive binaries otherwise 
\citep[with rare exceptions, such as in][]{quast2019} 
occurs on the much shorter thermal or even dynamical 
timescale, and it is much less likely to be observed. 

In Case\,A binaries, mass transfer first occurs on the 
thermal timescale of the donor star in a shrinking binary 
orbit. After the mass ratio is inverted, the orbit widens 
as a consequence of mass and angular momentum transfer. 
Thence, the mass transfer is driven by the slow expansion 
of the donor star due to core hydrogen burning. This gives 
rise to the nuclear timescale semi-detached phase, known 
as the slow Case\,A mass transfer, with a duration of a 
considerable fraction of the lifetime of the binary. Since 
all binaries in this phase previously experienced an intense 
phase of thermal timescale mass transfer (fast Case\,A), 
comparing models with observed semi-detached massive 
binaries is not only constraining the observable slow mass 
transfer phase, but also the much less understood thermal 
timescale mass transfer. 

Case\,A evolution has likely preceded most of the observed 
short period massive Wolf-Rayet+OB star binaries 
\citep{Massey1981,Niemela1982,Hucht2001,petrovic2005}
and massive black hole binaries such as Cygnus X-1, LMC-X1, 
and M33 X-7 \citep{Valsecchi2010a,qin2019,langer2020}.  
Recently, \cite{quast2019} have shown that nuclear timescale 
mass transfer can also occur in supergiant X-ray binaries 
\citep[see also][]{Pavlovskii2017,Marchant2021}. The short 
period of many of these post main sequence systems \citep{walter2015} 
implies that many of them evolved previously through 
Case\,A mass transfer. 

Massive Case\,A binaries are also thought to contribute 
to the double neutron star and double black hole binaries 
population in the Universe \citep[e.g.][]{kruckow2018}. 
At low metallicity (less than one-tenth of Solar) and 
high mass (above 40\,M$_{\odot}$) however, chemically 
homogeneous evolution may prevent expansion and mass 
transfer in short period binaries 
\citep{selma2009,pablo2016,mandel2016,hastings2020}, 
potentially leading to BH mergers observable by gravitational 
waves \citep{abbott2019}. 

Modern observations of massive star binaries allow to 
determine the binary and stellar properties of individual 
systems in great detail \citep{hilditch2005,torres2010,
martins2017,pavlovski2018,johnston2019,mahy2019b,mahy2019a,Janssens2021}. 
This provides ideal conditions to constrain the uncertain 
physics assumptions in binary evolution models \citep{ritchie2010,clark2014,abdul2019}. 
In particular, semi-detached double-lined systems provide 
a unique opportunity to derive the basic stellar properties 
of binaries with unprecedented precision, which can be 
used to test our models of stellar and binary evolution 
\citep{pols1997,nelson2001,selma2007,selma2009}.

Massive binaries with initial orbital periods below 
roughly 10\,d undergo Case\,A evolution. Based on the 
period distribution of massive binaries obtained by 
\citet{sana2012}, this implies that about one quarter
of all massive binaries will follow this path. 
Unfortunately, rapid binary evolutionary codes 
such as BSE \citep[Binary Star Evolution,][]{hurley2002}, 
Binary\_C \citep{Izzard2006}, StarTrack \citep{Belczynski2008}, 
COMPAS \citep[Compact Object Mergers: Population Astrophysics 
and Statistics,][]{Stevenson2017}, ComBinE \citep[][]{kruckow2018}, 
MOBSE \citep[Massive Objects in Binary Stellar Evolution,][]{Giacobbo2018}, 
SEVN \citep[Stellar EVolution for N-body,][]{Spera2019} 
and COSMIC \citep[Compact Object Synthesis and Monte Carlo 
Investigation Code,][]{Breivik2020} can treat Case\,A evolution 
only rudimentarily, since the above codes are based on single 
stars models, and on models of helium stars. However, the internal 
structure of mass donors in Case\,A binary models differs largely 
from both. Our detailed binary models include internal 
differential rotation and magnetic angular momentum 
transport in the individual stars, non-conservative 
mass and angular momentum transfer between the binary 
components, and time-dependent tidal coupling. 

Previous studies of the Case\,A mass transfer phase of massive binaries 
based on detailed binary evolution grids identified 
stellar binary models that can provide a good fit to 
the individual observed Algol binary systems \citep{nelson2001}, 
and constrain the underlying physics such as mass transfer 
efficiency as a function of the initial binary parameters 
\citep{selma2007}. The binary models of \citet[][BPASS- 
Binary Population and Spectral Synthesis]{Eldridge2017} 
also include Case\,A evolution. However, in their models the 
mass gainers are computed after the calculation of the evolution 
of the donor stars. Details of the Case\,A phase are not discussed. 

A detailed study of the expected 
observable characteristics of this phase is lacking 
in the literature. Others studies on the Case\,A mass 
transfer phase focussed on low and intermediate mass 
stars \citep{mennekens2017,negu2018}, where evidence 
for non-conservative mass transfer was found. In this 
work, we aim to provide a bridge between theory and 
observations by studying a grid of detailed binary 
evolution models and providing distributions of 
observable properties of massive binaries in this 
stage, that is, the slow Case\,A mass transfer phase. 

For binaries having shorter initial orbital periods, the accretors 
can also fill their Roche lobes during the slow Case\,A mass transfer 
phase, and reach a contact configuration \citep{pols1994,wellstein2001}, 
with the possibility of starting inverse mass transfer back to 
the donor while both components are burning hydrogen at their 
cores. Evolution during the contact phase is investigated 
in \cite{Menon2021}. Our study focusses on the semi-detached 
configuration during the Case\,A mass transfer phase.

Section \ref{method} gives a brief overview of the grid 
of detailed stellar evolution models and important physics 
assumptions. Section \ref{Case_A} describes the typical 
evolution of a Case\,A model and the mass transfer 
efficiency in our grid of models. Section \ref{slow_Case_A} 
describes the observable distributions of stellar 
parameters during the slow Case\,A mass transfer phase 
obtained from the binary model grid. The properties 
of our Case\,A models after core hydrogen depletion is reported 
in Sect.\,\ref{post_Case_AB}. We compare our model predictions 
with the observed massive Algol binaries in the Large Magellanic 
Cloud (LMC) and the Milky Way in Sect. \ref{observations_section}. In Sect. 
\ref{earlier_work}, we compare our work with relevant studies 
in the literature. We then briefly summarise our results and 
present our take-home messages in Sect. \ref{conclusions}.

\section{Method}
\label{method}
\subsection{The detailed binary evolution grid}

To study the properties of Case\,A mass transfer, we use a 
dense grid of detailed massive binary evolution models
\citep{pablothesis} with an initial metallicity and composition 
representative of young star-forming regions in the LMC. The 
models were calculated using the 1D stellar evolution code 
MESA\footnote{http://mesa.sourceforge.net/} (Modules for
Experiments in Stellar Astrophysics, \citealp{mesa11,
mesa13,mesa15,mesa18}, version 8845\footnote{Inlists to 
reproduce the models used in this work can be downloaded 
from github.com/orlox/mesa\_input\_data/tree/master/2016\_binary\_models}). 
The stellar and binary physics assumptions are described 
in detail by \cite{mesa15} and \cite{pablothesis}. The most 
important and relevant ones are discussed in the following 
paragraphs.

The chemical composition and wind mass loss rates are set 
as in \citet[][]{brott2011}: the hydrogen, helium 
and metal mass fractions used in our models are 0.7391, 
0.2562 and 0.0047 respectively. The initial abundances (in units of 12 + log [element/H])
for C, N, O, Mg, Si, Fe adopted are 7.75, 6.90, 8.35, 
7.05, 7.20 and 7.05 respectively. All other elements 
are solar abundances \citep{Asplund2005} scaled down 
by 0.4 dex. The physics of differential rotation 
and rotational mixing follows that of \cite{heger2000} 
while the magnetic angular momentum transport is as in 
\cite{heger2005}. The mass transfer rates are calculated 
following \cite{pablo2016}. When only one component 
of the binary overflows its Roche lobe, the mass 
transfer rate from the Roche lobe filling star is implicitly 
calculated so as to make the star just fill its Roche lobe. 
When both components of the binary 
fill their Roche lobes, the contact phase is calculated 
as described in Sect. 2.2 of the same work. 

The accretion of angular momentum from the transferred matter 
follows the implementation of \cite{selma2013}, which is based 
on the results of \citet{lubow1975} and \citet{ulrich1976}. A 
differentiation between disc and ballistic modes of accretion 
during the mass transfer phases is considered. To determine 
which mode of accretion occurs, we compare the minimum distance 
of approach of the accretion stream \citep{lubow1975,ulrich1976}
\begin{equation}
    R_{\rm min} = 0.0425 a \left( \frac{M_{\rm a}}{M_{d}} + \frac{M_{\rm a}^{2}}{M_{\rm d}^{2}} \right)^{1/4}
    \label{r_min}
\end{equation}
to the radius of the accretor star ($R_{\rm a}$). Here, $a$ 
is the orbital separation, $M_{\rm d}$ and $M_{\rm a}$ are 
the masses of the initially more massive donor star and initially 
less massive accretor star respectively. Accretion is assumed 
to be ballistic when $R_{\rm a} > R_{\rm min}$, with a specific 
angular momentum of $(1.7GM_{\rm a}R_{\rm min}$)$^{1/2}$. Otherwise, 
accretion is assumed via a Keplerian disc with a specific angular 
momentum of $(GM_{\rm a} R_{\rm a})^{1/2}$. 

Equation\,\ref{r_min} can be written in terms 
of orbital period, mass ratio and the mass of the accretor. 
We find that for an accretion disc to form, the orbital period 
of the binary during the mass transfer phase has to be longer 
than 20\,d, for a 10\,$M_{\odot}$ accretor when assuming a typical 
mass-radius relation \citep{Gorda1998} to estimate the accretor 
radius. Hence, our Case\,A models are not expected to form an 
accretion disc during the slow Case\,A mass transfer phase. On 
the other hand, there are some observed Algol binaries that have 
an accretion disc (Table\,\ref{table_galaxy}). Investigation 
into this discrepancy will be interesting, though beyond the 
scope of our present work. 

The rotation periods of both components of the binary are assumed to 
be synchronised to the orbital period at their zero-age main 
sequence (ZAMS). Tidal interactions are modelled in a time-dependent 
fashion following the implementation by \citet{detmers2008}, 
using a synchronization timescale associated with the dynamical 
tide model of \citet{zahn1977}, as appropriate for main sequence 
stars with radiative envelopes.  

We use the standard Mixing Length Theory \citep{bohm1960}
to model convection, with a mixing length parameter of 
$\alpha_{\mathrm{MLT}} = 1.5$. The treatment of semiconvection 
follows \cite{langer1991} using $\alpha_{\mathrm{SC}} = 0.01$ 
and the physics of thermohaline mixing is as in \cite{cantiello2010}.
We implement overshooting as a step function only to the 
top of the convective core up to 0.335 times the pressure 
scale height \citep{brott2011}. However, overshooting is 
applied only to layers inside a star which have a near-constant 
composition \citep{pablothesis}, implying that composition 
gradients are taken into account during the rejuvenation 
process of mass gaining stars \citep{braun1995}. The initial 
composition of the binary models also takes into account the 
nonsolar abundance ratios of the LMC, as in \cite{brott2011}. 
Unlike \cite{brott2011} however, we use custom made OPAL 
opacity tables \citep{iglesias1996} in agreement with the 
initial abundance ratios of the LMC. 

The initial donor mass ($M_{\mathrm{d}}$) of the binary 
models range from 10.0 to 39.8 $M_{\odot}$ in steps of 
log($M_{\mathrm{d}}$/$M_{\odot}$) = 0.05, where the donor 
star in a binary model is the initially more massive 
star. The accretor ($M_{\mathrm{a}}$) is the initially 
less massive star of the model. For every initial mass 
of the donor, models with different initial mass ratio 
(defined as $q$ = $M_{\mathrm{a}}$/$M_{\mathrm{d}}$) 
ranging from 0.275 to 0.975 (in steps of 0.025) and 
initial orbital period ($P_{\mathrm{i}}$) from $\sim$1.41\,d 
(log $P$/d = 0.150) to $\sim$3165\,d (log $P$/d = 3.500), 
in steps of $\Delta$log $P_{\mathrm{i}}$/d = 0.025, are 
calculated. This current work deals with the properties 
of models in the grid undergoing Case\,A mass transfer. 

\begin{figure*}
\centering
\includegraphics[width=\hsize]{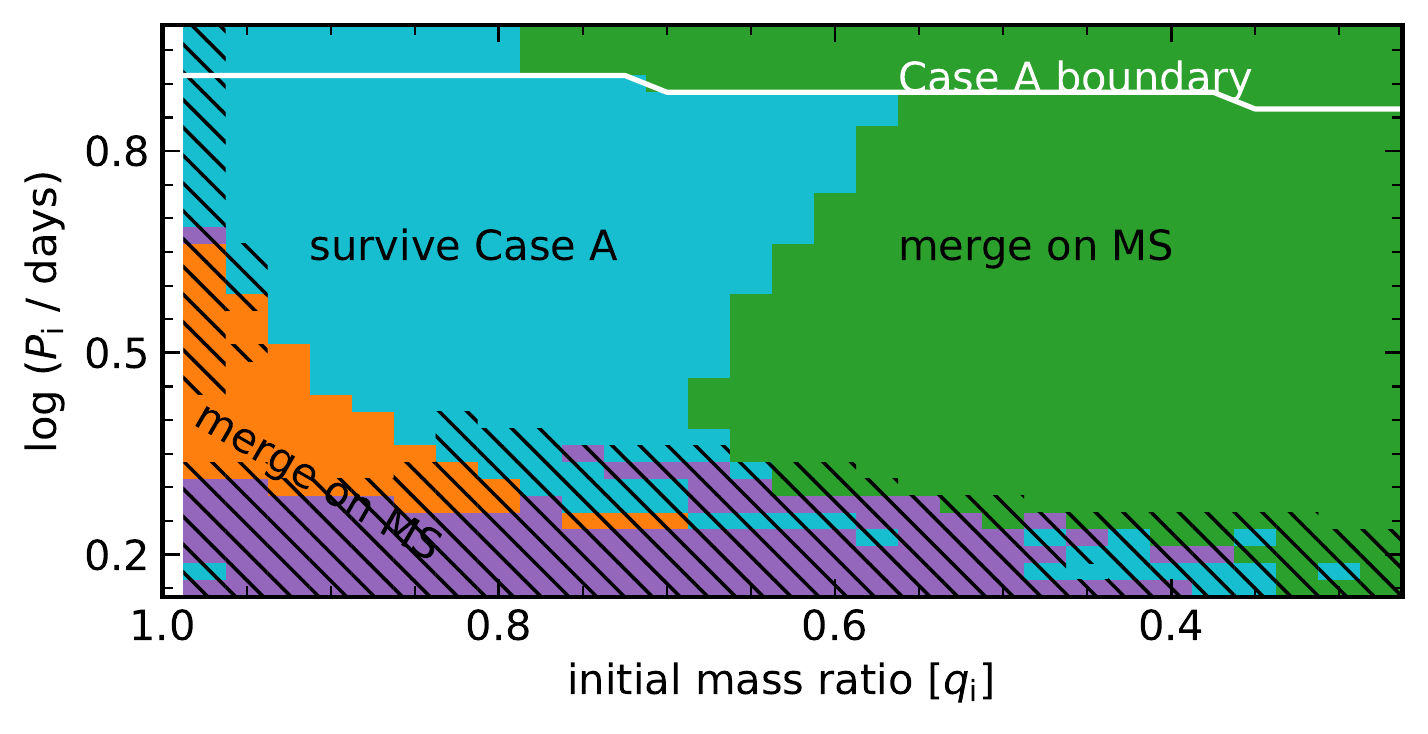}
\caption{Summary plot of binary models with an initial donor mass of 22.4 $M_{\odot}$, as a function of the initial binary orbital period and mass ratio. Each pixel in the plot represent the outcome of one binary model with the corresponding orbital period and mass ratio. The white line shows the maximum initial orbital period below which models undergo Case\,A mass transfer. The light blue coloured models survive the Case\,A mass transfer phase. The models marked in green are terminated during the fast Case\,A mass transfer phase. The purple models undergo mass overflow from the L2 Lagrangian point during core hydrogen burning. The orange models undergo inverse mass transfer onto the initially more massive Roche-lobe filling star while it is burning hydrogen, from the initially less massive component that has completed core hydrogen burning. Models that enter into a contact configuration, at any point of time during their main sequence evolution, are marked with hatching.}
\label{1.350_summary}
\end{figure*}

The models are evolved from the start of hydrogen burning 
and the orbit is assumed to be circular. 
The spin axes are assumed to be aligned to the orbital axis. 
This avoids the introduction of six free parameters at ZAMS, namely 
the initial rotation rates of each component. Admittedly, this 
assumption has no physical justification in nature, but it has 
been shown that moderate rotation does not significantly 
affect the evolution of individual stars \citep{brott2011,choi2016}
while the fastest rotating stars may be products of binary evolution
\citep{selma2013,wang2020}. More importantly, the initially 
short-period binary models, that is, most of the Case\,A models 
we are dealing with in our study, get tidally locked soon 
after the beginning of the evolution from ZAMS \citep{selma2009},
irrespective of their initial individual stellar spin and direction. 
The spins of the post Case\,A models investigated here are 
influenced by the mass transfer episode where the mass donor 
fills its Roche volume and the mass gainer is typically 
spun up due to mass accretion. For similar reasons, the assumption 
of a circular orbit is not expected to affect our results 
significantly as our work is focussed on the properties of 
short-period binaries. 

Mass transfer in the binary models is assumed to be conservative 
except for stellar winds until the accreting star spins up to 
critical rotation during the mass transfer process. However, 
\cite{packet1981} showed that stars only need to accrete a few 
percent of their total mass to spin up to critical rotation. 
Further accretion of matter is not 
possible unless angular momentum is lost by some mechanism. To 
model this loss of angular momentum when accretors reach critical 
rotation, the wind mass-loss rate of the accretor is increased to 
keep $\upsilon_{\rm rot}/\upsilon_{\rm crit}$ near $\sim$0.99, 
facilitating mass and angular momentum loss through the increased 
wind mass loss as described in \cite{n2003} and \cite{petrovic2005}. 
The amount of specific angular momentum removed from the 
orbit by the enhanced wind is equal to the specific orbital angular 
momentum of the accretor. At the same time, we remove the product 
of the specific spin angular momentum of the accretor at its surface 
times the mass removed from the system from the spin angular momentum 
of the accretor \citep{mesa15}. Here, $\upsilon_{\rm rot}$ and $\upsilon_{\rm crit}$ 
are the rotational and critical rotational velocities of the star 
as defined in \cite{langer1997}. 

Typically, the models where tidal 
effects can prevent the rapid spin-up of the accreting star 
lead to efficient mass transfer (i.e. very short orbital period 
models). In contrast, longer period models have comparatively 
much lower mass transfer efficiency. The efficiency of mass 
transfer is still an unsolved problem in binary evolution, with 
studies hinting at a decreasing efficiency with increasing orbital 
period \citep{selma2007}.

The assumption of increased mass-loss rates at close to critical 
rotation leads to a self-consistent way of determining the mass 
transfer efficiency, although it leads to very high mass-loss 
rates in models where tidal effects are unable to halt the rapid 
spin-up to critical rotation. To assess the feasibility of such 
high mass-loss rates in a model, an upper limit to the mass-loss 
rate ($\dot{M}_{\rm upper}$) is calculated. For a radiation-driven 
wind, the upper limit is found by assuming that the mass that has 
to be lost is removed from a Keplerian disc whose radius is at 
most equal to that of the Roche lobe radius of the mass accreting 
star (with mass $M_{\mathrm{a}}$). Ignoring the donor (with mass 
$M_{\mathrm{d}}$), the energy per unit mass (or gravitational 
potential) required to remove mass from the Roche lobe radius 
of the accretor is $GM_{\mathrm{a}}/R_{\mathrm{RL,a}}$. Assuming 
that the luminosity of both stars ($L_{\mathrm{d}}$ and 
$L_{\mathrm{a}}$) is used to drive this mass loss, the resulting 
mass-loss rate is
\begin{equation}
\mathrm{log}\frac{\dot{M}_{\mathrm{upper}}}{M_{\odot}/ \rm yr} = -7.19 + \mathrm{log}\frac{L_{\mathrm{d}}+L_{\mathrm{a}}}{L_{\odot}} - \mathrm{log}\frac{M_{\mathrm{a}}}{M_{\odot}} + \mathrm{log}\frac{R_{\mathrm{RL,a}}}{R_{\odot}}\:.
\label{upper-mdot-limit}
\end{equation}
For a typical model in our grid, having $L_{\rm d}$, 
$L_{\rm a}$, $M_{\rm a}$ and $R_{\rm RL,a}$ of 4570 
$L_{\odot}$, 3631 $L_{\odot}$, 6.25 $M_{\odot}$ and 
10 $R_{\odot}$ respectively, the maximum mass loss 
rate ($\dot{M}_{\rm upper}$) is $\sim10^{-3}\,M_{\odot}/ \rm yr$. 

When the mass-loss rate required in the model exceeds this 
maximum value that can be powered by the photon energy of 
both stars, the evolution of the model is stopped. The 
evolution of contact binaries are modelled as in \cite{pablo2016} 
and they are stopped if mass overflow occurs from the L2 
Lagrangian point during the contact phase. We also stop the 
evolution of a model if inverse mass transfer occurs in a 
binary model with a post main sequence component. We assume 
that the mass transfer in such models will not be stable and 
will lead to a common envelope evolution and merge. Otherwise, 
models are evolved till core carbon exhaustion if they have 
helium core masses less than 13 $M_{\odot}$ at the end of 
core helium burning, and till core helium depletion for 
models that have helium core mass greater than 13 $M_{\odot}$ 
(due to numerical issues faced in modelling the more massive 
stars all the way to core carbon depletion). Our study deals 
with the properties of binaries where both components are on 
the main sequence. 

\label{methods}

\subsection{Initial binary distribution function}
\label{Wm}

For our population synthesis predictions, each binary 
model m in our grid is assigned a weight factor 
$W_{\rm m}$ that depends on the initial donor mass 
$M_{\mathrm{d}}$, the initial mass ratio $q_{\mathrm{i}}$ 
and the initial orbital period log $P_{\mathrm{i}}$. 
For the initial donor mass distribution, we use the 
Salpeter initial mass function \citep{salpeter1955}, 
and the \cite{sana2012} exponents for the initial mass 
ratio and orbital period. While the latter have been 
derived using a modest sample of $\sim$40 O-type binaries 
in the Milky Way, more recent studies of O and B type 
stars both in the Milky Way and the LMC have failed to 
reveal any statistically significant differences 
\citep{sana2013,Dunstall2015,Villasenor2021,Banyard2021}. 
Hence, our adopted values can also be taken as 
representative of the OB star population at LMC metallicity. 
As such, we define $W_{\rm m}$ as
\begin{equation}
    \label{eq_Wm}
    W_{\rm m} = (\mathrm{log\:}M_{\mathrm{d,i}}/M_{\odot})^{-1.35} * q_{\mathrm{i}}^{-0.10} * (\mathrm{log\:}P_{\mathrm{i}}/\rm d)^{-0.55} \:.
\end{equation}

We obtain histograms for the distribution of an 
observable $O_{\mathrm{obs}}$, defined as the value 
of the stellar parameter during the contact (cnt) or 
semi-detached (SD) phase, weighing with the amount of 
time spent in the cnt or SD phase and the initial binary 
distribution functions. The number fraction ($f_{\rm obs,cnt\:or\:SD}$) 
of an observable stellar parameter in a given bin with 
bin edges [$O_{\rm 1}$,$O_{\rm 2}$] is given by,
\begin{equation}
f_{\rm obs,cnt\:or\:SD} \: (O_{\rm 1} < O_{\rm obs} < O_{\rm 2}) = \frac{\sum_{\rm m=1}^{\rm N} \delta_{O_{\rm 1}O_{\rm 2},\rm m}\:W_{\rm m} \: \Delta t_{\rm cnt\:or\:SD,m}}{\sum_{\rm m=1}^{\rm N} \:W_{\rm m} \, \Delta t_{\rm cnt\:or\:SD,m}}\:,
\label{observed_period_eqn}
\end{equation}
where m is the model number of a particular model in 
our grid and N is the total number of models that 
undergo Case\,A mass transfer (Case\,A models). 
$\Delta t_{\rm cnt\:or\:SD,m}$ is the total time spent in 
the contact or semi-detached configuration by model m.
$\delta_{O_{\rm 1}O_{\rm 2},\rm m}$ is equal to 1 
when the value of the observable for the model m is 
between $O_{\rm 1}$ and $O_{\rm 2}$ during the contact 
or semi-detached phase and zero otherwise.

\subsection{Binary parameter space}
\label{general_props}

To help understand the general properties of the 
models undergoing Case\,A mass transfer and aid in 
the interpretation of the results of our population 
synthesis, we briefly describe a slice of our 
parameter space. Figure\,\ref{1.350_summary} shows 
the evolutionary outcomes up to core hydrogen 
depletion for models with an initial donor mass 
of 22.4 $M_{\odot}$ and different initial orbital 
periods and mass ratios. Models with log ($P_{\rm i}$/d) 
$\lesssim$ 0.9 undergo Case\,A mass transfer. The 
orbital period cut-off below which models undergo 
Case\,A mass transfer depends on the initial donor 
mass, with longer periods binaries able to undergo 
Case\,A mass transfer for higher initial donor masses 
(c.f. Fig.\,\ref{1.100_summary}). 

Notably, we find by inspection that all our Case\,A 
accretors undergo ballistic mass accretion from the 
donors (see discussion of Eq.\,\ref{r_min}). We see 
that only a part of the parameter space survives the 
Case\,A mass transfer phase (marked by the lightblue 
colour). Models with low mass ratios (green) are 
terminated (and assumed to merge) during the fast 
Case\,A mass transfer phase because the combined 
luminosity of both of stars is insufficient to drive the excess 
mass-loss rate required to hinder over-critical 
rotation of the mass accretor, that is, the mass 
transfer rate in the model exceeds $\dot{M}_{\rm upper}$ 
(given by Eq.\,\ref{upper-mdot-limit}). They spend $\sim$5000 
yrs in the fast Case\,A mass transfer phase before 
terminating and hence are not expected to significantly 
change the distribution of the observable properties 
of semi-detached binaries. 

Models with very short orbital periods (purple)
enter into a contact configuration during the slow 
Case\,A mass transfer and eventually experience L2 
overflow. Such binaries are assumed to merge, due 
to the very low orbital periods. In the orange models, 
the initially less massive star overtakes the evolution 
of the mass donor during the slow Case\,A mass transfer 
phase and completes core hydrogen burning before the 
initially more massive star \citep{pols1994,wellstein2001}. 
As such, a reverse mass transfer is initiated from 
initially less massive star (that is expanding rapidly 
after core hydrogen burning), on to the initially more 
massive Roche-lobe filling star that is still in its 
main sequence. This inverse mass transfer will likely 
be unstable and the binary is assumed to merge. 

However, the above two types of models (purple and 
orange) spend up to $\sim$10 Myrs in the slow 
Case\,A mass transfer phase before they undergo 
L2 overflow. Hence, the contribution from these short 
period binaries to the observable properties of 
Algol binaries cannot be neglected. Similar figures 
for other initial donor masses are provided in the 
Appendix (Fig.\,\ref{1.100_summary}). Since the radii 
of more massive donor stars are larger, binaries can 
undergo Case A mass transfer at longer orbital periods 
for greater initial donor masses. Moreover, since the 
luminosity of stars rises steeply with mass, the 
upper limit to the maximum mass transfer rate (Eq.\,\ref{upper-mdot-limit}) 
also increases for higher initial donor masses. Since 
this upper limit to the maximum mass transfer rate determines the boundary between 
systems that merge vs systems that survive the fast 
Case\,A mass transfer, more systems survive the Case\,A 
mass transfer at higher initial donor masses. 

\section{Case\,A mass transfer}
\label{Case_A}
\subsection{A Typical example}
\label{CaseA_example}
\begin{figure}
\centering
\includegraphics[width=\hsize]{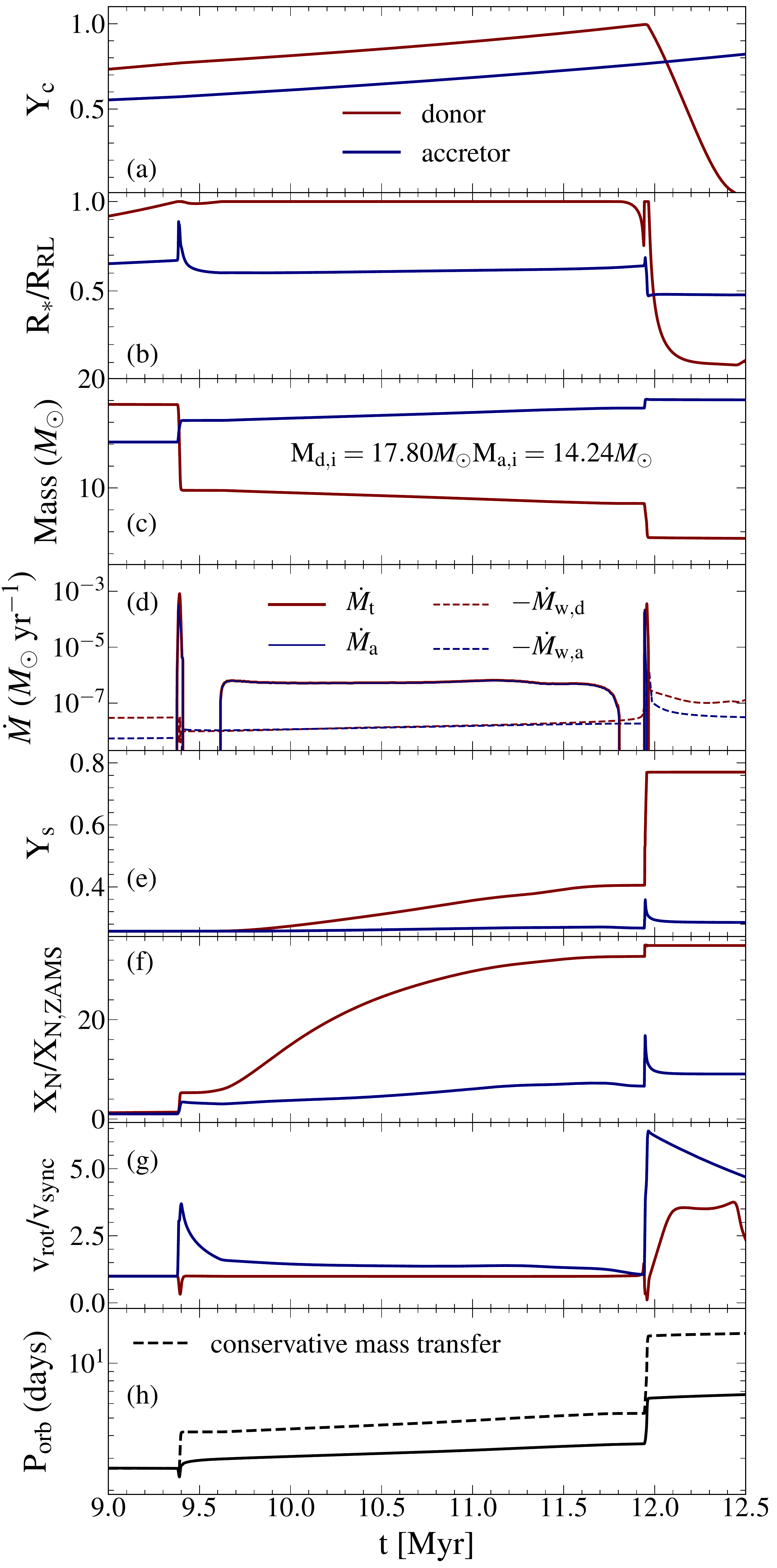}
\caption{
Example of a Case\,A and Case\,AB mass transfer episode for a typical binary model in our grid. The initial donor mass, mass ratio and orbital period of the model are 17.8  $M_{\odot}$, 0.8 and $\sim$2.7\,d respectively. Various quantities are plotted as function of time, with
$t=0$ corresponding to the ZAMS stage of both stars. \textit{(a):} Central helium mass fraction of donor (red) and accretor (blue). \textit{(b):} Ratio of  donor and accretor radius to their Roche lobe radii. \textit{(c):} Donor and accretor mass. \textit{(d):} Mass transfer rate $\dot{M}_{\rm t}$ (red solid line), effective mass accretion rate $\dot{M_{\rm a}}$ (blue solid line), and wind mass loss
rates of donor ($-\dot{M}_{\rm w,d}$, red dotted line) and accretor ($-\dot{M}_{\rm w,a}$, blue dotted line), respectively. \textit{(e):} Surface helium mass fraction. \textit{(f):} Surface nitrogen  enhancement factor. \textit{(g):} Ratio of rotational to orbital angular velocity. \textit{(h):} Orbital period (solid black line), and the orbital period our model would have obtained if the mass transfer would have been fully conservative (dashed black line). 
}
\label{typical_Case_A}
\end{figure}

\begin{figure}
\includegraphics[width=\linewidth]{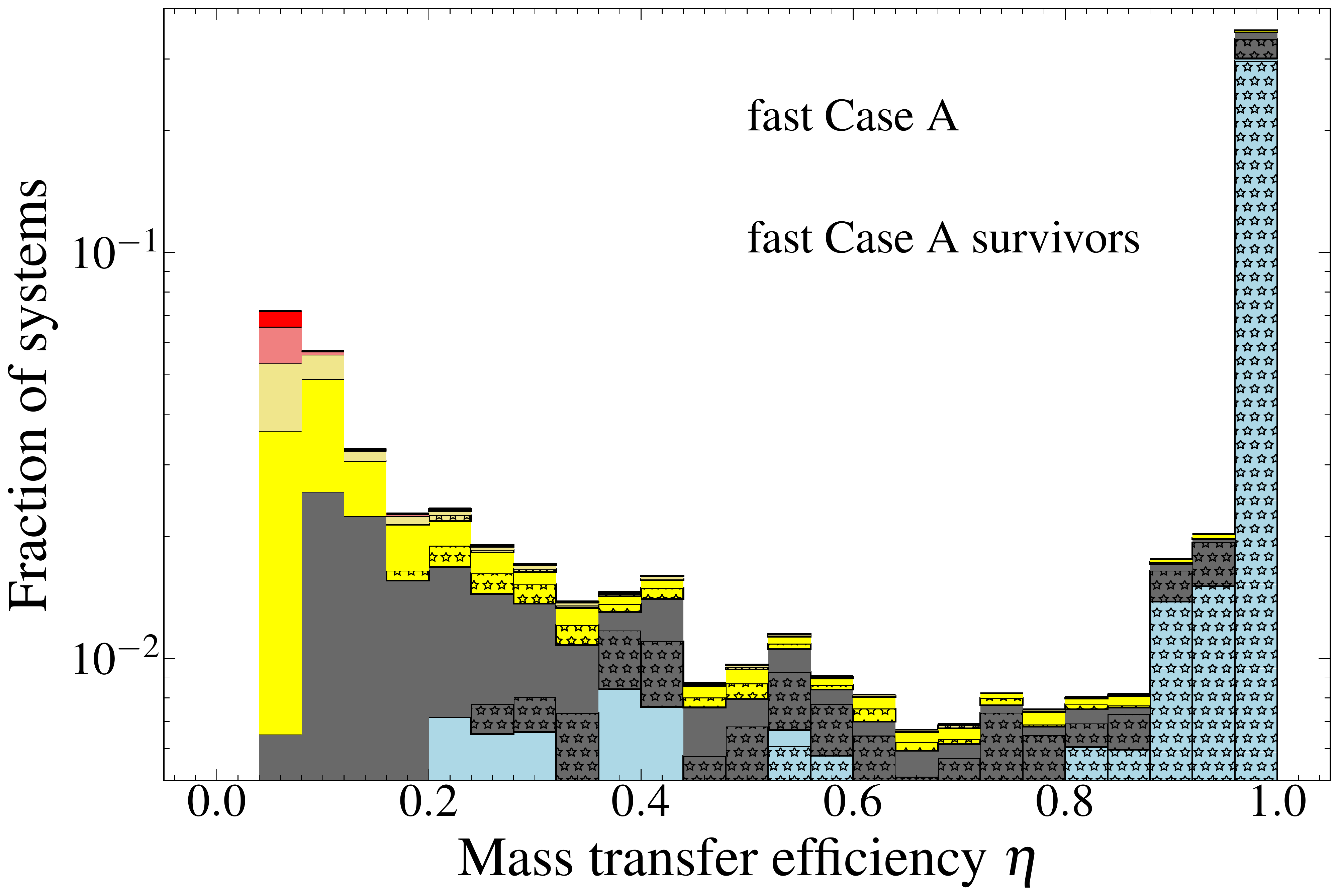}
\includegraphics[width=\linewidth]{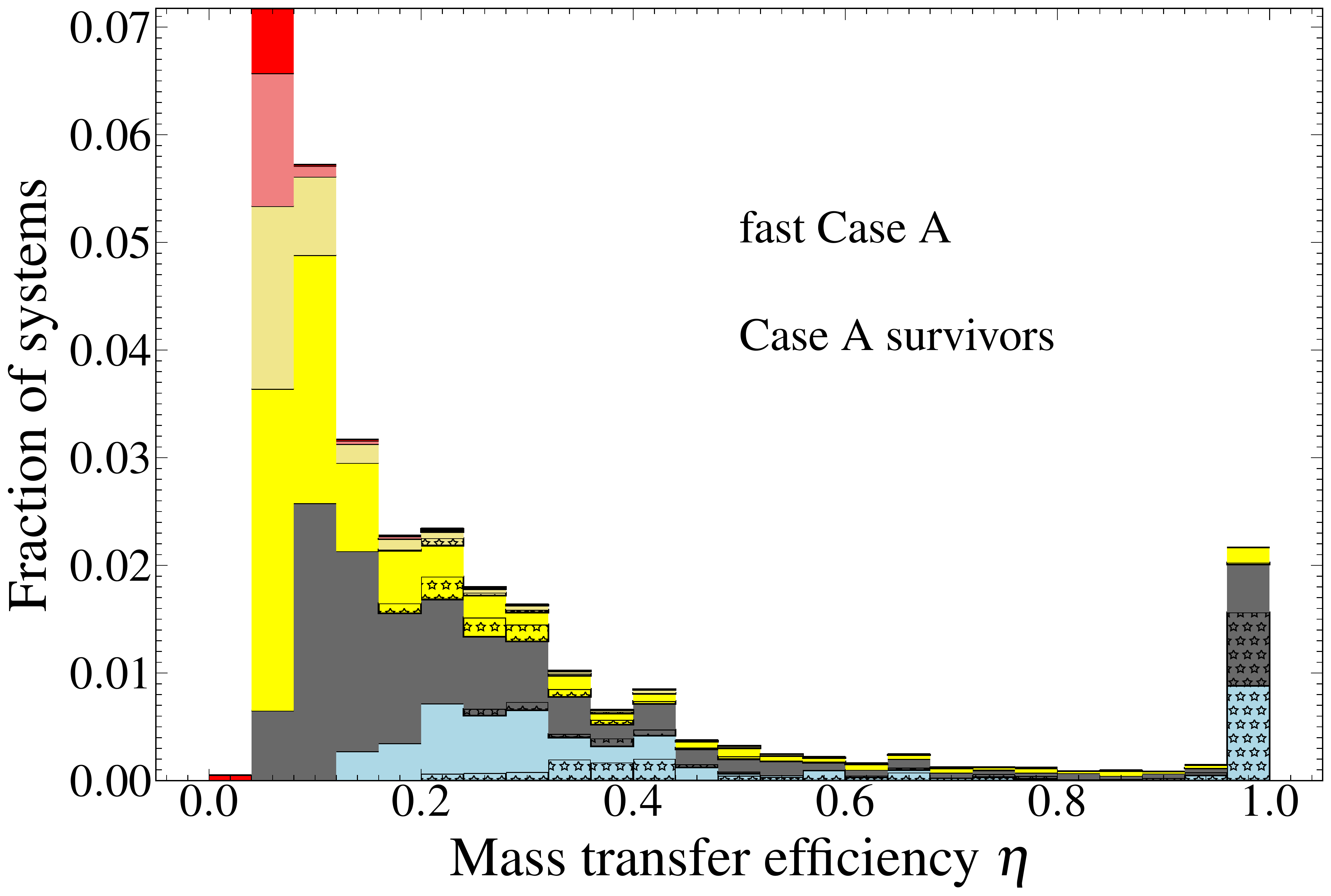}    
\includegraphics[width=\linewidth]{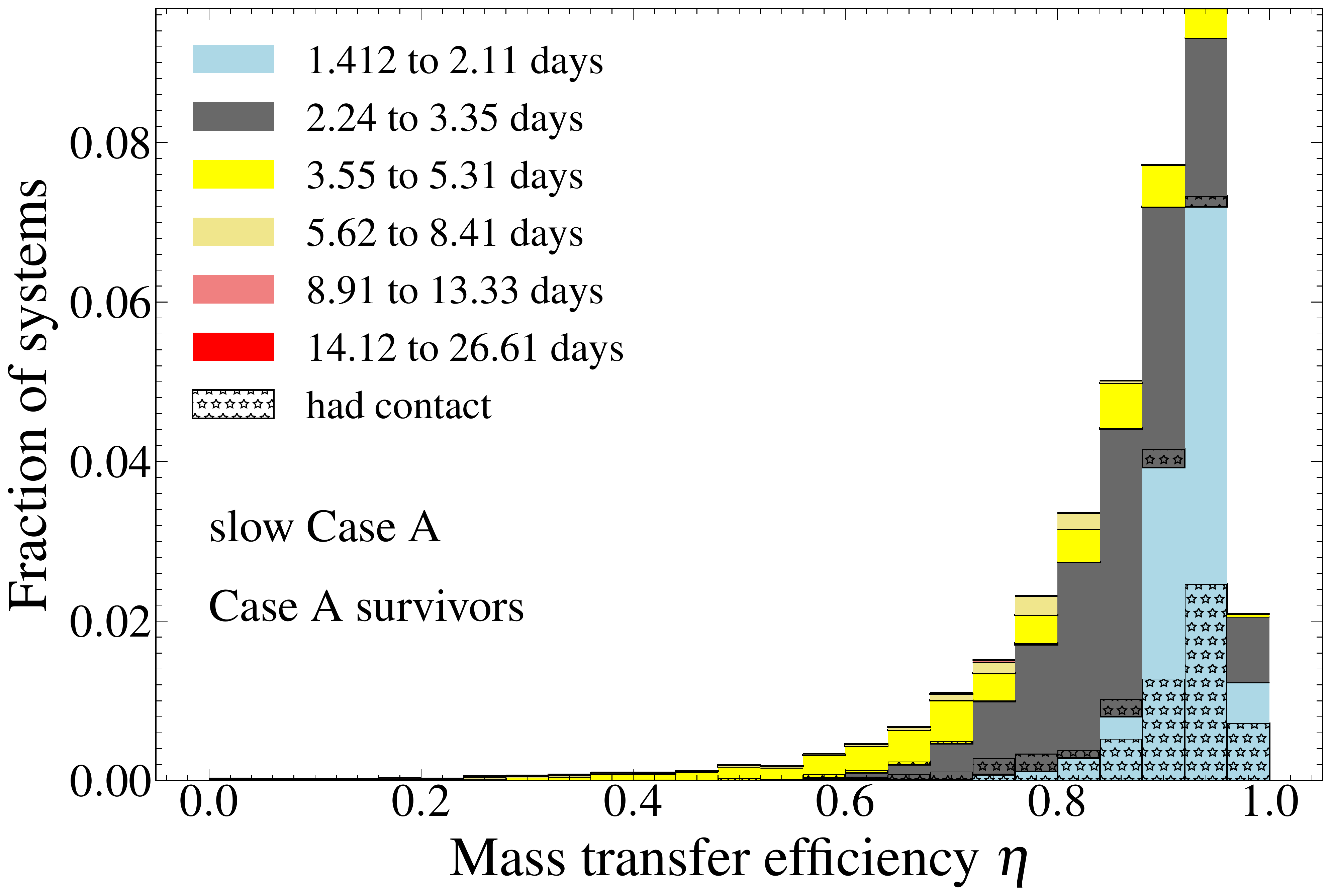}
\caption{
Distribution of the time-averaged mass transfer efficiency during fast (top two panels) and slow (bottom panel) Case\,A mass transfer. The top plot includes all models that survive fast Case\,A mass transfer phase (i.e. light blue, orange and purple models in Fig.\,\ref{1.350_summary}), while the middle panel includes only those models that survive also the slow Case\,A mass transfer phase (light blue models in Fig.\,\ref{1.350_summary}). The colour coding is with respect to the initial orbital period of the binary model. The histograms are weighted with the birth probability of the binaries, and with the amount of time spent in the respective Case\,A mass transfer phase. The ordinate values are normalised to unity such that the value for each bin gives the weighted fraction of models that have their mass transfer efficiency between those bin edges during fast/slow Case\,A mass transfer. The fraction of models in any bin that undergo a contact phase are marked with star hatching.
}
\label{mass_trans_eff_fast_period}
\label{mass_trans_eff_slow_period}
\end{figure}

Figure\,\ref{typical_Case_A} shows the typical evolution 
of a short period ($\sim$2.7 d) massive binary with 
initial donor and accretor mass of 17.80 $M_{\odot}$ and 
14.24 $M_{\odot}$, respectively. It encounters mass 
transfer while both stars are burning hydrogen 
in their cores (Panel\,`a'). The initially more massive star 
fills its Roche lobe at $\sim$9.4 Myr which initiates mass 
transfer via Roche-lobe overflow (panel\,`b'). The mass 
transfer rate rises above 10$^{-4}$ $M_{\odot}$yr$^{-1}$ 
(panel\,`d') during the so-called fast Case\,A mass transfer 
phase which occurs at the thermal timescale of the donor. 
We see that in this model, the fast Case\,A mass transfer 
is not purely conservative as the accretor quickly spins 
up to critical rotation (evidenced by the sharp increase 
in the ratio of rotational velocity to synchronous velocity 
in Panel\,`g'). Only about 20\% of the mass lost by the donor 
is accreted by the companion (Panel\,`c'). The orbital period 
(solid black line, panel\,`h'), and in turn the orbital 
separation, first decreases and then increases as the mass 
ratio gets inverted and keeps diverging from unity. 

For comparison, we also show the evolution of the orbital 
period if the mass transfer was fully conservative (black 
dashed line). We compute the orbital period ($P$) evolution in the 
conservative case following \citep{Soberman1997} 
\begin{equation}
    \frac{P}{P_{\mathrm{i}}} = \left( \frac{M_{\mathrm{d,i}}M_{\mathrm{a,i}}}{M_{\mathrm{d}} M_{\mathrm{a}}} \right)^3 \:,
    \label{pcons}
\end{equation}
where $P_{\mathrm{i}}$, $M_{\mathrm{d,i}}$ and $M_{\mathrm{a,i}}$ 
are the initial orbital period of the binary, donor mass 
and accretor mass respectively. $M_{\mathrm{d}}$, $M_{\mathrm{a}}$ 
and $P$ are the masses of the donor, the accretor, and 
the orbital period at any stage during the conservative 
mass transfer process. To compute the orbital period under 
the assumption of conservative mass transfer, we take 
$M_{\mathrm{d}}$ from our simulations and compute 
$M_{\mathrm{a}}$ assuming the mass transfer process was 
conservative. Since the mass transfer is non-conservative 
during the fast Case\,A phase in this model, the increase 
in orbital period due to the fast Case\,A mass transfer 
is lower than for the conservative mass transfer case. 

Following the fast Case\,A mass transfer episode, the 
binary enters into a nuclear timescale mass transfer 
phase where the mass transfer rate is of the order of 
the nuclear timescale. This phase is known as the slow 
Case\,A phase, or the Algol phase (for a more detailed 
discussion, see \citealp{wellstein2001}). In our work, we 
distinguish between the fast and slow Case\,A mass 
transfer phase based on the mass transfer rate in the 
binary, with the boundary at 10$^{-5}$ $M_{\odot}$yr$^{-1}$.

There is a gradual increase in the surface abundances of 
helium and nitrogen during the slow Case\,A phase (panels 
`e' and `f') as the mass transfer exposes the deeper layers 
of hydrogen burning processed material of the donor. We 
also note that most of the mass from the donor is lost during the fast 
Case\,A phase (panel\,`c') and there is a slow loss and 
gain in mass of the donor and accretor respectively during 
the slow Case\,A phase. The amount of mass accreted depends 
on, amongst other factors, the spin-up of the accretor star 
in response to the mass accretion. 

At the end of core hydrogen exhaustion, the remaining hydrogen 
envelope in the mass donor starts to expand and refill its Roche 
lobe. This leads to the onset of another thermal timescale mass 
transfer phase, that is, the Case\,AB phase, where most of the 
remaining hydrogen envelope is removed from the mass donor and the 
surface nitrogen abundance reaches the CNO equilibrium value 
and the surface helium mass fraction reaches $\sim$0.77. Since 
the mass ratio increasingly diverges away from unity to higher 
values, the orbital period of the binary also increases during 
this mass transfer phase.

We define Case\,A mass transfer to be occurring in our 
binary models when there is a non-zero mass transfer 
rate in the binary model and the Roche lobe filling star 
is still burning hydrogen at its centre. This means that, 
in addition to the fast and slow Case\,A phases, main 
sequence contact models and inverse mass transfer during 
core hydrogen burning from the initially less massive 
component to the initially more massive star will also 
be part of the Case\,A mass transfer phase. The corresponding 
configuration of a binary model where one star (both stars) 
fills its (their) Roche lobe(s) is called the semi-detached 
(contact) configuration.

One important point to keep in mind, however, is that for 
inverse slow Case\,A mass transfer from an initially less 
massive accretor, we do not change the notation of `donor' 
and `accretor' in our results. As such, we use the word 
`donor' to denote the initially more massive star that fills 
its Roche lobe first, and `accretor' to denote the initially 
less massive accretor star. Nevertheless, we point out 
whenever the contribution from inverse mass transfer binaries 
to the general population of slow Case\,A binaries is present. 

\subsection{Mass transfer efficiency in our models}
\label{mass_transfer_efficiency_section}

Here, we define and analyse the mass transfer efficiency 
during the fast and slow Case\,A mass transfer in our binary 
models. To avoid misunderstandings, it needs to be carefully 
defined, since both binary components may also lose mass to 
a stellar wind. Notably, the mass transfer efficiency in our 
models is a function of time, and below, we evaluate their 
time-averaged mass transfer efficiency during fast and slow 
Case\,A mass transfer.

When we designate the donor's wind mass-loss rate as 
$\dot{M}_{\rm w,d}$,  and the rate of mass transfer 
via Roche-lobe overflow as $\dot{M}_{\rm t}$, 
the total mass-loss rate from the donor stars is
\begin{equation}
    \dot{M}_{\rm d} = \dot{M}_{\rm w,d} - \dot{M}_{\rm t},
\end{equation}
where we define the mass transfer rate as a positive quantity 
and wind mass loss rate as a negative quantity. During fast 
Case\,A mass transfer, it is $\dot{M}_{\rm t}\gg -\dot{M}_{\rm w,d}$, 
such that the wind mass-loss rate from the donor is 
insignificant (c.f., Panel\,`d' of Fig.\,\ref{typical_Case_A}). 
During the slow Case\,A mass transfer phase however,
wind mass loss and mass transfer rate can become comparable 
for the most massive donors in our grid. For a model 
with ($M_{\rm d,i}$, $P_{\rm i}$, $q_{\rm i}$) 
$= (39.8\,\mso, 6.3\,{\rm d}, 0.800)$, we find that $\sim 0.5 
\mso$ of mass are lost from the donor via its stellar 
wind during the slow Case\,A mass transfer phase, 
while $\sim$2 $M_{\odot}$ of mass are transferred via 
the first Lagrangian point during the same time. 

The mass change of the accretor is obtained
by subtracting its wind mass loss from the amount transferred,
that is, $ \dot{M}_{\rm a} = \dot{M}_{\rm t} + \dot{M}_{\rm w,a}$, 
where $ \dot{M}_{\rm a} $ is the total rate of change of mass 
of the accretor and $\dot{M}_{\rm w,a}$ is the wind mass-loss 
rate of the accretor. Here, the steep increase in wind mass 
loss for accretors rotating near critical rotation can become 
important. As explained in Sect.\,\ref{methods}, this effect 
can lead to highly non-conservative mass transfer in our models,
such that the term $\dot{M}_{\rm w,a}$ can exceed the wind 
mass-loss rate of slowly rotating models by orders of magnitude.
This happens for spun-up mass gainers near critical rotation,
where the mass loss which prevents it from rotating faster than 
critical. For fully spun-up accretors, it is $\dot{M}_{\rm w,a} \simeq - \dot{M}_{\rm t}$.

As we model inefficient accretion like a stellar wind, we do 
not differentiate between the ordinary stellar wind and the 
mass loss required to prevent over-critical rotation in defining 
our mass transfer efficiency $\eta$, and set 
\begin{equation}
    \eta = \frac{\dot{M}_{\rm a}}{\dot{M}_{\rm t}}.
\end{equation}
In this way, the mass transfer efficiency can range from 0 for 
critically rotating accretors to $\sim$1 for tidally locked 
and/or slowly rotating accretors. 

Figure\,\ref{mass_trans_eff_fast_period} shows the time-averaged 
mass transfer efficiencies during fast (top and 
middle panel) and slow Case\,A (bottom panel). The top 
panel shows the mass transfer efficiencies of all our 
systems that survive fast Case\,A mass transfer as a binary. 
It excludes models that we expect to merge very soon after 
the onset of the fast mass transfer (models marked in green 
in Fig.\,\ref{1.350_summary}). We see that mass transfer 
efficiencies, from fully conservative to fully non-conservative 
are realised by our models. While the distribution is 
rather flat for most of the mass transfer efficiency range, 
it shows distinct peaks near $\eta=1$ and $\eta=0.05$. 

About $\sim$35\% of our models undergo nearly conservative 
fast Case\,A mass transfer. The initial period information 
(colour coding in Fig.\,\ref{mass_trans_eff_fast_period}) 
reveals that this happens for the shortest period binaries 
in our model grid. These systems remain effectively tidally 
locked at all times, such that their accretors are not spun 
up. On the other hand, models with initial orbital periods 
larger than $\sim$4\,d show mass transfer efficiencies below 
40\%, and the initially widest binaries evolve with accretion 
efficiencies below $\sim$15\%. This reflects the decreasing 
strength of the tidal interaction for larger orbital separations.

The middle panel of Fig.\,\ref{mass_trans_eff_fast_period} 
shows the mass transfer efficiency during fast Case\,A for 
models that also survive the slow Case\,A mass transfer 
phase without merging. Of these, we find that 99.2\% also 
survive the ensuing Case\,AB mass transfer. Models in our 
grid that survive the entire Case\,A mass transfer typically 
have low average fast Case\,A mass transfer efficiencies. 
Comparing the top and middle panels, we see that a large 
number of models that undergo conservative fast Case\,A 
mass transfer merge during the slow Case\,A mass transfer 
phase. These models, originating from very short initial 
periods that eventually undergo L2 overflow, actually spend 
a considerable amount of time in the slow Case\,A mass 
transfer phase before merging (compare peak at mass transfer 
efficiency near unity between the left and right panel). 
We find that some of these models also go through a 
nuclear-timescale contact phase before merging \citep[see also][]{Menon2021}. 
Quantitatively, the weighted fraction of Case\,A models 
contributing to the middle panel is $\sim$63\% smaller than 
that in the top panel. 

The time-averaged mass transfer efficiency during slow 
Case\,A mass transfer, for the models which survive the 
slow Case\,A (bottom panel of Fig.\,\ref{mass_trans_eff_fast_period}), 
is generally high. The slow Case\,A mass transfer occurs 
at the nuclear timescale and the mass transfer rate is 
also much lower, of the order of 10$^{-7}$ $M_{\odot}$/yr. 
For the model with ($M_{\rm d,i}$, $P_{\rm i}$, $q_{\rm i}$) 
= (39.8\,$\mso$, 6.3\,d, 0.800), we find that $\sim$0.3 
$M_{\odot}$ of mass are lost from the accretor via stellar 
winds during the slow Case\,A mass transfer phase. Hence, 
the mass accreted by the mass gainer during the slow Case\,A 
mass transfer phase in this model is $\sim$1.7 $M_{\odot}$. 
The mass transfer efficiency of this binary during slow 
Case\,A is about 85\%. We show later that the majority 
of the accretors in the slow Case\,A phase are tidally 
synchronised. A comparison of time spent in the slow 
Case\,A phase between models that merge during the slow 
Case\,A phase vs models that survive the slow Case\,A 
phase reveals that our models predict $\sim$70\% of Algols 
binaries are expected to merge during their main sequence.

Fig.\,\ref{eta_fast_slow_ab} shows the mass transfer efficiency 
of individual models in our grid for an initial donor mass 
of $\sim$16 $M_{\odot}$ and $\sim$40 $M_{\odot}$. We see that 
the mass transfer efficiency of the thermal timescale fast 
Case\,A and Case\,AB phase is low for the most of the 16 
$M_{\odot}$ models. Owing to the increased tidal strength 
of higher mass donors, the mass transfer efficiency is near 
unity for the shortest period models of the 40 $M_{\odot}$ 
slice. We also see the dividing boundary as a function of 
orbital period and mass ratio at which the tidal strength 
is unable to counteract the spin-up of the accretor star.
A lot more number of models undergo conservative slow 
Case\,A mass transfer, owing to the nuclear timescale mass 
transfer rate. Nevertheless, we again see a reasonably 
clear boundary between efficient and nonefficient mass 
transfer in the slow Case\,A mass transfer phase too. 

We note that the amount of mass lost and gained during the 
slow Case\,A phase can be up to $\sim$6 $M_{\odot}$ for the 
models (Fig.\,\ref{slow_Case_A_mass_lost_gained}) with an 
initial donor mass of 40 $M_{\odot}$, due to efficient mass 
accretion during the slow Case\,A phase. Another $\sim$4\,$M_{\odot}$ 
of mass can be lost during the Case\,AB mass transfer phase 
(Fig.\,\ref{fast_Case_AB_mass_lost_gained}) in our models. 
Also, a very small number of models with highest initial 
donor masses can undergo efficient Case\,AB mass transfer. 

\subsection{Life-time of Case\,A mass transfer \label{sec:tau_case_a} }

\begin{figure}
\centering
\includegraphics[width=\hsize]{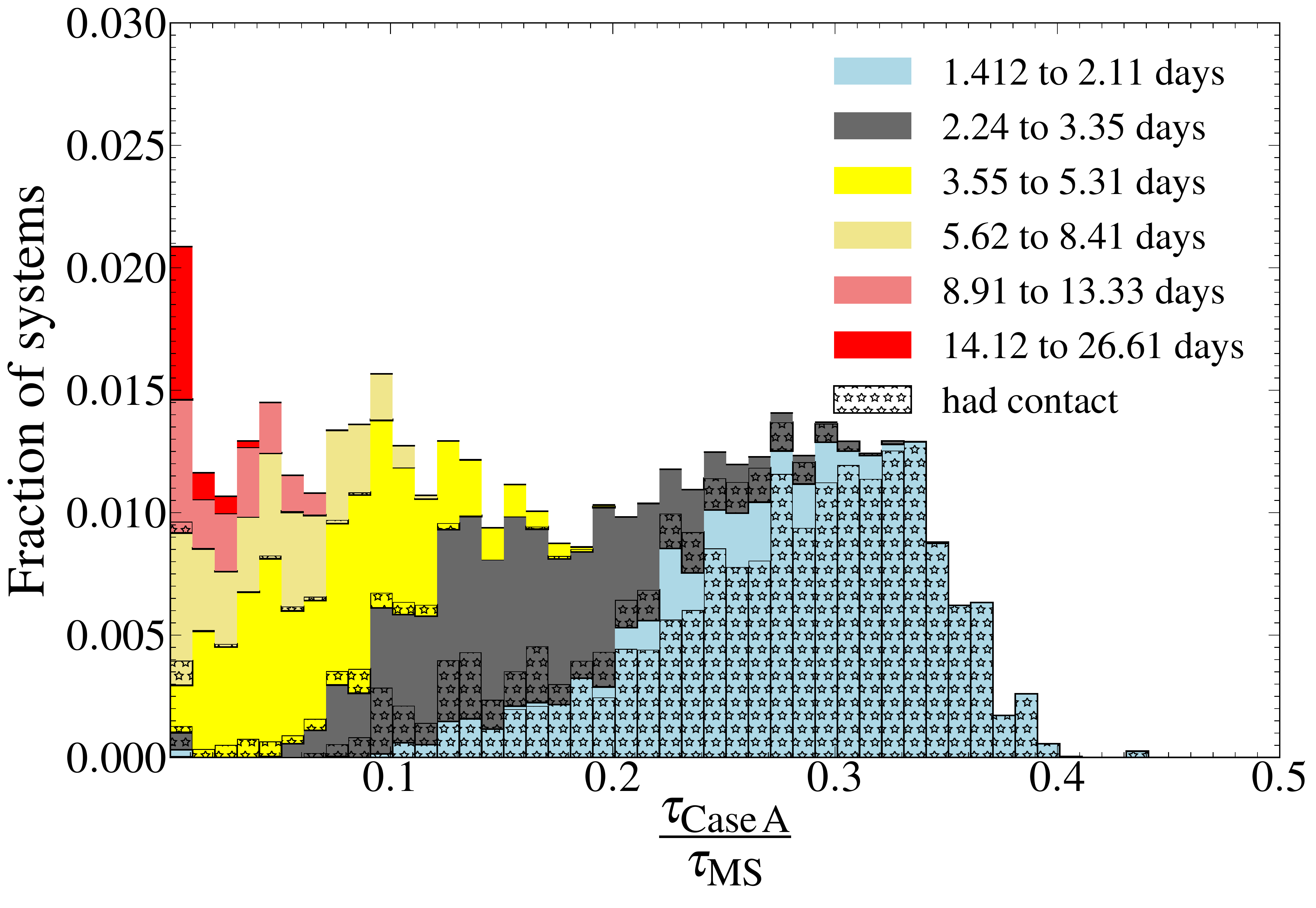}
\caption{Distribution of the fractional main sequence lifetime spent in the Case\,A mass transfer phase ($\tau_{\rm Case\,A} / \tau_{\rm MS}$; see text for the exact definition) of models 
with $\tau_{\rm Case\,A} / \tau_{\rm MS} > 0.001$.
The colour coding differentiates between models arising from different ranges of initial orbital periods. The star hatching in each fractional main sequence lifetime bin denotes the relative number fraction of models in that bin that go through a contact phase during Case\,A mass transfer.}
\label{F2_distribution}
\end{figure}

Here, we look at the amount of time spent by the binary models in the 
Case\,A mass transfer phase. Figure\,\ref{F2_distribution} assesses the 
fraction of the main sequence lifetime our models spent in the Case\,A 
mass transfer phase, $\tau_{\rm Case\,A}$, with the Case\,A mass transfer 
phase as defined in Sect.\,\ref{CaseA_example}. Here, we define the 
main sequence lifetime ($\tau_{\rm MS}$) of a binary model as the 
hydrogen burning lifetime of the binary component which completes 
hydrogen burning first, or, for those binary models which merge during 
Case\,A evolution, the main sequence lifetime of a single star with 
half the total mass of the binary model.  The number fraction $f$ in 
each fractional main sequence ($\tau_{\rm Case\,A}/\tau_{\rm MS}$) 
bin [$a,\,b$] is given as
\begin{equation}
f(a<\tau_{\rm Case\:A}/\tau_{\rm MS}<b) = \frac{\sum_{\rm m=1}^{\rm N} \delta_{\rm ab,m} W_{\rm m}}{\sum_{\rm m=1}^{\rm N} W_{\rm m}}\:,
\label{F2_dist_eqn}
\end{equation}
where $\delta_{\rm ab,m}$ = 1 if the fraction of the 
main sequence lifetime spent by the model in Case\,A 
mass transfer phase is between $a$ and $b$, and 
$\delta_{\rm ab,m}$ = 0 otherwise.

Figure\,\ref{F2_distribution} only includes models that spend 
more than 0.1\% of their main sequence lifetime in the Case\,A 
mass transfer phase, which is of the order of the thermal 
timescale of the mass donors. This removes about 60\% of all 
models that undergo Case\,A mass transfer, whose fate is
to merge during fast mass transfer (green models in Fig.\,\ref{1.350_summary}). 
Figure\,\ref{F2_distribution} shows that the shorter period 
binaries spend more time in the Case\,A mass transfer phase. 
Many of these models also go through a contact phase as 
denoted by the star hatching. The hatched versus unhatched 
part of each bin in this figure indicates the fraction of 
the models in any bin go through a contact phase, but is not 
indicative of the time spent in the respective phases. 

We find that Case\,A models can spend up to a third 
of their main sequence lifetime in the semi-detached configuration. Usually, 
once Roche-lobe overflow occurs during core hydrogen burning in 
binary, the binary spends most of the remaining main sequence 
lifetime of the donor in the semi-detached configuration 
(Fig.\,\ref{typical_Case_A}). However, we find that binaries 
in which Roche-lobe overflow occurs very early in their main sequence lifetime, 
the binary enters into a contact configuration and 
eventually merges via L2 overflow. This is why 
we do not find our Case\,A models to spend more than 40\% 
of their main sequence lifetimes in the slow Case\,A 
phase. We assess the number of semi-detached massive binaries 
in the LMC which is expected from our models in Sect.\,\ref{number}.

\section{Observable properties of semi-detached models}
\label{slow_Case_A}

Here, we evaluate the distributions of observable properties of 
our binary models while they are in a semi-detached configuration. 
Besides our theoretical predictions, many of the plots in this 
section already include information about observed semi-detached 
binaries. However, we perform a comparison of our results with 
observations separately, in Sect.\,\ref{observations_section}. 

\subsection{Orbital period and mass ratio distribution}
\label{pq}

\begin{figure*}[t]
\centering
\includegraphics[width=0.47\hsize]{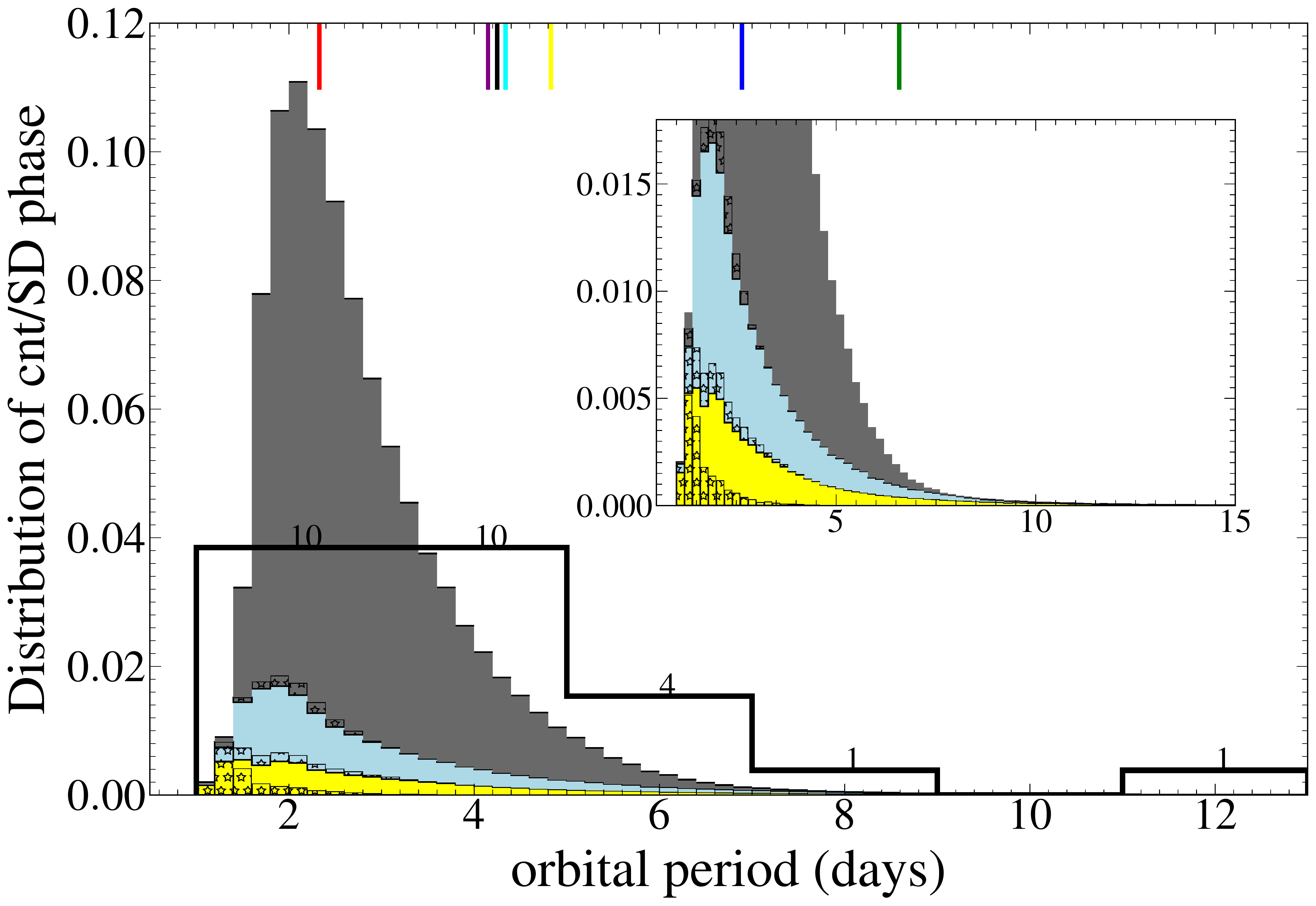}
\includegraphics[width=0.47\hsize]{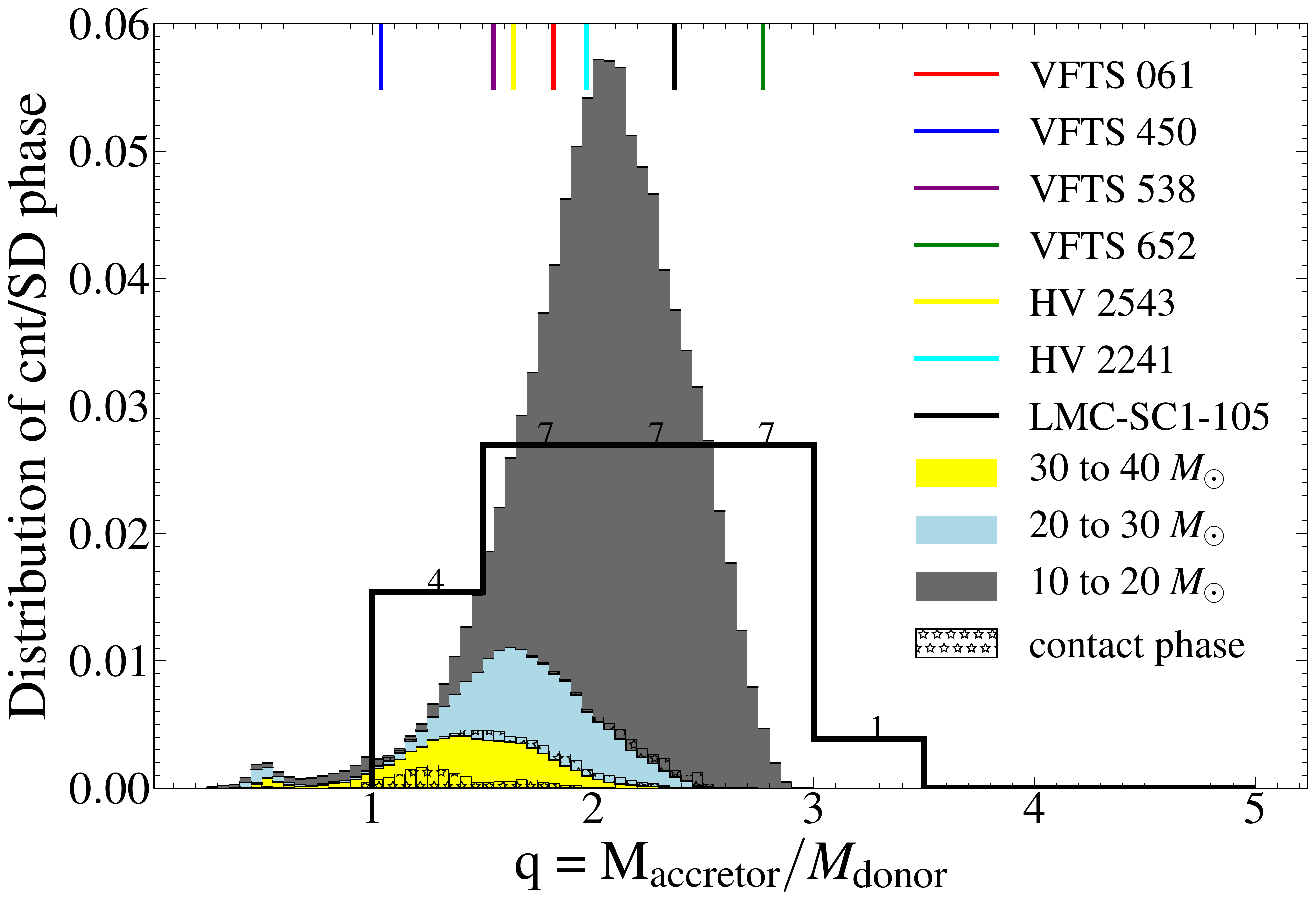}
\caption{Distribution of orbital periods (left panel) and mass ratios (right panel) of our models during the semi-detached phase (unhatched) or contact phase (hatched with stars) of Case\,A mass transfer. Colour coding marks three different initial mass ranges, as indicated.
The ordinate values are normalised such that the sum of the number fractions in all bins equals unity.
Vertical lines at the top of the plots denote the orbital periods and mass ratios 
of observed Algol systems in the LMC (Table\,\ref{table}), identified through their colour (see legend, same for both panels). The inset in the left panel shows a zoomed-in distribution of the yellow and blue models. The black step-histogram denotes the distribution of orbital periods and mass ratios of 26 observed massive Algol systems in the Milky Way (Table \ref{table_galaxy}). It is normalised such that the area under both the histograms is same. The number of stars contributing to each bin is given on top of each bin. 
}
\label{observed_period}
\label{observed_q}
\end{figure*}

Figure\,\ref{observed_period} shows the predicted orbital 
period distribution of our Case\,A binary models during the 
semi-detached (unhatched) and contact (hatched with stars) 
phases of their evolution. The histogram is weighted by the 
binary birth probability (see Sect.\,\ref{Wm}), and their 
time spent in each orbital period bin during their semi-detached 
or contact phase. Comparing the contributions from the 
semi-detached phase and the contact phase shows that $\sim$96\% 
of the interacting Case\,A binaries in our grid are in the 
semi-detached configuration. However, our assessment of the 
contact binary parameter space is incomplete as our lowest 
initial orbital period is $\sim$1.4\,d. Including models with 
shorter initial periods leads to a larger predicted fraction 
of contact systems \citep[see Sect. \ref{number} and][]{Menon2021}. 

The drop in the period distribution of the semi-detached systems 
towards the shortest periods in Fig.\,\ref{observed_period} is,
however, not due to our lower initial period cut-off. We see from 
Fig.\,\ref{1.350_summary} that the shortest period models in our 
grid do not survive the Case\,A mass transfer phase as binaries. 
Amongst these presumable mergers, the models with shortest orbital 
periods spend increasingly more time in the contact than in the 
semi-detached configuration. Consequently, our semi-detached model 
population is essentially complete, for the investigated mass range. 

For models with longer orbital periods, Case\,A mass transfer starts 
comparatively later during the main sequence evolution of the donor. Once 
the slow Case\,A mass transfer starts, it continues until the end of 
the main sequence evolution of the donor. Since the histogram is weighted by 
the lifetime of the models in the Case\,A mass transfer phase, the 
longer period models do not contribute to the same extent as 
the shorter period models. Moreover, fewer long orbital period models 
undergo Case\,A mass transfer (depending on the initial donor mass, 
c.f. Fig.\,\ref{1.100_summary}). Hence, the distribution drops off at 
higher orbital periods. The contribution from models in the contact 
phase keep increasing towards shorter periods and our grid is not ideal 
to investigate this phase. We predict to find most semi-detached 
systems around an orbital period of $\sim$1.5-4\,d. 

Figure\,\ref{observed_q} (right panel) shows the predicted distribution of the 
mass ratios of our Case\,A models during the semi-detached and 
contact phases. Since the mass ratio is inverted during  
fast Case\,A, the 
models do not spend much time at mass ratios below unity. A significant 
fraction of time is spent in the slow Case\,A mass transfer phase, 
where the mass ratio has already inverted. Hence, we see a 
large peak at mass ratios of $\sim$2. For more massive systems, the peak moves to smaller
mass ratios due to the larger core mass fraction at higher mass.

The smaller peak near $q=0.5$ arises from the shortest period models 
where a thermal timescale contact occurs, followed by a nuclear timescale 
inverse slow Case\,A mass transfer. For example, a system with parameters 
$(M_{\rm 1,i}, q_{\rm i}, P_{\rm orb,i}) = (28.3\mso, 0.800, 1.41\,{\rm d})$ 
evolves as follows. The initially more massive star undergoes fast 
Case\,A mass transfer, followed by slow Case\,A. During the slow 
Case\,A mass transfer episode, the then more massive accretor fills 
its Roche lobe and contact occurs, which initiates inverse mass 
transfer. Shortly after the onset of contact, the initially more 
massive star shrinks below its Roche lobe radius, and the inverse 
slow Case\,A mass transfer occurs in a semi-detached configuration. 
Their expected number is small. However, they open a small chance 
to find Algol systems with mass flow from the currently more 
massive to the currently less massive star \citep[e.g. VFTS 176, see][]{mahy2019b}. 

\begin{figure}
\centering
\includegraphics[width=\hsize]{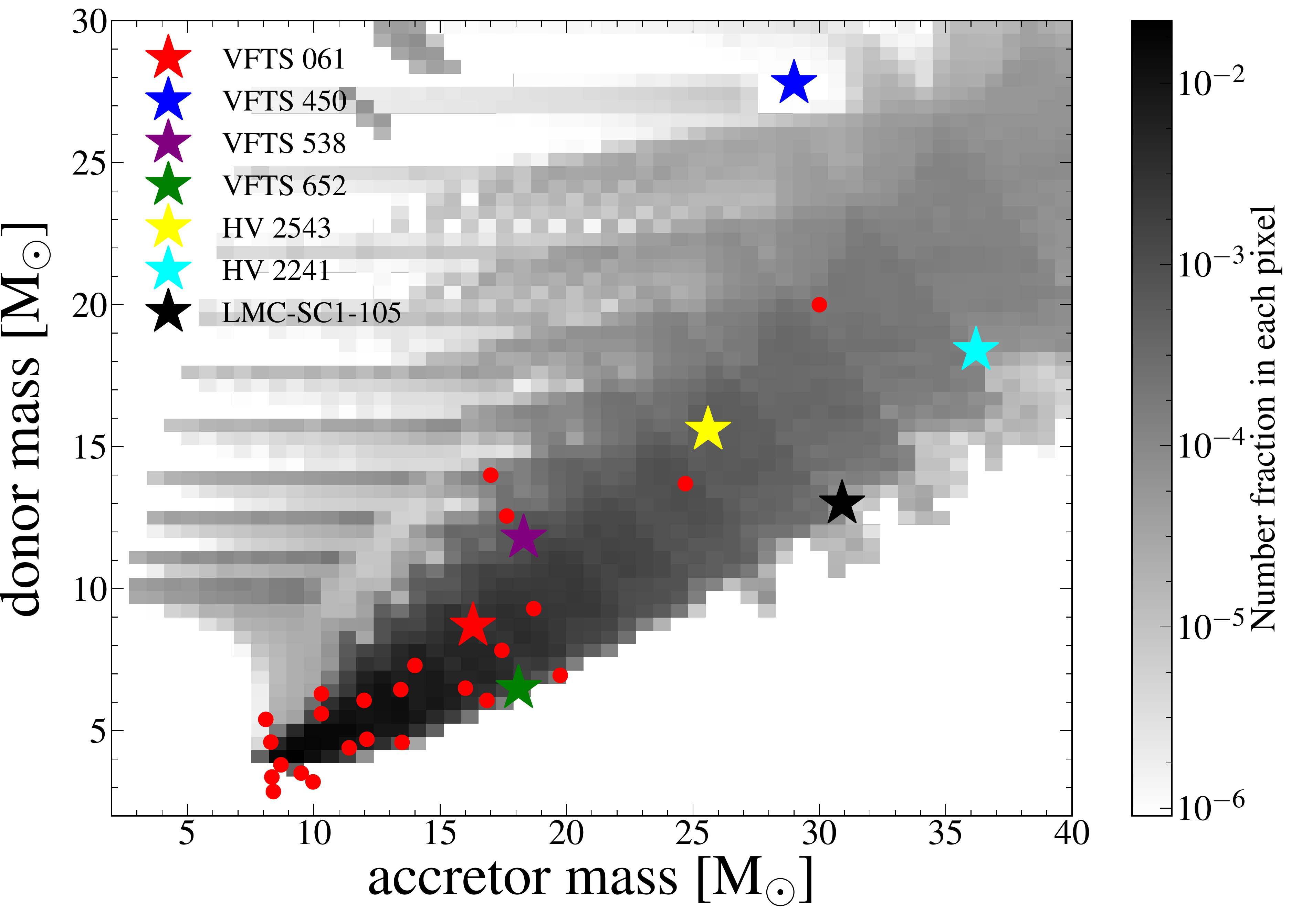}
\caption{Probability distribution of the donor and accretor masses that is predicted to be observed in the semi-detached configuration of the Case\,A mass transfer phase based on the model grid. The different coloured `stars' denote the position of observed semi-detached systems in the LMC (Table \ref{table}). Grey-scale: See description in Fig.\,\ref{hessHe}. The red circles denote the parameters for the Galactic Algol systems (Table \ref{table_galaxy}).}
\label{observed_masses}
\end{figure}

Figure\,\ref{observed_masses} shows distribution of the absolute  
donor and accretor masses during the semi-detached phase. We see 
that the donor masses are significantly lower than the accretor 
masses, which is consistent with 
the fact that mostly the Roche-lobe filling star is the lower mass 
star of the binary. The highest probabilities for the lowest 
masses come from the combined effect of the IMF weight on 
the distribution functions and the fact that the lower mass 
donors spend more time in the Case\,A mass transfer phase. We see 
a broader range of accretor masses for a given donor mass since more 
massive systems avoid merging during Case\,A evolution for a larger
initial mass ratio range (cf., Fig.\,\ref{1.100_summary}).

Notably, the increase in orbital period after the Case\,AB phase 
(Fig.\,\ref{periodchange}) is highest in the lower initial donor 
mass models because of the smaller core size of the lower mass 
donors. Owing to the higher envelope to core ratio, a comparatively 
larger fraction of mass is transferred in the lower mass models, 
which increases the mass ratio of the binary and the orbital 
period also increases more. 

\subsection{Surface abundances}
\label{sec: surf_abun_models}

\begin{figure*}
{\centering
\includegraphics[width=0.48\hsize]{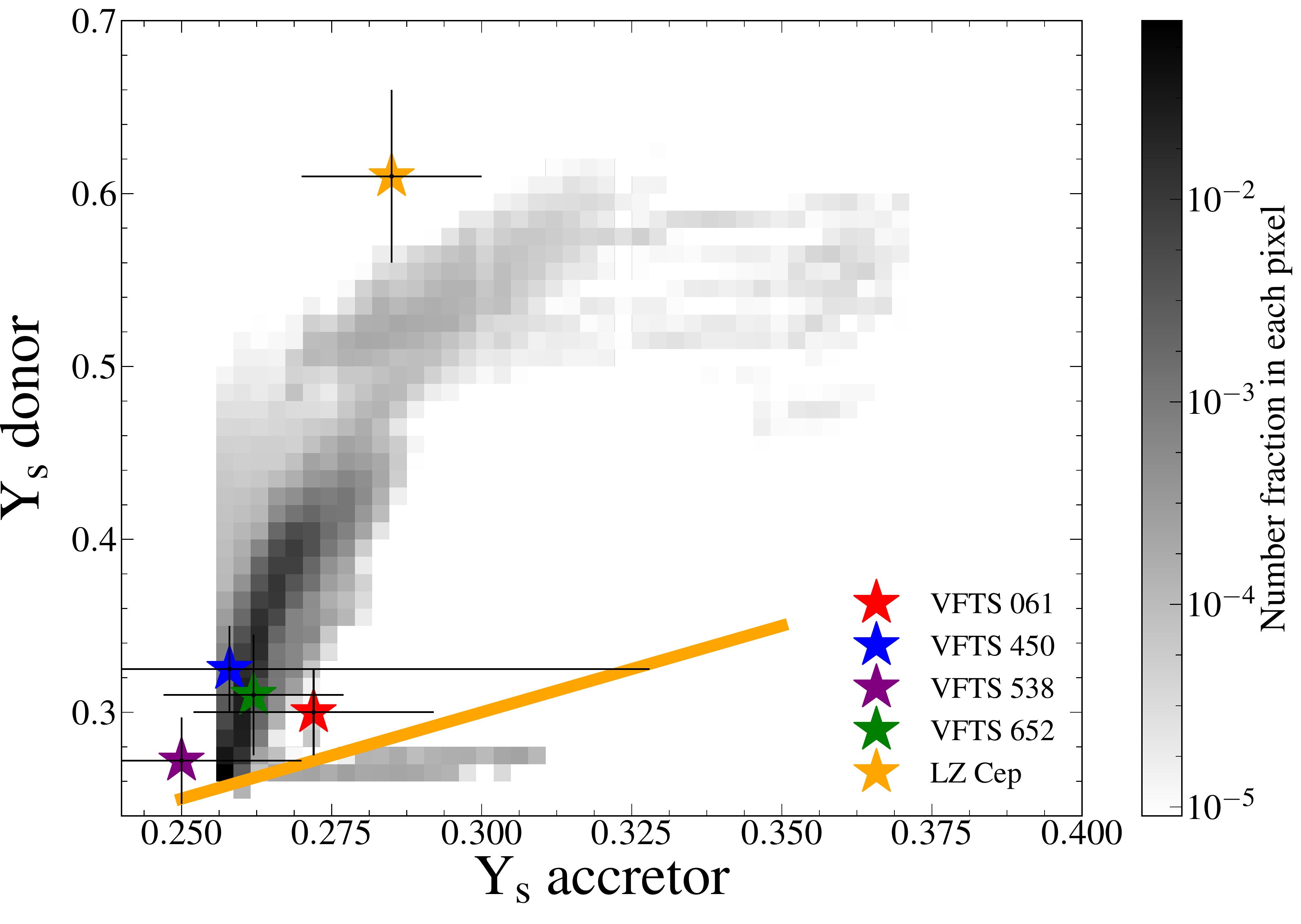}
\includegraphics[width=0.48\hsize]{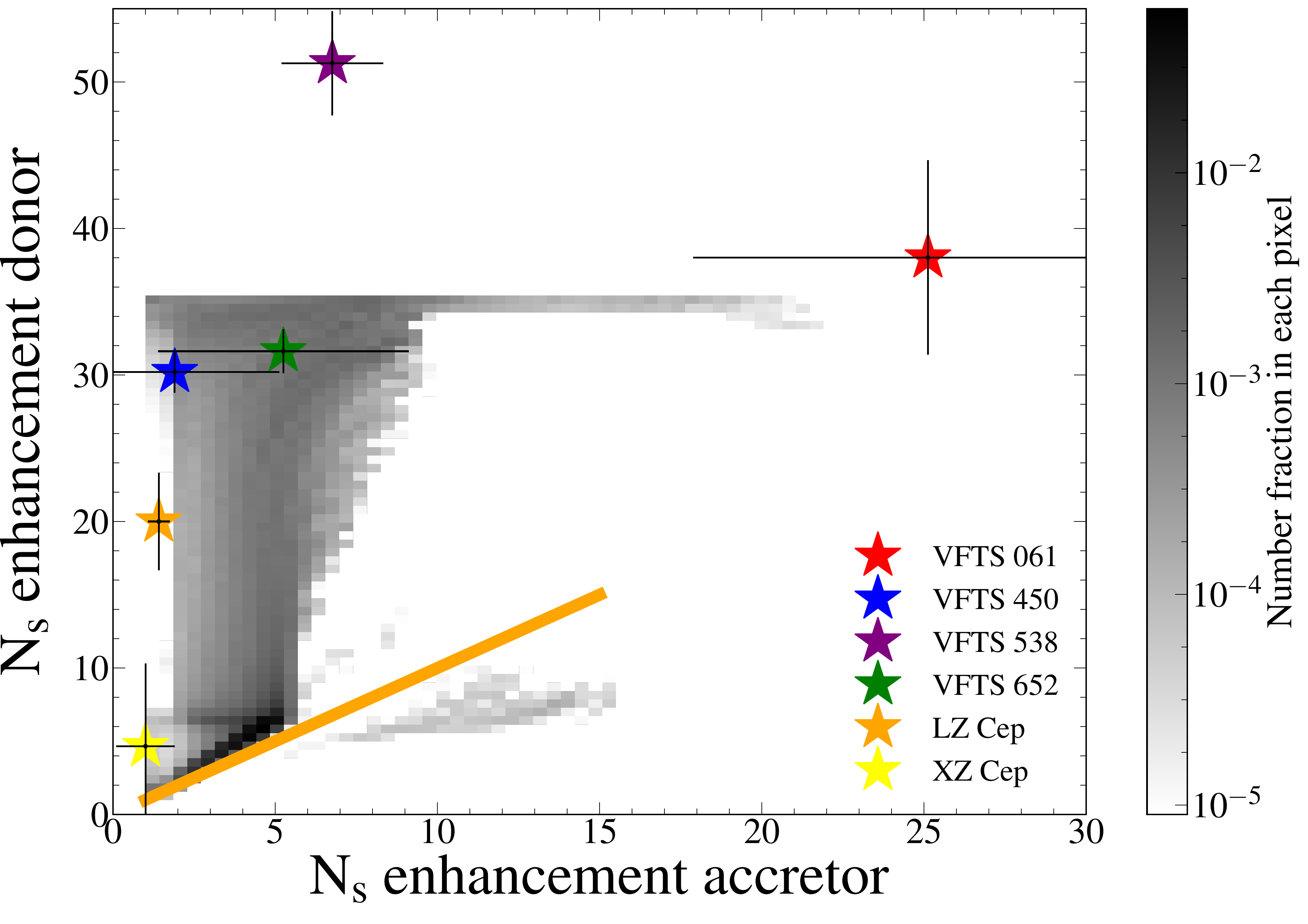}}
\caption{Probability distribution of the surface helium mass fraction (Y$_{\rm s}$, left panel) and surface nitrogen mass fraction (N$_{\rm s}$) enhancement (right panel, see text) of donor vs the accretor that is predicted to be observed in the semi-detached configuration of the Case\,A mass transfer phase based on the model grid. The different coloured `stars' with error bars denote the position of the semi-detached systems of the TMBM survey and Galaxy. The surface nitrogen mass fraction enhancement of the LMC systems are evaluated w.r.t. to the LMC nitrogen abundance baseline, while the enhancement for LZ Cep and XZ Cep are evaluated w.r.t to the Solar baseline. The grey-scale gives the probability fraction in each pixel. The total probability is normalised such that the integrated sum over the entire area is 1. The orange line indicates where the surface helium mass fraction or surface nitrogen enrichment of the donor and accretor is the same.}
\label{hessHe}
\end{figure*}

\begin{figure*}
{\centering
\includegraphics[width=0.48\hsize]{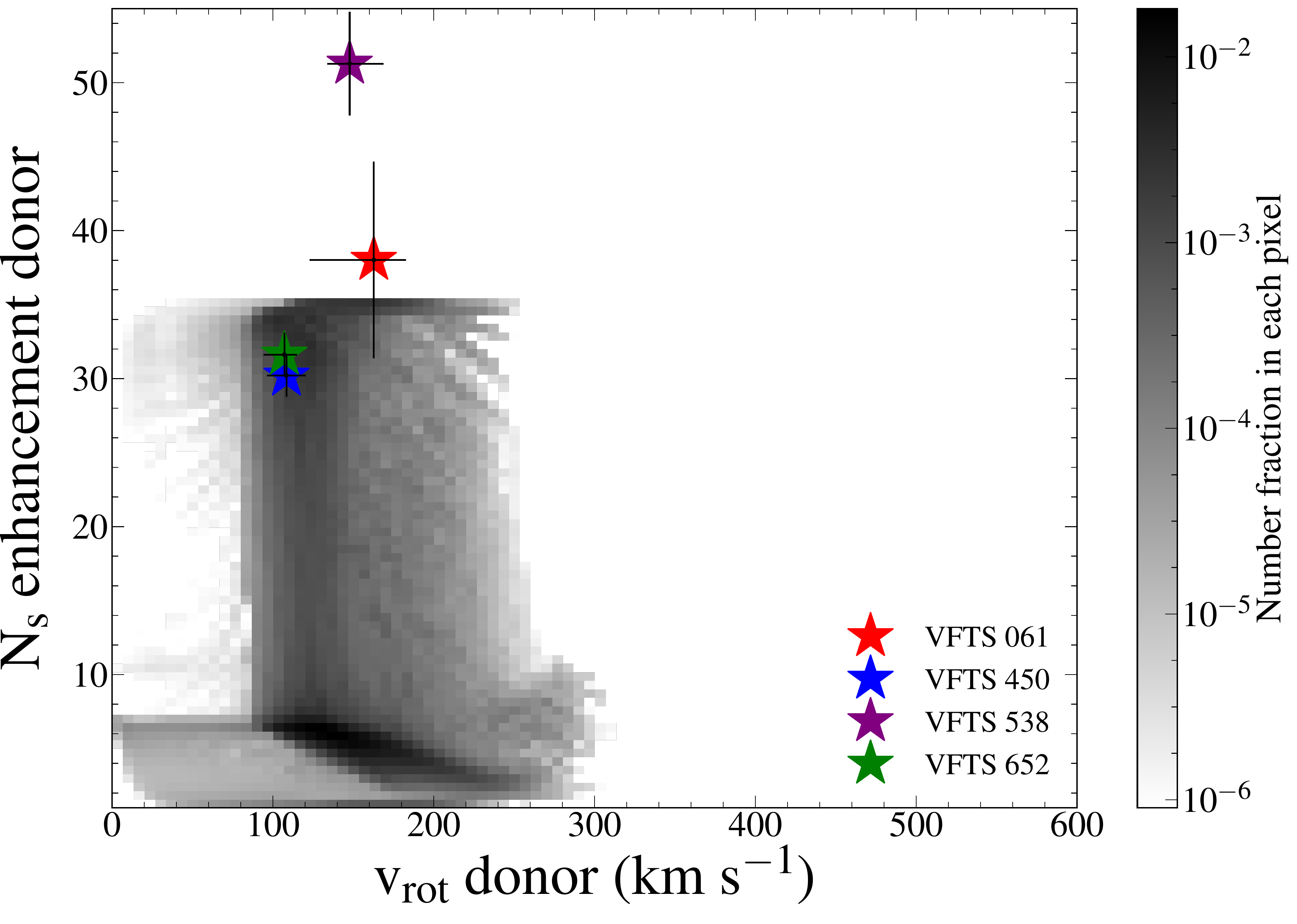}
\includegraphics[width=0.48\hsize]{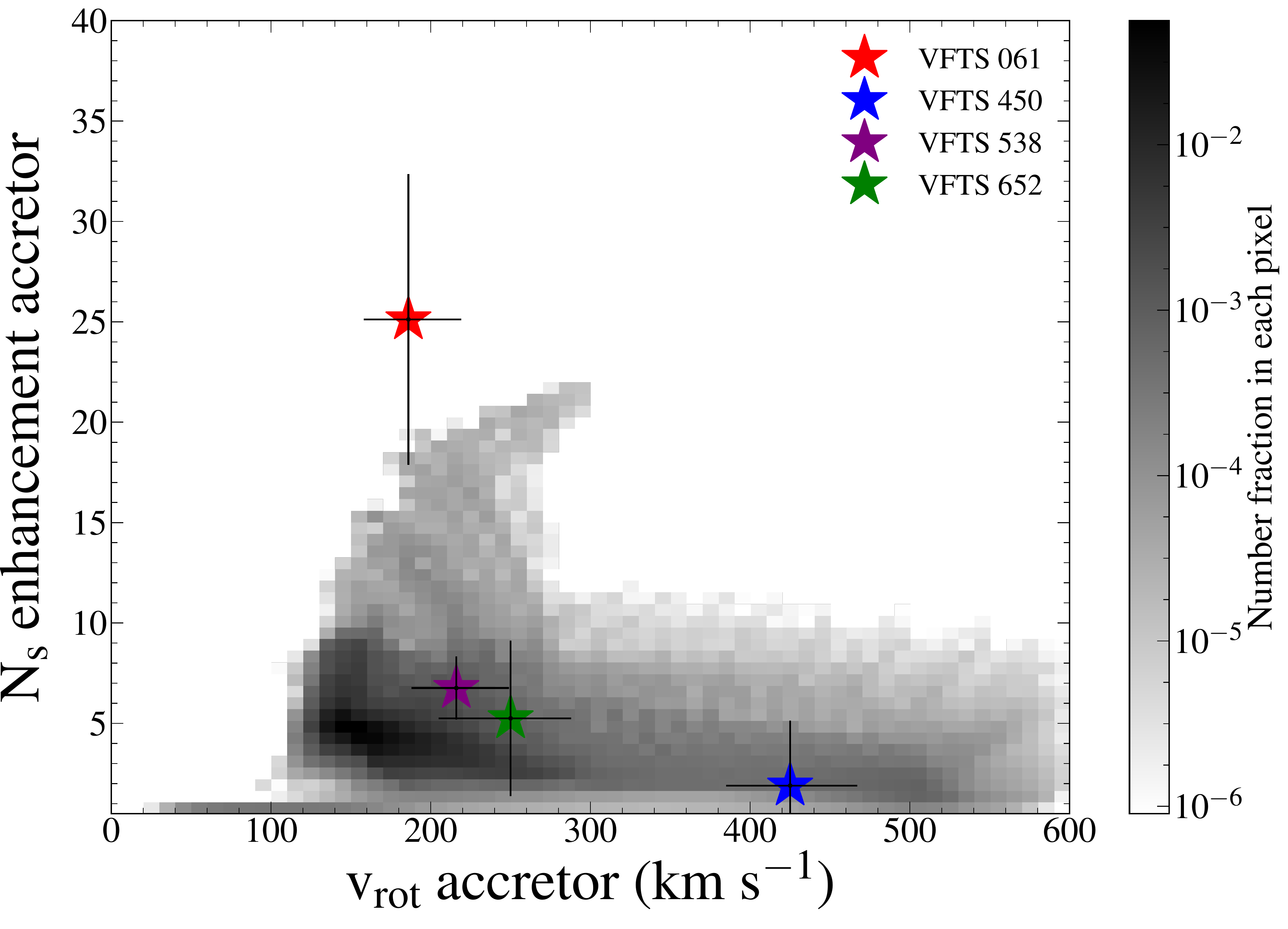}}
\caption{Probability distribution of the surface nitrogen mass fraction enhancement vs rotational velocity of the donor (\textit{left panel}) and accretor (\textit{right panel}) during the semi-detached phase. The different coloured stars with error bars denote the position of the semi-detached systems of the TMBM survey (Table \ref{table_velocities}). Grey-scale: See description in Fig.\,\ref{hessHe}.}
\label{HessNPrimVsVrotPrim}
\end{figure*}

In massive stars, the enrichment of helium and nitrogen is related, 
because both are products of the CNO-cycle \citep{Kippenhahn1990}. 
Within the convective core, CNO equilibrium is established quickly, 
such that nitrogen obtains its CNO equilibrium value while only 
little hydrogen has been converted into helium. However, the convective 
core of massive main sequence stars is receding with time, and leaves 
a transition region within which the hydrogen abundance drops from 
its initial value to the value currently inside the convective core. 
In stars above $\sim 15\mso$, this transition layer is slowly mixed 
with overlying unprocessed matter by semiconvection, which extends 
the transition layers to regions above the initial convective core, 
and leads to layers in which both, helium and nitrogen are enhanced, 
but with a nitrogen abundance below the CNO equilibrium value 
\citep{langer1991}. In addition to that, the temperature above even 
the extended transition layer is high enough to allow for incomplete 
CNO burning, where some carbon is transformed into nitrogen, but 
oxygen is not. In our donor stars, all these layers appear successively 
at the surface and determine the abundances of the matter accreted 
by the accretor star, whose entire envelope undergoes thermohaline 
mixing as soon as even a small helium enrichment is present in the 
accreted matter. 

Figure\,\ref{hessHe} shows the surface helium mass fraction (Y$_{\rm s}$, 
left panel) and the ratio of the surface nitrogen mass fraction (N$_{\rm s}$) 
to the initial surface nitrogen mass fraction (right panel, which we call 
surface nitrogen mass fraction enhancement), of our donor and accretor 
models in the semi-detached configuration. It shows that the mass donors 
emerge from fast Case\,A mass transfer with essentially all of the unenriched part of 
their envelopes removed. A mild nitrogen enrichment is present in almost 
all donors, which results from CN-processing occurring above the H/He-transition 
layers in the donor star. At the beginning of the semi-detached phase, 
helium is still practically unenriched at the mass donor's surface (cf., 
Panel\,`e' of Fig.\,\ref{typical_Case_A}).  

During the further evolution, mass transfer from the donor during the 
slow Case\,A phase gradually brings layers from the hydrogen-helium 
gradient region of the donor to its surface, which raises its surface 
helium mass fraction appreciably (see also Fig.\,\ref{typical_Case_A}). 
However, the helium surface mass fraction of the donor stars usually remain 
below $Y_{\rm s}\simeq0.45$. Correspondingly, the surface nitrogen 
mass fraction enhancement increases and eventually reaches the CNO 
equilibrium value of $\sim$35 for the LMC during the slow Case\,A phase. 
This indicates that during slow Case\,A mass transfer, layers start to be 
uncovered which were part of the convective core earlier in the evolution.

The enrichment of the surface of the mass gainers is mediated by 
thermohaline mixing, and therefore remains significantly smaller 
than that of the donor. We see from Fig.\,\ref{hessHe} that their 
helium mass fraction remains mostly below $Y_{\rm s}\simeq0.29$ 
(the initial value is 0.256), whereas the nitrogen enrichment does 
not exceed a factor of 10 in the vast majority of cases. Core 
hydrogen burning in the accretors has already created a strong 
mean molecular weight gradient by the time the Roche-lobe overflow 
begins, which prevents a strong nitrogen enrichment due to rotational 
mixing even in rapidly rotating accretors \citep{wang2020}.

In a small fraction of our models, the mass gainers reach rather high enrichment,
that is, helium surface mass fractions in the range $0.30\dots 0.38$, and
nitrogen mass fraction enhancements of up to 22 (Fig.\,\ref{hessHe}). In 
Appendix\,\ref{surface_abundance_additional_figures}, 
we provide similar plots, but separately for three mass bins (Fig.\,\ref{hessHeN_10-20-30-40}).
This shows that the highly enriched mass gainers are restricted to systems
with larger initial masses. 

From investigating the Case\,A models that survive or eventually 
merge during slow Case\,A, separately, we find that the latter 
contain mass donors with surface nitrogen mass fraction enhancement 
below $\sim$8 for $\sim$75\% of the time. Models that survive the 
slow Case\,A phase have donors whose surface nitrogen mass 
fraction enhancement is above $\sim$8 for $\sim$90\% of the 
time they spend as semi-detached systems. Since both groups of 
binaries experience a nuclear timescale mass transfer phase, 
we expect observational counterparts to both these kinds of 
systems. In Sect.\,\ref{mass_transfer_efficiency_section}, 
we had shown that models that eventually merge during slow 
Case\,A phase contribute to $\sim$70\% of the predicted 
observable binaries in the semi-detached phase. 

Now, the fraction of models that survive vs merge during the 
slow Case\,A phase depends on the response of accretor radius 
to mass accretion. We know that accretors with inefficient 
semiconvective mixing (as in our models) are larger in radius 
than accretors with efficient semiconvective mixing \citep[c.f. 
model 47 and 48 of][]{wellstein2001}. This implies that our 
models are more likely to enter contact during slow Case\,A 
mass transfer than models with efficient semiconvection \citep[
such as the models of][]{Menon2021}. Hence, the predicted 
fraction of mergers during the slow Case\,A phase depends on, 
amongst other factors, the semiconvective efficiency. 

Our binary models predict that in a sample of 100 observed 
massive Algol binaries,  at least 52.5 (75\% of 70\% of 100 
binaries) of the donors should show surface nitrogen mass 
fraction enhancement less than a factor of 8. On the other 
hand, at most 27 (90\% of 30\% of 100) donors can show surface 
nitrogen mass fraction enhancement greater than a factor of 
8. If these number cut-offs (`at least 52.5' and `at most 27') 
in the observed sample are largely different from our predictions, 
we can use this as another observational constraint to constrain 
the efficiency of semiconvection used in stellar models, via 
the close massive binary evolution channel \citep[see also][]{schootemeijer2019}. 

The small group of models below the orange lines in Fig.\,\ref{hessHe} arises
from the small number of models undergoing inverse slow mass transfer (cf.,
Sect.\,\ref{pq}). In these models, mass flows back to the original donor star
from its currently more massive and less enriched companion.
These binaries arise from very 
short period models and eventually merge via L2 overflow after a nuclear 
timescale inverse slow Case\,A mass transfer phase followed by a contact phase. 

\subsection{Effects of stellar rotation on nitrogen mass fraction enhancement}

\begin{figure*}
\begin{minipage}{0.5\linewidth}
\includegraphics[width=\hsize]{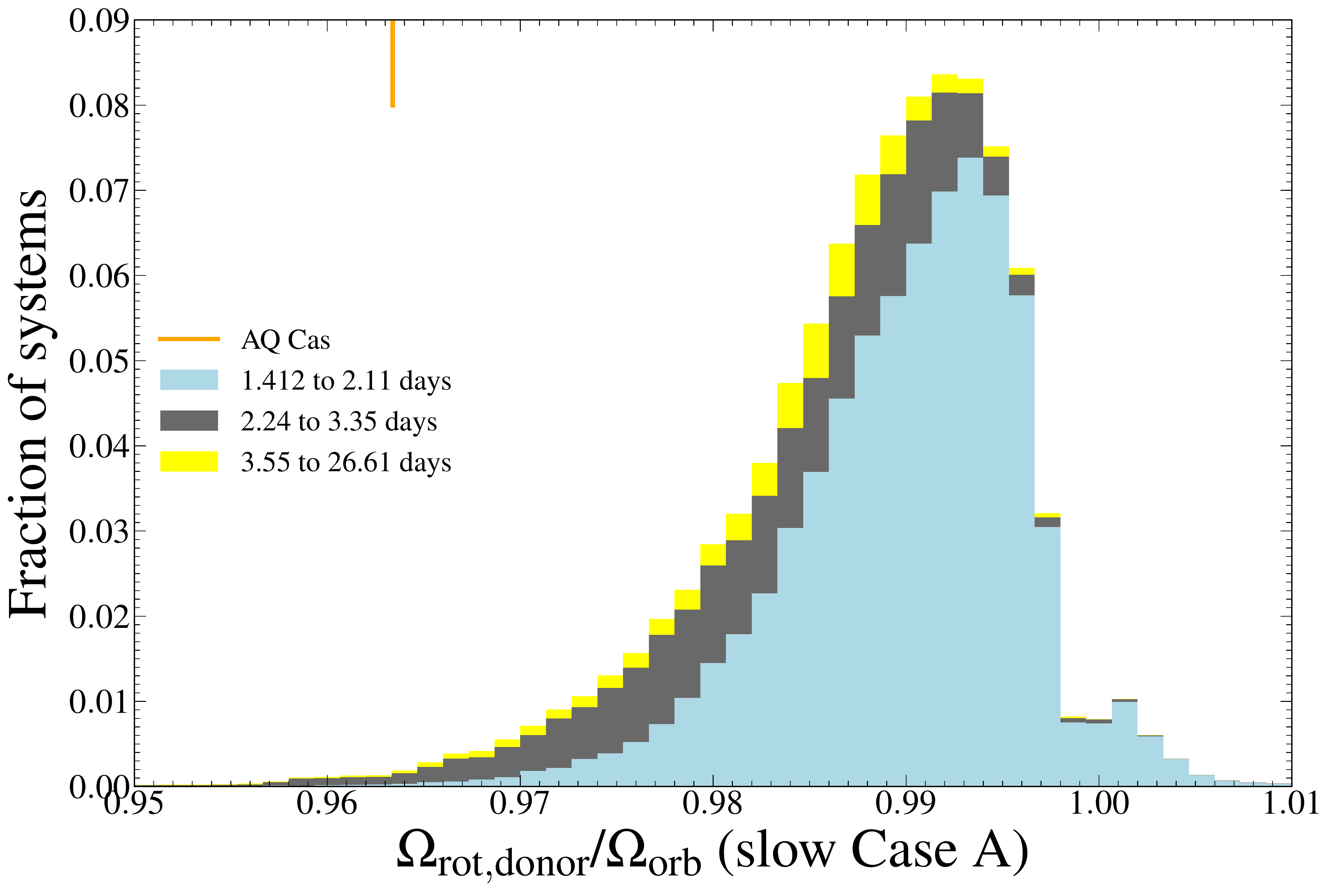} 
\end{minipage}
\begin{minipage}{0.5\linewidth}
\includegraphics[width=\hsize]{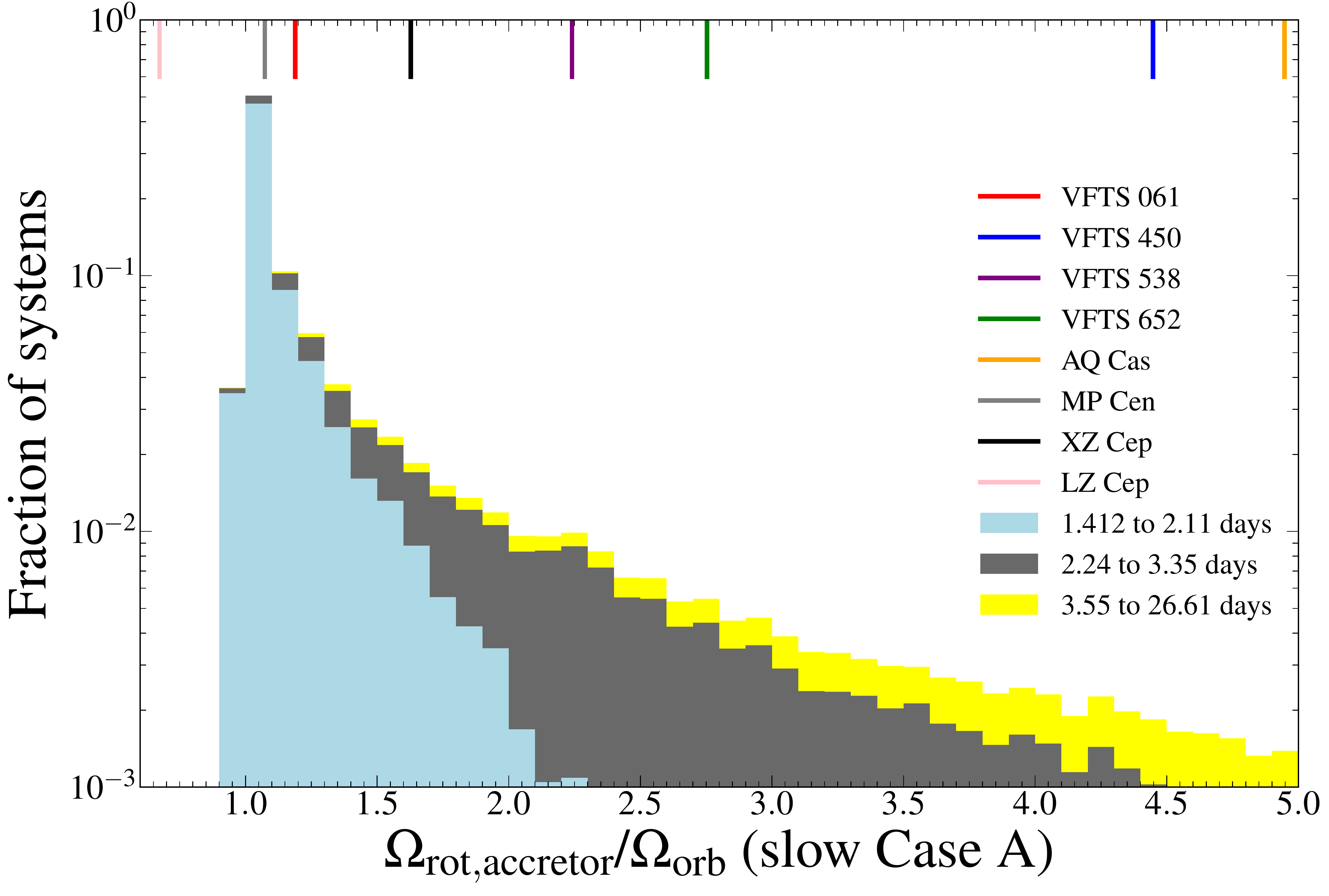}
\end{minipage}
\caption{Probability distribution of the ratio of spin to orbital angular velocity 
for the mass donor (left panel) and the mass accreting star (right panel) in our models
during the semi-detached configuration, weighted by the birth probability of each system and 
the lifetime in the respective bin. The colour coding indicate the initial orbital period
ranges indicated in the legend. The coloured vertical lines at the top of the right plot
indicate the ratio of spin to orbital angular velocity of some of the observed massive Algol binaries where both parameters were found in the literature (Table\,\ref{table_velocities}). The Y-axis of the right panel is in log scale.
}
\label{sync_donor}
\end{figure*}

In single stars, the enhancement of surface nitrogen mass 
fraction gets higher with the increase in rotational velocity 
and mass of the rotating star, as internal mixing is more 
efficient in more massive and faster rotating stars \citep[for 
a review, see][]{Maeder2000}. In binaries however, the donors 
can undergo envelope stripping which can expose layers of CNO 
processed material to the surface, increasing the surface 
abundance of nitrogen. Hence, it is essential to understand 
and differentiate between the contribution from internal mixing 
and envelope stripping to the observed surface nitrogen mass 
fraction enhancement in massive binaries. 

Figure\,\ref{HessNPrimVsVrotPrim} shows the distribution of the 
surface nitrogen mass fraction enhancements of donor and accretor
during the semi-detached phase as function of their surface 
rotational velocity. The surface rotational velocity of the donors 
is set by the orbital motion of the binaries, as their rotation is 
synchronised to the orbital revolution. 

The shortest orbital period binaries have the highest surface 
rotational velocities in our grid, owing to tidal locking at 
short orbital periods and the fact that we initialise the initial 
spins of the binary components to be equal to that of the orbit. 
In contrast to the single star picture, the donors rotating the 
fastest do not show the highest surface nitrogen mass fraction enhancement 
because the nitrogen enhancement in these binary models is not 
an outcome of rotational mixing but of envelope stripping. The 
donors with the highest rotational velocities (300 km s$^{-1}$) 
during the semi-detached phase arise from binaries with the shortest 
initial orbital period, go into a contact phase with an episode 
of inverse slow Case\,A mass transfer and eventually merge during 
their main sequence lifetime. It is hence likely that these models 
will not undergo significant envelope stripping by Case\,A mass 
transfer and do not show very high nitrogen enrichment until they 
merge.

Donors with surface rotational velocities between $\sim$100-250 
km s$^{-1}$ have relatively higher initial orbital periods, where 
the binaries survive the slow Case\,A mass transfer and the 
envelope of the donors are efficiently stripped and the surface 
nitrogen mass fraction enhancement reaches CNO equilibrium values 
towards the end of the slow Case\,A phase (see vertical column of increasing 
surface nitrogen abundance near X $\simeq$ 150-250 km s$^{-1}$). 
We see two islands of higher probability of surface nitrogen 
enhancement, one at ordinate values of $\simeq$ 2-6 and the other 
near the CNO equilibrium value for the LMC (i.e. $\simeq$ 35). 
As already discussed, the lower enhancement 
peak comes from models that eventually merge during slow Case\,A 
and the high surface enhancement peak comes from models that 
survive the entire Case\,A mass transfer phase. Hence, we 
expect the lower enhancement to be the result of rotational 
mixing \citep[see also][]{Maeder2000}, while the enhancement 
all the way up to CNO equilibrium is from envelope stripping. 

Surface rotational velocities less than 90 km s$^{-1}$ occur 
for a very brief time during the fast Case\,A mass transfer 
phase in our models when a large amount of mass is lost by 
the donor while tidal forces cannot instantaneously synchronise 
the rotation to the orbit. The light grey shading indicates 
that the amount of time spent in this region of the parameter 
space is very low. We also see two islands of surface nitrogen 
enhancement where the rotational velocity is below 90 km s$^{-1}$. 
This dichotomy arises typically from low and high mass models 
whereby in lower mass models, the envelope is not very efficiently 
stripped, leading to low surface nitrogen mass fraction enhancement (c.f. 
Panel\,`f' in Fig.\,\ref{typical_Case_A}) during and 
just after the fast Case\,A phase. In higher mass models, the 
envelope stripping is much more efficient, owing to the larger 
core to envelope mass, such that the surface nitrogen enrichment 
becomes very high right from the fast Case\,A mass transfer phase. 

The right panel of Fig.\,\ref{HessNPrimVsVrotPrim} shows 
the distribution of the surface nitrogen mass fraction enhancement of 
the accretor versus their equatorial rotation velocity. As we 
shall see (Fig.\,\ref{sync_donor}, right panel), most 
of the accretors in our grid are tidally locked 
during the slow Case\,A phase. The increase in surface 
rotational velocity around 200-250 km s$^{-1}$ with the 
increase in surface nitrogen mass fraction enhancement from 15-22 indicates 
that these accretor stars only spin up slightly, not up to 
critical rotation, towards the end of slow Case\,A mass 
transfer phase. These models having surface nitrogen 
enhancement > 15 have orbital periods less than $\sim$4 
days and survive the Case\,A mass transfer phase. Hence, 
the secondaries are able to accrete the N-rich matter 
transferred by the donor while not spinning up all the 
way to critical rotation during the late stages of the 
slow Case\,A mass transfer phase.

Accretors with rotational 
velocities above 300 km s$^{-1}$ are not found to be 
tidally locked. Hence they undergo inefficient slow 
Case\,A mass transfer (see Fig.\,\ref{eta_fast_slow_ab}) 
and the surface nitrogen abundance is not as highly 
enhanced as for accretors that are rotating synchronously. 
We note that the probability fraction of accretors that 
are tidally synchronised is much higher than the accretors 
that are not tidally locked, consistent with Fig.\,\ref{sync_donor} later. 

\subsection{Tidal synchronization}

Our mass transfer efficiency depends on the extent to which 
tidal forces can halt the spin-up of the mass accreting star 
(Sect.\,\ref{method}). For very short period binaries, tides 
can be strong enough to keep the mass accreting star rotating 
synchronously with the orbital velocity, enabling conservative 
mass transfer in our model grid. The strength of tidal forces 
implemented in our models can be tested by comparing the degree 
of synchronisation of our accretors during the slow Case\,A 
mass transfer phase as a function of orbital period to the 
degree of synchronisation of the accretors in observed massive 
Algol systems. 

The tidal synchronization timescale 
of a mass donor decreases as the star expands to fill its 
Roche lobe. Figure\,\ref{sync_donor} (left panel) shows the 
distribution of the ratio between spin and orbital angular velocity of the mass-donating star 
during the semi-detached phase. As these stars persist to fill their Roche lobes
during this phase, we see that all our donors are effectively rotating 
fully synchronised during the entire semi-detached phase.

The right panel of Fig.\,\ref{sync_donor} shows the distribution 
of the degree of synchronization of the mass accreting star 
during the semi-detached phase. We see that most of the mass 
accreting stars are rotating synchronously since the biggest 
contribution to the semi-detached phase comes from short period 
models where tidal interaction is strong. For higher orbital 
periods where tidal synchronization is inefficient, the accretors 
rotate super-synchronously. As expected, the degree of super-synchronicity
is higher for longer orbital periods. 

\subsection{Orbital period derivatives}

\begin{figure}
\centering
\includegraphics[width=\hsize]{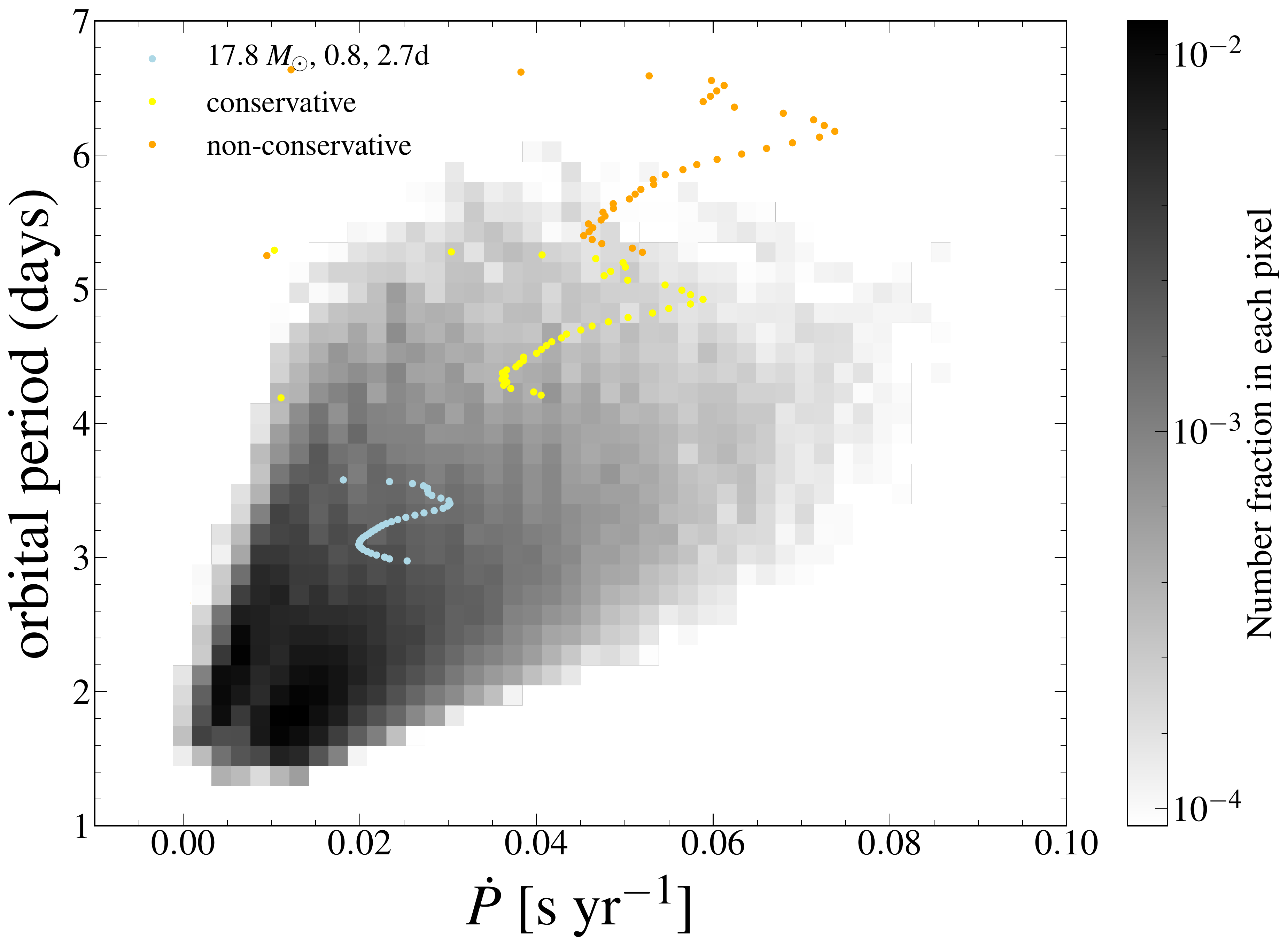}
\caption{Probability distribution of the orbital period and its derivative that is predicted to be observed in the semi-detached configuration of the Case\,A mass transfer phase based on the model grid. The lightblue coloured dots show the evolution of the orbital period and its derivative during the semi-detached phase for our example model in Fig.\,\ref{typical_Case_A}.  The yellow and orange dots show the evolution for the same model if the mass transfer phase was modelled as totally conservative (Eq.\,\ref{pcons}) or totally non-conservative with the mass lost via isotropic re-emission \citep[eq. 29 of][]{quast2019}, respectively. The dots are placed at intervals of 50000 years with the first dots of each colour having the lowest orbital period. Grey-scale: See description in Fig.\,\ref{hessHe}.}
\label{observed_Pdot}
\end{figure}

The orbital period derivative helps us assess the degree to which 
mass and angular momentum are lost from the binary and thus determine 
the efficiency of mass transfer in the binary system. 
Figure\,\ref{observed_Pdot} shows the distribution of the orbital 
period and its time derivative during the slow Case\,A mass transfer 
phase. The period derivative value is very tightly constrained 
between 0.004-0.03 secs/year. This value of the orbital period 
derivative is related to the mass and angular momentum lost from 
the binary and the mass transfer efficiency during the slow Case\,A 
phase. We have seen that the mass transfer efficiency during the 
slow Case\,A phase in our models is largely conservative 
(Fig.\,\ref{mass_trans_eff_slow_period}). 

The orbital period and its derivative if the entire Case\,A mass 
transfer was modelled as purely conservative or purely non-conservative 
with isotropic re-emission from the surface of the accretor can be 
analytically derived \citep[eq. 30 or 29 of][respectively]{quast2019}, 
with deviations coming from the mass and angular momentum lost due to stellar 
winds. Both the orbital period and its derivative is higher for the 
case of totally conservative mass transfer (yellow dots) or totally 
non-conservative mass transfer with isotropic re-emission (orange dots) 
than our models. 

We note that we implement non-conservative mass transfer in our models 
as an increase in wind mass loss rather than isotropic re-emission 
from the accretor \citep[c.f. eq. 14 and 15 of][]{quast2019}. As such, 
the orbital period in our models do not increase to the same extent as 
for the case of isotropic re-emission from the accretor. Comparing 
eq. 29 and 30 of \citet{quast2019} we can see that the orbital 
period derivative for the case of non-conservative mass transfer 
via isotropic re-emission becomes positive before the mass ratio 
is inverted and is higher than for the conservative case, as is seen 
in Fig.\,\ref{observed_Pdot}.

Hence, one can expect to test the mass transfer efficiency during the 
slow Case\,A phase and the implementation of non-conservative mass 
transfer for the overall Case\,A mass transfer phase in our models 
from measurements of the orbital period derivative of observed Algol 
systems over a few years. We note that the orbital period measurements 
are more constrained \citep{mahy2019b} than the values of orbital 
period derivatives we find during the semi-detached phase. For the 
mass dependence of this orbital period derivative during the slow 
Case\,A phase, see Fig.\,\ref{hess_Pdot_diff_ranges}.  

\section{Binary properties after Case\,AB mass transfer}
\label{post_Case_AB}

After the slow Case\,A mass transfer, which is ter\-mi\-na\-ted by 
core hydrogen exhaustion of the donor star (Fig.\,\ref{typical_Case_A}), 
the donor star's envelope still contains several solar masses of 
hydrogen-rich matter (Fig.\,\ref{fast_Case_AB_mass_lost_gained}). 
The contracting helium core therefore leads to an envelope expansion, 
which initiates another thermal timescale mass transfer phase, 
named Case\,AB. We find that 99.2\% of the models that survive 
the slow Case\,A mass transfer also survive the Case\,AB mass 
transfer.

While this paper focusses on the semi-detached phase of evolution, 
we also briefly discuss the properties of our binary models 
after the Case\,AB mass transfer here. At this stage, they consist 
of a main sequence star with a stripped-envelope star as companion. 
Observations of such binaries are difficult, since the stripped 
star is very hot and hard to detect next to the brighter OB\,companion 
\citep{wellstein2001,schootemeijer2018,gotberg2020,Wang2021}. Only for the most 
massive systems in our model grid, it is expected that the stripped 
stars develop an optically thick wind and appear as Wolf-Rayet stars. 
According to Pauli et al. (2021, in prep), who analyse the WR+O star 
phase for a model grid with primary masses of up to $\sim 90\mso$, 
this occurs for primary masses above $\sim 28\mso$ \citep[see also,][]{shenar2020} 
at the metallicity of the LMC. In the following, we do not distinguish 
stripped stars with and without optically thick winds. 

\begin{figure*}
{\centering
\includegraphics[width=0.5\hsize]{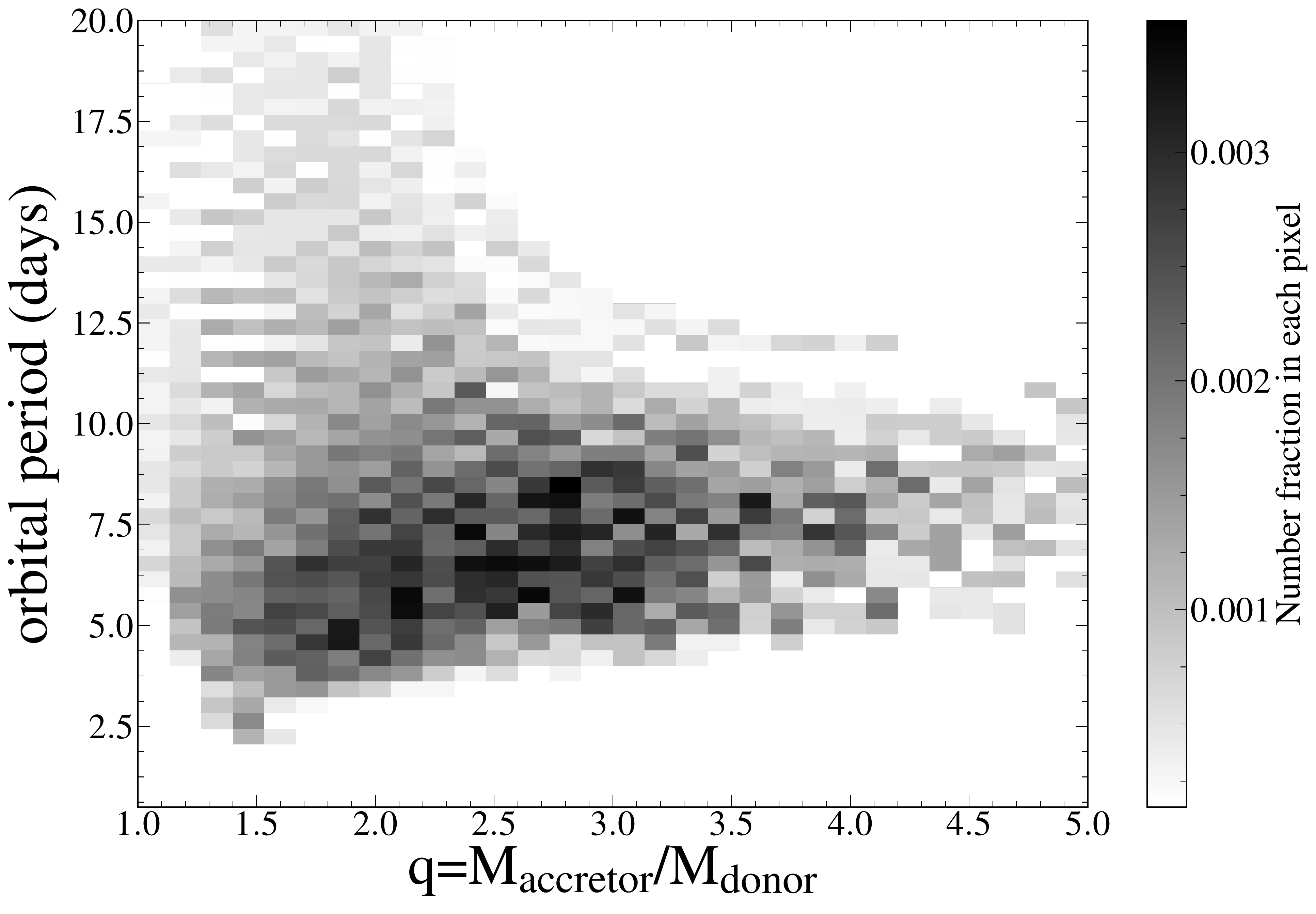} 
\includegraphics[width=0.5\hsize]{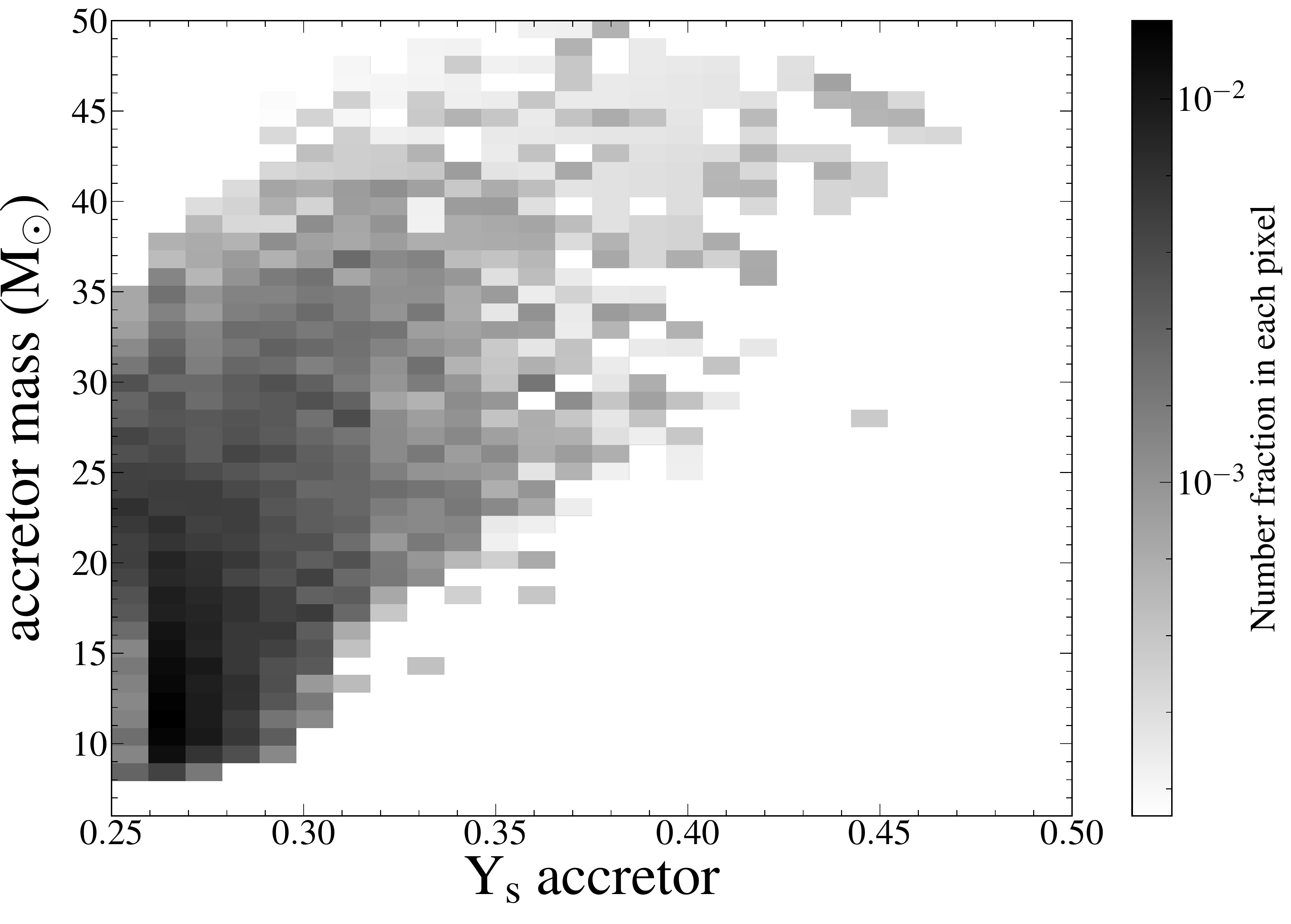}}
{\centering
\includegraphics[width=0.5\hsize]{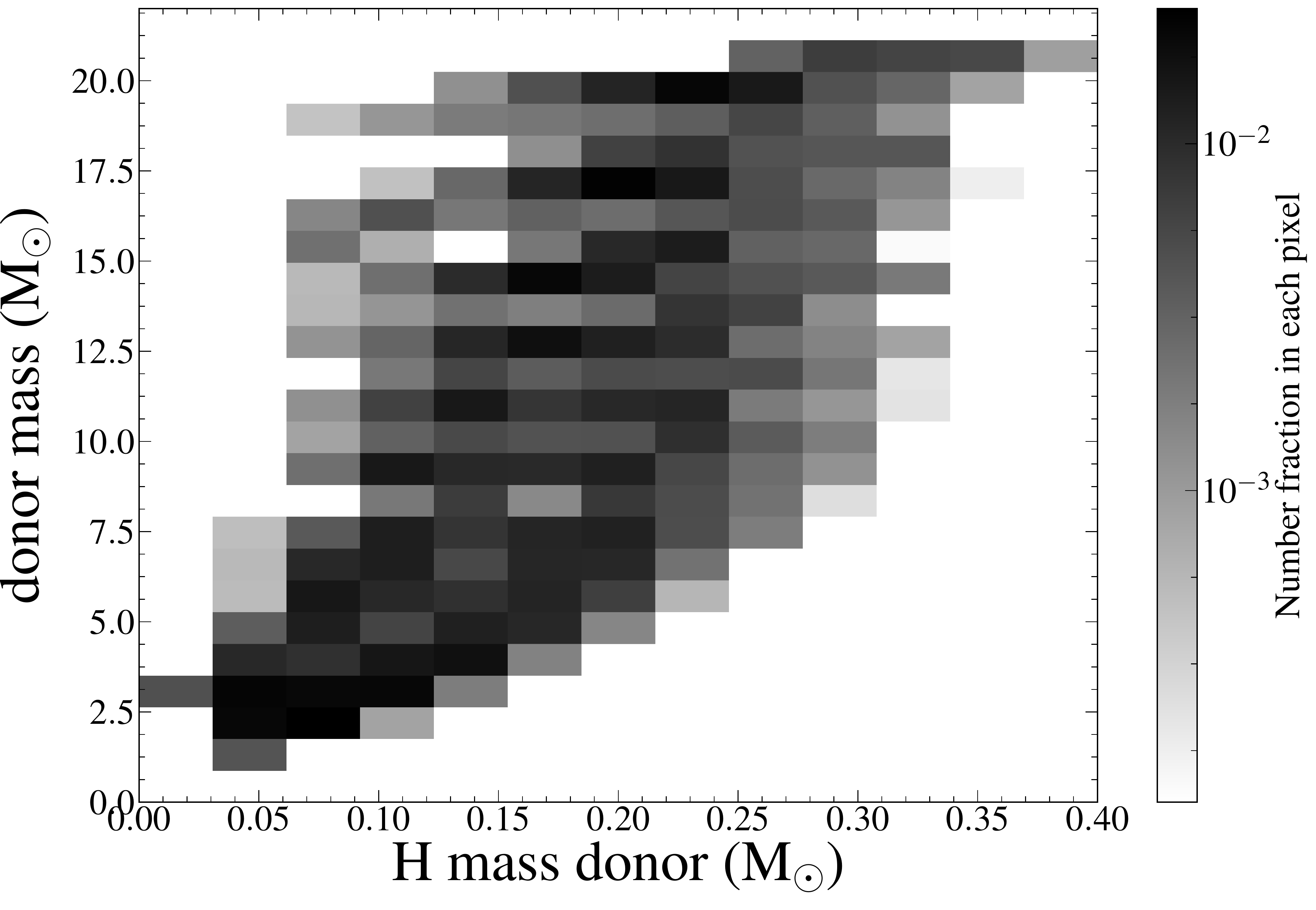}
\includegraphics[width=0.5\hsize]{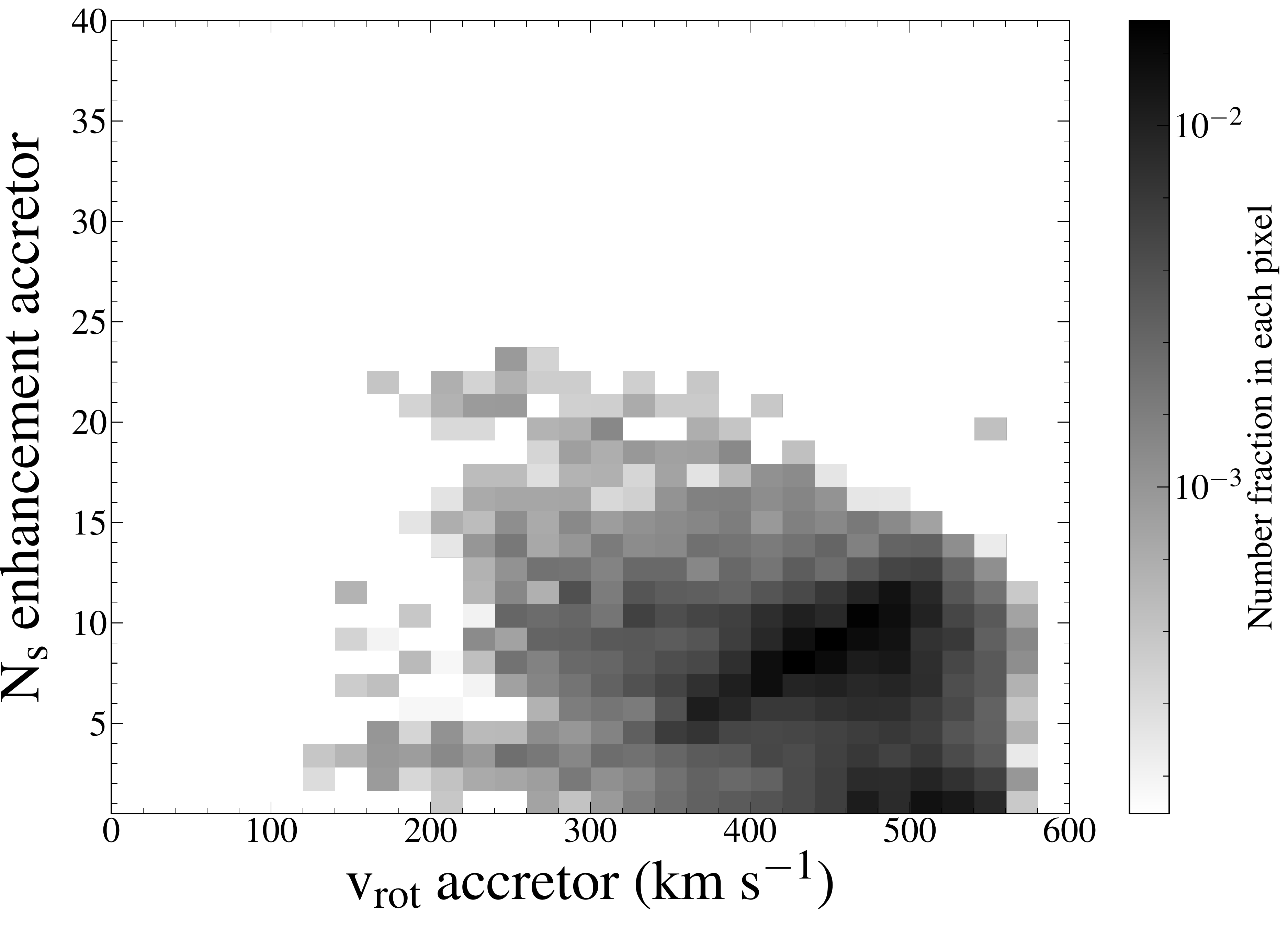}}
\caption{Probability density distribution of several properties of our binary models
after Case\,AB mass transfer, when the central helium mass fraction of the donor has decreased to 0.90:
orbital period versus mass ratio (top left), surface helium abundance and mass of the mass gainer
(top right), leftover hydrogen mass in the donor and donor mass (bottom left), and
surface nitrogen mass fraction enhancement versus rotational velocity of the mass gainer (bottom right).
}
\label{obs_period_CaseAB}
\label{Ys_accretor_CaseAB}
\label{Nenh_vrot_donors_CaseAB}
\label{hydrogen_mass_CaseAB}
\end{figure*}

In our models, no mass transfer occurs during core helium 
burning and hence their orbital period and mass ratio does 
not change significantly throughout the core helium burning 
phase of the mass donor (c.f. Fig.\,\ref{typical_Case_A}, 
panel\,`b' and `g'). The orbital period increases slightly 
due to wind mass-loss while the mass ratio change remains 
small. Also, the surface properties of the individual binary 
components undergo only moderate changes. For our analysis, 
we sample the properties of our post Case\,AB models at a 
time when the core helium mass fraction of the mass donor 
(the stripped star) has decreased to 0.90 due to helium 
burning (except for Fig.\,\ref{hydrogen_mass_CaseAB1d-1}). 

Figure\,\ref{obs_period_CaseAB} (top left panel) shows the 
orbital period and mass ratio distribution of our models 
after the Case\,AB mass transfer phase. Due to the additional 
mass transfer, the mass ratios of our post Case\,AB models 
are larger than during the slow Case\,A phase (cf., Fig.\,\ref{observed_period}), 
with values up to $q$ = 5. Since during Case\,AB, mass flows 
from  a less massive to a more massive star (see panel\,`h' 
of Fig.\,\ref{typical_Case_A}), the orbital period also 
increases significantly compared to models in the slow Case\,A 
phase, to values of $\sim$15\,days and more. At the same time, 
since the Case\,AB mass transfer in our models is highly 
non-conservative (cf., Fig.\,\ref{fast_Case_AB_mass_lost_gained}), 
orbital periods and mass ratios in our models remain significantly 
smaller than in fully conservative models \citep[e.g.][]{wellstein2001}.

The top right panel of Fig.\,\ref{Ys_accretor_CaseAB} displays the 
distribution of the surface helium mass fraction and stellar mass of 
our mass gainers. A comparison with Fig.\,\ref{hessHe} shows that 
the helium mass fraction is increased by a few percent compared to 
the semi-detached phase. Still, most mass gainers contain only a 
mild surface helium enrichment. However, more massive mass gainers 
tend to be more helium enriched, although there is a large spread 
for every considered mass. This is so because during core hydrogen 
burning, more massive stars have larger convective core mass fractions. 

The surface helium mass fraction of the donors at the considered 
time is 0.70-0.90, and is most drastically reduced for the most 
massive systems, in which the donors become Wolf-Rayet stars (cf., 
Pauli et al. 2021 in prep). Figure\,\ref{hydrogen_mass_CaseAB} (bottom left 
panel) shows the distribution of remaining hydrogen mass in the post 
Case\,AB mass donors, as function of their mass. We compute 
the remaining hydrogen mass ($M_{\rm H}$) in the donor as $M_{\rm H} = 
\int_{0}^{M} X(m)\,dm$. At most $\sim$0.4 
$M_{\odot}$ of hydrogen remains in any of our models. We see the 
overall trend that the remaining hydrogen mass increases with the 
donor mass \citep[see also,][]{yoon2017}. The wind stripping during 
the core helium burning phase of the mass donor is illustrated by 
Fig.\,\ref{hydrogen_mass_CaseAB1d-1}. A comparison with 
Fig.\,\ref{hydrogen_mass_CaseAB} shows that many of the most massive 
donors manage to lose all their hydrogen, whereas some hydrogen 
remains in all donors below $\sim 12\mso$. 

After Case\,AB mass transfer, many of the mass gainers may 
observationally appear as single stars because the high mass 
ratios and longer orbital periods make binary detection more 
elusive). Hence, it is interesting to consider the distribution 
of their surface nitrogen mass fraction enhancements and 
rotational velocities (Fig.\,\ref{Nenh_vrot_donors_CaseAB}, 
bottom right panel). The surface nitrogen mass fraction enhancement is higher 
then it was in the Algol phase (cf., Fig.\,\ref{HessNPrimVsVrotPrim}), 
but far from its CNO equilibrium value of $\sim 35$. Whereas the 
matter that is transferred during Case\,AB does contain CNO 
equilibrium abundances, the mass transfer efficiency is low 
(cf., Fig.\,\ref{mass_trans_eff_fast_period}), and thermohaline 
mixing dilutes the small amounts of accreted matter in the massive 
envelope of the accretor.

Compared to the semi-detached phase (Fig.\,\ref{HessNPrimVsVrotPrim}), 
where most of the mass accretors are tidally locked, 
Fig.\,\ref{Nenh_vrot_donors_CaseAB} shows that the Case\,AB mass 
transfer spins them up considerably. This is so because the orbital 
periods increase rapidly during the Case\,AB phase, rendering tidal 
braking inefficient in most cases. However, whereas many of our accretor 
models end Case\,AB mass transfer very close to critical rotation, 
tides remain strong enough to spin down most of our models over 
their core helium burning evolution (see Fig.\,\ref{Nenh_vrot_accretors_CaseAB1d-1}). 
For our most massive models, stellar winds also help to spin down 
the accretors. Therefore, our models provide only a small contribution 
to the overall population of Be stars, in particular compared to 
massive binary models undergoing mass transfer after core hydrogen 
burning \citep[Case\,B, e.g. see][]{wang2020}. We also note that the 
parameter space for Case\,A mass transfer is smaller than that for 
Case\,B mass transfer \citep[Fig. 2 of][]{langer2020}. 

\section{Comparison with observations}
\label{observations_section}

\subsection{Observed Algol binaries in the LMC}
\label{obs_LMC}
The Tarantula Massive Binary Monitoring (TMBM) survey 
\citep{almeida2017,mahy2019b,mahy2019a} investigated 
$\sim$100 massive binaries in the Tarantula region 
of the LMC. \cite{mahy2019b,mahy2019a} classified six 
of them as semi-detached. From those, we find four 
to be in the Algol configuration, where the less massive 
star transfers mass to its more massive companion, 
and shows surface helium and nitrogen enrichment. 

The other two systems (VFTS\,176 and VFTS\,094) have 
their more massive stars filling its Roche lobe, 
and are therefore not in the Algol configuration. 
Furthermore, none of the components shows any surface enrichment. 
Both systems have a very short orbital period (1.77\,days 
and 2.25\,days, respectively).  It appears plausible that 
VFTS\,176 and VFTS\,094 evolved previously through a thermal 
timescale contact phase and are currently undergoing inverse 
nuclear timescale slow Case\,A mass transfer (see discussion 
in Sect. \ref{pq} for VFTS 176 and `System 2' 
type evolution of contact binaries in \citealp{Menon2021} 
for VFTS 094). Due to the lack of a concrete understanding 
of the current evolutionary stage of these two systems, 
we left these two systems out of our comparison with 
the predictions obtained from the model grid of binary 
models for the Algol phase. We list the observed properties 
of the classical TMBM Algol 
systems in Table\,\ref{table}, along with those of three 
massive Algols in the LMC described in the literature.

\begin{table*}[t]
\centering
\caption{Semi-detached double-lined eclipsing binaries observed in the LMC with at least one component having a mass greater than 8 $M_{\odot}$. }
\label{table}
\begin{tabular}{c c c c c c c c c c c c c c}
\hline
\hline
Name (colour & Period & $M_{\rm a}$ & $M_{\rm d}$ & q & R$_{\rm a}$ & R$_{\rm d}$ & T$_{\rm a}$ & T$_{\rm d}$ & References\\    
in plots)   & days & $M_{\odot}$ & $M_{\odot}$ &  & R$_{\odot}$ & R$_{\odot}$ & kK & kK & \# \\    
\hline

VFTS 061 (red)     & 2.33 & 16.3$\pm$1.4 &  8.7$\pm$0.6 & 1.87 &  7.2$\pm$0.2 & 7.30$\pm$0.3 & 33.5$\pm$0.9  & 32.9$\pm$0.7   & [1]  \\
VFTS 652 (green)   & 8.59 & 18.1$\pm$3.9 &  6.5$\pm$1.1 & 2.78 & 15.4$\pm$0.7 & 16.8$\pm$0.7 & 32.1$\pm$0.9  & 23.9$\pm$0.5   & [1]  \\
VFTS 538 (purple)  & 4.15 & 18.3$\pm$1.9 & 11.8$\pm$1.4 & 1.55 &  7.9$\pm$0.6 & 14.7$\pm$1.0 & 35.6$\pm$1.7  & 32.0$\pm$0.3   & [2]  \\
HV 2543 (yellow)   & 4.83 & 25.6$\pm$0.7 & 15.6$\pm$1.0 & 1.64 & 15.5$\pm$0.4 & 14.0$\pm$0.4 & 35.3$\pm$0.6  & 28.7$\pm$0.5   & [3]  \\
VFTS 450 (blue)    & 6.89 & 29.0$\pm$4.1 & 27.8$\pm$3.9 & 1.04 & 13.0$\pm$3.0 & 22.2$\pm$0.4 & 33.8$\pm$2.3  & 28.3$\pm$0.3   & [1]  \\
SC1-105 (black)    & 4.25 & 30.9$\pm$1.0 & 13.0$\pm$0.7 & 2.37 & 15.1$\pm$0.2 & 11.9$\pm$0.2 & 35.0$\pm$2.5  & 32.5$\pm$2.5   & [4]  \\
HV 2241 (cyan)     & 4.34 & 36.2$\pm$0.7 & 18.4$\pm$0.7 & 1.96 & 14.9$\pm$0.4 & 13.7$\pm$0.4 & 38.4$\pm$1.4  & 29.5$\pm$1.2   & [5]  \\

\hline
\end{tabular}
\tablebib{
(1) \citet{mahy2019b}; (2) \citet{almeida2017,mahy2019a}; (3) \citet{ostrov2000}; (4) \citet{bonanos2009}; (5) \citet{ostrov2001};.
}
\tablefoot{Subscripts `a' and `d' denote the accretor and donor star respectively. The systems are ordered by increasing mass of the mass gainer.}
\end{table*}

\label{observations}

\subsection{Observed Algol binaries in the Milky Way}

In order to enhance the statistical basis of the comparison 
of our models with observations, we also consider observed 
Galactic massive Algol systems. We do this despite the 
metallicity difference between Milky Way and LMC, because 
the physics of mass transfer has no known direct metallicity 
dependence. However, we caution that an indirect impact of 
metallicity is possible, due to three effects. Firstly, stars 
of higher metallicity are slightly less compact \citep{brott2011}, 
such that longer orbital period binaries can contribute to 
the semi-detached phase at higher metallicity. Secondly, 
stellar winds are stronger in the Milky Way compared to the 
LMC \citep{mokiem2007}, which may be relevant in particular 
for the most massive binaries discussed here. And thirdly, 
the most massive stars in our samples may be sufficiently 
close to the Eddington limit that envelope inflation may 
occur \citep{sanyal2015,sanyal2017}. While the first effect 
remains at the level of $\sim 10$\% \citep{Brott2011models}, the other two effects 
may be larger, especially for the most massive binaries 
discussed here. No metallicity dependence in the orbital 
period-mass ratio distribution of massive contact binaries 
has been predicted by \cite{Menon2021} between the LMC and 
SMC, although they expect envelope inflation to play a role 
in explaining some of the observed long period ($\geq5\,d)$ 
massive contact binaries in the Milky Way. 

\cite{malkov2020} performed a comprehensive literature survey 
of semi-detached double line eclipsing binaries in the Milky 
Way. He listed 119 semi-detached systems, of which 32 have at 
least one component with a stellar mass above 8\,$M_{\odot}$ 
(Table\,\ref{table_galaxy}). Investigation into the individual 
systems (see systems with footnotes in Table \ref{table_galaxy}) 
reveals that GT Cep is in triple system, RZ Sct and BY Cru have 
an accretion disc that makes accurate determination of their 
masses very challenging, the mass ratio of TU Mon is highly 
debated in the literature, V453 Cyg is now considered as a 
detached system, V729 Cyg is a potential contact system and 
BY Cru shows evidence of very heavy interactions and contain 
an F type supergiant donor. As such, we do not include these 
six systems in our analysis. These systems will be very 
interesting for follow-up observations. 

\begin{table*}[t]
\centering
\caption{Semi-detached double-lined eclipsing binaries observed in the Galaxy with at least one component exceeding 8\,$M_{\odot}$. }
\label{table_galaxy}
\begin{tabular}{l r r r r r r r r r}
\hline
\hline
Name & Period & $M_{\rm a}$\;\;\;\;\;\;\;\;\, & $M_{\rm d}$\;\;\;\;\;\;\;\;\, & $q$ & R$_{\rm a}$\;\;\;\;\;\;\;\;\, & R$_{\rm d}$\;\;\;\;\;\;\;\;\, & T$_{\rm a}$\;\;\;\;\;\;\;\;\, & T$_{\rm d}$\;\;\;\;\;\;\;\;\, & References\\    
     & d & $M_{\odot}$\;\;\;\;\;\;\;\;\, & $M_{\odot}$\;\;\;\;\;\;\;\;\, &  & R$_{\odot}$\;\;\;\;\;\;\;\;\, & R$_{\odot}$\;\;\;\;\;\;\;\;\, & kK\;\;\;\;\;\;\;\;\, & kK\;\;\;\;\;\;\;\;\, & \# \\    
\hline
TT Aur  & 1.33 & 8.10\;\;\;\;\;\;\;\;\, & 5.40\;\;\;\;\;\;\;\;\, & 1.50 & 3.90\;\;\;\;\;\;\;\;\, & 4.20\;\;\;\;\;\;\;\;\, & 24.8\;\;\;\;\;\;\; & 18.2\;\;\;\;\;\;\;& [1]  \\
$\mu_{1}$ Sco & 1.44 &  8.30$\pm$1.00 & 4.60$\pm$1.00 & 1.80 & 3.90$\pm$0.30 & 4.60$\pm$0.40 & 24.0$\pm$1.0 & 17.0$\pm$0.7  & [2] \\
SV Gem   & 4.00 & 8.34\;\;\;\;\;\;\;\;\, & 3.37\;\;\;\;\;\;\;\;\, & 2.47 & 4.87\;\;\;\;\;\;\;\;\, & 6.70\;\;\;\;\;\;\;\;\, &                &                & [3] \\
V454 Cyg & 2.31 & 8.40\;\;\;\;\;\;\;\;\, & 2.86\;\;\;\;\;\;\;\;\, & 2.93 & 4.99\;\;\;\;\;\;\;\;\, & 4.64\;\;\;\;\;\;\;\;\, &                &                & [1]  \\
BF Cen   & 3.69 &  8.70\;\;\;\;\;\;\;\;\, & 3.80\;\;\;\;\;\;\;\;\, & 2.29 & 5.10\;\;\;\;\;\;\;\;\, & 7.10\;\;\;\;\;\;\;\;\, & 12.9\;\;\;\;\;\;\; & 9.4\;\;\;\;\;\;\; & [1]  \\
BM Ori   & 6.47 & 9.50\;\;\;\;\;\;\;\;\, & 3.51\;\;\;\;\;\;\;\;\, & 2.70 & 4.47\;\;\;\;\;\;\;\;\, & 8.94\;\;\;\;\;\;\;\;\, &                &                & [3] \\
IZ Per   & 3.68 &  9.97$\pm$0.55 & 3.20$\pm$0.17 & 3.11 & 7.51$\pm$0.13 & 6.90$\pm$0.12 & 19.0$\pm$0.5  & 9.5$\pm$0.5   & [4] \\
AI Cru   & 1.41 & 10.30$\pm$0.20 & 6.30$\pm$0.10 & 1.63 & 4.95$\pm$0.06 & 4.43$\pm$0.05 & 24.2$\pm$0.5  & 17.7$\pm$0.5  & [5] \\
SX Aur   & 1.21 & 10.30$\pm$0.40 & 5.60$\pm$0.30 & 1.84 & 5.17$\pm$0.09 & 3.90$\pm$0.07 &                &                & [3] \\
MP Cen   & 2.99 & 11.40$\pm$0.40 & 4.40$\pm$0.20 & 2.59 & 7.70$\pm$0.10 & 6.60$\pm$0.10 & 18.7$\pm$0.4  & 12.4$\pm$0.1  & [6]  \\
IU Aur   & 1.81 & 11.99$\pm$0.08 & 6.07$\pm$0.04 & 1.97 & 6.30$\pm$0.20 & 5.20$\pm$0.20 & 33.3$\pm$0.1  & 28.8$\pm$0.2  & [7]  \\
V356 Sgr & 8.89 & 12.10\;\;\;\;\;\;\;\;\, & 4.70\;\;\;\;\;\;\;\;\, & 2.57 & 7.40\;\;\;\;\;\;\;\;\, & 14.00\;\;\;\;\;\;\;\;\, & 19.1\;\;\;\;\;\;\; & 8.8\;\;\;\;\;\;\; & [1]  \\
V Pup\tablefootmark{a}    & 1.45 & 12.85$\pm$0.50 & 6.33$\pm$0.30 & 2.03 & 5.48$\pm$0.18 & 4.59$\pm$0.06 & 28.2$\pm$1.0  & 26.6$\pm$1.0  & [8] \\
V498 Cyg & 3.48 & 13.44\;\;\;\;\;\;\;\;\, & 6.45\;\;\;\;\;\;\;\;\, & 2.08 & 6.81\;\;\;\;\;\;\;\;\, & 8.12\;\;\;\;\;\;\;\;\, &                &                & [1]  \\
GN Car   & 4.34 & 13.49\;\;\;\;\;\;\;\;\, & 4.59\;\;\;\;\;\;\;\;\, & 2.94 & 7.14\;\;\;\;\;\;\;\;\, & 8.29\;\;\;\;\;\;\;\;\, &                &                & [1]  \\
LZ Cep   & 3.07 & 16.00$\pm$9.80 & 6.50$\pm$2.40 & 2.46 &11.70$\pm$3.30 & 9.40$\pm$2.60 & 32.0$\pm$1.0 & 28.0$\pm$1.0   & [9]  \\
$\delta$ Pic  & 1.67 & 16.30\;\;\;\;\;\;\;\;\, & 8.60\;\;\;\;\;\;\;\;\, & 1.89 & 7.62\;\;\;\;\;\;\;\;\, & 5.05\;\;\;\;\;\;\;\;\, & 25.2\;\;\;\;\;\;\; & 21.4\;\;\;\;\;\;\;& [10] \\
XX Cas   & 3.06 & 16.85\;\;\;\;\;\;\;\;\, & 6.07\;\;\;\;\;\;\;\;\, & 2.77 & 8.28\;\;\;\;\;\;\;\;\, & 7.19\;\;\;\;\;\;\;\;\, &                &                & [1]  \\
HH Car   & 3.23 & 17.00\;\;\;\;\;\;\;\;\, &14.00\;\;\;\;\;\;\;\;\, & 1.21 & 6.10\;\;\;\;\;\;\;\;\, &10.70\;\;\;\;\;\;\;\;\, & 35.5\;\;\;\;\;\;\; & 29.9\;\;\;\;\;\;\; & [1]  \\
V337 Aql & 2.73 & 17.44$\pm$0.31 & 7.83$\pm$0.18 & 2.22 & 9.86$\pm$0.06 & 7.48$\pm$0.04 & 28.0$\pm$0.5  & 23.6$\pm$0.5  & [11]  \\
AQ Cas   &11.70 & 17.63$\pm$0.91 &12.50$\pm$0.81 & 1.41 &13.40$\pm$0.64 &23.50$\pm$0.73 & 27.0$\pm$1.0  & 16.7$\pm$0.4  & [12]  \\
XZ Cep   & 5.09 & 18.70$\pm$1.30 & 9.30$\pm$0.50 & 2.01 &14.20$\pm$0.10 &14.20$\pm$0.10 & 28.0$\pm$1.0  & 24.0$\pm$3.0  & [13]  \\
29 CMa\tablefootmark{b}   & 4.39 & 19.00\;\;\;\;\;\;\;\;\, &16.00\;\;\;\;\;\;\;\;\, & 1.20 & 10.00\;\;\;\;\;\;\;\;\, & 13.00\;\;\;\;\;\;\;\;\, & 29.0\;\;\;\;\;\;\; & 33.7\;\;\;\;\;\;\; & [14] \\
AB Cru   & 3.41 & 19.75$\pm$1.04 & 6.95$\pm$0.65 & 2.84 &10.50$\pm$0.32 & 8.85$\pm$0.32  & 35.8\;\;\;\;\;\;\; & 27.2\;\;\;\;\;\;\; & [15] \\
V448 Cyg & 6.51 & 24.70$\pm$0.70 &13.70$\pm$0.70 & 1.80 & 7.80$\pm$0.20 &16.30$\pm$0.30 & 30.5$\pm$0.1  & 20.3$\pm$0.1  & [16] \\
QZ Car\tablefootmark{c}   & 5.99 & 30.00$\pm$3.00 &20.00$\pm$3.00 & 1.50 &10.00$\pm$0.50 &20.00$\pm$1.00 & 36.0\;\;\;\;\;\;\; & 30.0\;\;\;\;\;\;\; & [17]  \\

\hline
\textbf{*}GT Cep\tablefootmark{d}   & 4.90 &10.70$\pm$0.50  & 2.58$\pm$0.14 & 4.14 & 6.34$\pm$0.19 & 6.98$\pm$0.11 & 22.4$\pm$1.0  & 10.9$\pm$0.3  & [18] \\
\textbf{*}RZ Sct\tablefootmark{e}   &15.20 & 11.70\;\;\;\;\;\;\;\;\, & 2.49\;\;\;\;\;\;\;\;\, & 4.68 & 15.00\;\;\;\;\;\;\;\;\,          & 15.90\;\;\;\;\;\;\;\;\,         & 19.0\;\;\;\;\;\;\; & 6.5\;\;\;\;\;\;\; & [1] \\
\textbf{*}TU Mon\tablefootmark{f}   & 5.04 & 12.00\;\;\;\;\;\;\;\;\, & 2.50\;\;\;\;\;\;\;\;\, & 4.80 &5.50\;\;\;\;\;\;\;\;\, &7.20\;\;\;\;\;\;\;\;\, & 19.0\;\;\;\;\;\;\; & 8.1\;\;\;\;\;\;\; & [1] \\
\textbf{*}V453 Cyg\tablefootmark{g} & 3.89 & 14.36$\pm$0.20 &11.10$\pm$0.13 & 1.29 & 8.55$\pm$0.06 & 5.49$\pm$0.06 & 26.6$\pm$0.5  & 25.5$\pm$0.8  & [1] \\
\textbf{*}V729 Cyg\tablefootmark{h} & 6.60 &36.00$\pm$3.00 & 10.00$\pm$1.00 & 3.60 &27.00$\pm$1.00 &15.00$\pm$0.60 &               &               & [19]\\
\textbf{*}BY Cru\tablefootmark{i}  &106.40 & 9-11\;\;\;\;\;\;\;\;\; & 1.80\;\;\;\;\;\;\;\;\, & 5.30 &        &               &                &                & [20] \\
\hline

\end{tabular}
\tablebib{
(1) \citet{surkova2004}; (2) \citet{budding2015}; (3) \citet{budding2004}; (4) \citet{hilditch2007};
(5) \citet{bell1987}; (6) \citet{terrell2005}; (7) \citet{surina2009}; (8) \citet{stickland1998}; 
(9) \citet{mahy2011}; (10) \citet{evans1974}; (11) \citet{tuysuz2014}; 
(12) \citet{ibanoglu2013}; (13) \citet{martins2017}; (14) \citet{bagnuolo1994} (15) \citet{lorenz1994}; 
(16) \citet{djurasevi2009}; (17) \citet{walker2017}; (18) \citet{cakrli2015}; (19) \citet{rauw1999}; (20) \citet{daems1997}.
}
\tablefoot{ Subscripts `a' and `d' refer to the mass accretor and mass donor, respectively. The list is adopted from the work of \cite{malkov2020} and \cite{surkova2004}. Systems without effective temperature estimates have only photometric orbit solutions and mass estimates. Comments on individual systems marked with superscripts: 
\tablefoottext{a}{Masses and radii taken from \cite{budding2021}.}
\tablefoottext{b}{May be a contact system (Mahy et al. in prep.).}
\tablefoottext{c}{Is a part of SB2+SB2 system, see \citet{Morrison1979,Blackford2020}.}
\tablefoottext{d}{Possibly a triple system, see discussion in text.}
\tablefoottext{e}{The more massive star is rapidly rotating and has an accretion disc. See \cite{wilson1985}.}
\tablefoottext{f}{Debate about the mass ratio of this system exists in the literature. See the discussion in \cite{cester1977}.}
\tablefoottext{g}{Recent studies now consider this as a detached system. See \cite{southworth2004}.}
\tablefoottext{h}{A potential contact system \citep{yasarsoy2014} and a part of a hierarchical triple system \citep{Rauw2019}.}
\tablefoottext{i}{A highly interacting binary with an F type supergiant donor, with the accretor invisible due to an accretion disc around it. The long orbital period suggests it may be undergoing thermal timescale Case\,AB or Case\,B mass transfer.} As such, we do not compare our model predictions with six (d,e,f,g,h,i) observed systems marked with an asterisk. 
}
\end{table*}

For four galactic Algol systems, an estimate of the projected 
rotational velocities of the component stars and for the orbital 
inclination was available in the literature. We list their 
parameters,  together with those of the four classical TMBM 
Algols, in Table\,\ref{table_velocities}. We assume that the 
inclination of the orbital plane is the same as the inclination 
of the stellar spin axes. As such, we calculate the rotational 
velocity of the binary components from the projected rotational 
velocity and the orbital inclination (see Fig.\,\ref{HessNPrimVsVrotPrim}). 
The orbital inclinations for the Algol binaries are taken from 
the references given in Table\,\ref{table_galaxy}.

\begin{table*}[t]
\centering
\caption{Orbital velocities, rotational velocities, and luminosities of the semi-detached double-lined eclipsing binaries, whenever available. }
\label{table_velocities}
\begin{tabular}{l r r r r r r r r r r r r}
\hline
\hline
Name & Period & v$_{\rm rot,a}$ sin \textit{i} & v$_{\rm rot,d}$ sin \textit{i} & $\Omega_{\rm rot,a}/\Omega_{\rm orb}$ & $\Omega_{\rm rot,d}/\Omega_{\rm orb}$ & log L$_{\rm a}$ & log L$_{\rm d}$ & Inclination \textit{i} \\    
     &   days &                           km/s &                           km/s &                 &                 &    L$_{\odot}$ &    L$_{\odot}$ &                degrees \\    

\hline
\vspace{0.1cm}
VFTS 061      & 2.33  & 174   $^{+33}_{-28}$ & 152  $^{+20}_{-40}$ & 1.18 & 1.02 & 4.77$\pm$0.04  & 4.75$\pm$0.05 & 69.1$^{+0.9}_{-0.9}$ \\
\vspace{0.1cm}
MP Cen        & 2.99  & 140   $^{+12}_{-12}$ & 95   $^{+15}_{-15}$ & 1.07 & 0.85 & 3.80$\pm$0.20  & 3.00$\pm$0.20 & 82.2$^{+0.2}_{-0.2}$ \\ 
\vspace{0.1cm}
LZ Cep        & 3.07  & 130   $^{+10}_{-10}$ & 80   $^{+10}_{-10}$ & 0.67 & 0.51 & 5.11$\pm$0.19  & 4.69$\pm$0.19 & 48.1$^{+2.0}_{-0.7}$ \\
\vspace{0.1cm}
VFTS 538      & 4.15  & 158   $^{+28}_{-33}$ & 108  $^{+14}_{-21}$ & 2.24 & 0.82 & 4.96$\pm$0.08  & 5.31$\pm$0.06 & 47.0$^{+5.0}_{-5.0}$ \\
\vspace{0.1cm}
XZ Cep        & 5.09  & 230   $^{+15}_{-15}$ & 110  $^{+12}_{-12}$ & 1.62 & 0.78 & 5.05$\pm$0.06  & 4.79$\pm$0.22 & 80.0$^{+2.0}_{-2.0}$ \\ 
\vspace{0.1cm}
VFTS 450      & 6.89  & 380   $^{+42}_{-40}$ & 97   $^{+12}_{-12}$ & 4.44 & 0.66 & 5.30$\pm$0.21  & 5.46$\pm$0.04 & 63.5$^{+1.7}_{-1.2}$ \\
\vspace{0.1cm}
VFTS 652      & 8.59  & 224   $^{+38}_{-45}$ & 96   $^{+10}_{-13}$ & 2.75 & 1.08 & 5.35$\pm$0.06  & 4.92$\pm$0.06 & 63.7$^{+0.9}_{-4.8}$ \\
\vspace{0.1cm}
AQ Cas        & 11.70 & 287   $^{+10}_{-10}$ & 98   $^{+11}_{-11}$ & 4.94 & 0.96 & 4.94$\pm$0.08  & 4.59$\pm$0.05 & 84.4$^{+4.0}_{-4.0}$ \\ 

\hline
\end{tabular}
\tablefoot{The references to each individual system are the same as in Table\,\ref{table} and \ref{table_galaxy}. The subscript `a' denotes the accreting star and `d' denotes the Roche lobe filling mass donating star. }
\end{table*}

\subsection{The number of semi-detached binaries in the LMC}
\label{number}
To compare the expected number of semi-detached binaries 
from our grid of binary models with observations, 
we note that the Tarantula Massive Binary Monitoring (TMBM) 
survey \citep{almeida2017,mahy2019b,mahy2019a} has observations 
of 102 O type massive binary candidates with orbital 
periods in the range 1-1000\,days. The TMBM sample was 
based on binary detections in the Very Large Telescope 
Flames Tarantula survey \citep{Evans2011}. \citet{sana2013} 
showed that the binary detection probability is a function 
of the orbital period. It goes from 95\% at 2\,d to 87\% at 
10\,d to 70\% at 100\,d to 25\% at 1000\,d. It is however 
beyond the scope of this paper to find this in our analysis. 
Hence, we consider all models in our grid that have initial 
orbital periods of less than 1000\,days. This includes many 
models that do not undergo Case\,A mass transfer. To filter 
out the semi-detached models, we only consider those in which 
mass transfer during core hydrogen burning lasts for more than 
three times the thermal timescale of the mass donor (calculated 
at the onset of mass transfer). This way we find that 16\% of 
all models with orbital periods less than 1000\,d to undergo 
nuclear timescale Case\,A mass transfer. 

We then find the sum of the weighted duration of the Case\,A 
mass transfer phase relative to the main sequence lifetime for 
each model as 
\begin{equation}
F2 = \frac{\sum_{\rm m=1}^{N_{\rm g}}\tau_{\rm Case\:A}W_{\rm m}}{\sum_{\rm m=1}^{\rm N_{\rm g}}\tau_{\rm MS}W_{\rm m}}\:,
\end{equation}
where $\tau_{\rm MS}$ is as defined for Eq.\,\eqref{F2_dist_eqn}
(cf., Sect.\,\ref{sec:tau_case_a} and Fig.\,\ref{F2_distribution}),
$N_{\rm g}$ is the number of models in the grid with initial 
orbital periods below 1000\,days, and $W_{\rm m}$ represents the birth 
probability of each system according to Eq.\,\eqref{eq_Wm}.
We find that $F2\simeq 0.03$, which implies that we expect 
3 out of every 100 main sequence binaries originating from 
initial donor masses between 10-40 $M_{\odot}$ and orbital 
periods of 1.41-1000\,days to be observed in the slow Case\,A 
mass transfer phase. This is in good agreement with the 
VFTS-TMBM survey, where we find four binaries to be in 
the Algol configuration amongst the 102 observed binaries. 

Table \ref{f2_masses} gives the contribution from each 
initial donor mass in our grid to the total of 3\% (2nd 
column), in comparison to the expected contribution if 
it would simply scale with the Salpeter IMF (3rd column). 
It shows that while more massive systems contribute less 
to the total synthetic Algol population than less massive 
ones, their contribution is larger than expected according 
to the IMF, with 40$\mso$ binaries contributing three 
times as much. This estimate needs to be amended by the 
fact that 10$\mso$ stars live about five times longer than 
40$\mso$ stars, which implies a $\sim 15$ times larger 
contribution of our most massive systems than naively 
expected. It reflects the fact that, for a given initial 
primary mass, the period and mass ratio range of surviving 
Case\,A binary models is much larger for larger initial 
primary mass, as can be seen when comparing the blue coloured 
areas in the upper and lower panels of Fig.\,\ref{1.100_summary}.

\begin{table}[t]
\centering
\caption{Contribution from models at each initial donor mass ($M_{\rm d,i}$) to the total percentage (3\%) of all massive main sequence binaries that is predicted to be found in the semi-detached configuration. }
\label{f2_masses}
\begin{tabular}{c c c c}
\hline
\hline
$M_{\rm d,i}$ ($M_{\odot}$) & fraction (\%) & IMF (\%) & Relative excess\\    
\hline

10.0 & 0.33 & 0.33 & 1.0 \\
11.2 & 0.31 & 0.28 & 1.1 \\
12.6 & 0.29 & 0.24 & 1.2 \\
14.1 & 0.27 & 0.20 & 1.3 \\
15.8 & 0.26 & 0.18 & 1.4 \\
17.8 & 0.26 & 0.15 & 1.7 \\
19.9 & 0.24 & 0.13 & 1.8 \\
22.4 & 0.21 & 0.11 & 1.9 \\
25.1 & 0.20 & 0.09 & 2.2 \\
28.2 & 0.18 & 0.08 & 2.3 \\
31.6 & 0.17 & 0.07 & 2.4 \\
35.5 & 0.16 & 0.06 & 2.7 \\
39.8 & 0.15 & 0.05 & 3.0 \\

\hline
\end{tabular}
\tablefoot{The third column gives the relative IMF weights for the initial donor masses, normalised to 0.33\% for 10 $M_{\odot}$. The last column gives the relative excess (column 2/column 3) of binaries predicted in the semi-detached configuration relative to the case where the contribution from the different masses would simply scale with the Salpeter IMF.}
\end{table}

About 25\% of all massive stars in the LMC are associated with 
the Tarantula region, with the estimated number of O-type stars 
being 570 \citep{,doran2013,crowther2019}. Extrapolating, we expect about 2000 
O-type stars in the LMC. This accounts for a total of $\sim$1000 
O-type star binaries in the LMC, assuming a binary fraction of 
50\% \citep{sana2013}. We found in our model grid that 3\% of 
all massive main sequence binaries are in the semi-detached 
phase. Therefore, we expect around 30 binaries that contain 
an O-type star in the semi-detached phase in the LMC, from our 
binary parameter space.

At the same time, in our parameter space, the ratio of models 
expected to be observed in the contact phase to the semi-detached 
phase is $\sim$1/19 (see discussion of Fig.\,\ref{observed_period}). 
This implies that we expect $\sim$1.6 O-type contact binaries in 
the LMC originating from our initial parameter space. However, our 
initial orbital period range does not cover the shortest period 
contact systems, which have the longest contact lifetimes \citep{Menon2021}. 
Therefore, our number is smaller than the result of \citet{Menon2021}, 
who, based on their evolutionary grids of contact binary models, 
and for a binary fraction of 50\%, predict $\sim$8 O-type contact 
binaries in the LMC. B stars above $\sim$10 $M_{\odot}$ live 
roughly twice as long as average O stars, and accounting for a 
Salpeter initial mass function, we expect about 60 Algol binaries 
and 3 contact binaries amongst the $\sim$4000 B stars above 10 
$M_{\odot}$ in the LMC. 

\subsection{Orbital periods and mass ratios}

Figure\,\ref{observed_period} compares the observed orbital 
periods of the semi-detached binaries with the predicted 
period distribution. Whereas observed and predicted period 
ranges agree well, there are more system observed at larger 
period than expected from the theoretical distribution, in 
particular in the LMC sample. We note however that, while the 
distribution in Fig.\,\ref{observed_period} is dominated by 
the lower mass systems due to the steepness of the IMF, the 
observed Algol binaries in the LMC are very massive (see 
Fig.\,\ref{observed_masses}), and the predicted orbital 
period distributions stretch to significantly larger values 
for higher masses (Fig.\,\ref{hessSD_mass_dependence}). 
On the other hand, we find the peak of the distribution 
of the Galactic Algol systems to be in good agreement,
but again have more high period systems than predicted.
While the significance of this mismatch remains unclear
due to the unknown observational biases, we find one 
systems in the LMC (VFTS\,652) and two in the Galaxy 
(V356 Sgr and BM Ori) each with orbital periods above 
8\,d but with accretor masses below $20\,$M$_{\odot}$.
Concerning the mass ratios, we find a reasonable match  
with most of the observed semi-detached binaries in the 
LMC and the Galaxy (Fig.\,\ref{observed_q}). There is 
one systems (IZ\,Per) with a mass ratio above\,3 that is 
not explained by our model grid. 

\begin{figure}
\centering
\includegraphics[width=0.98\hsize]{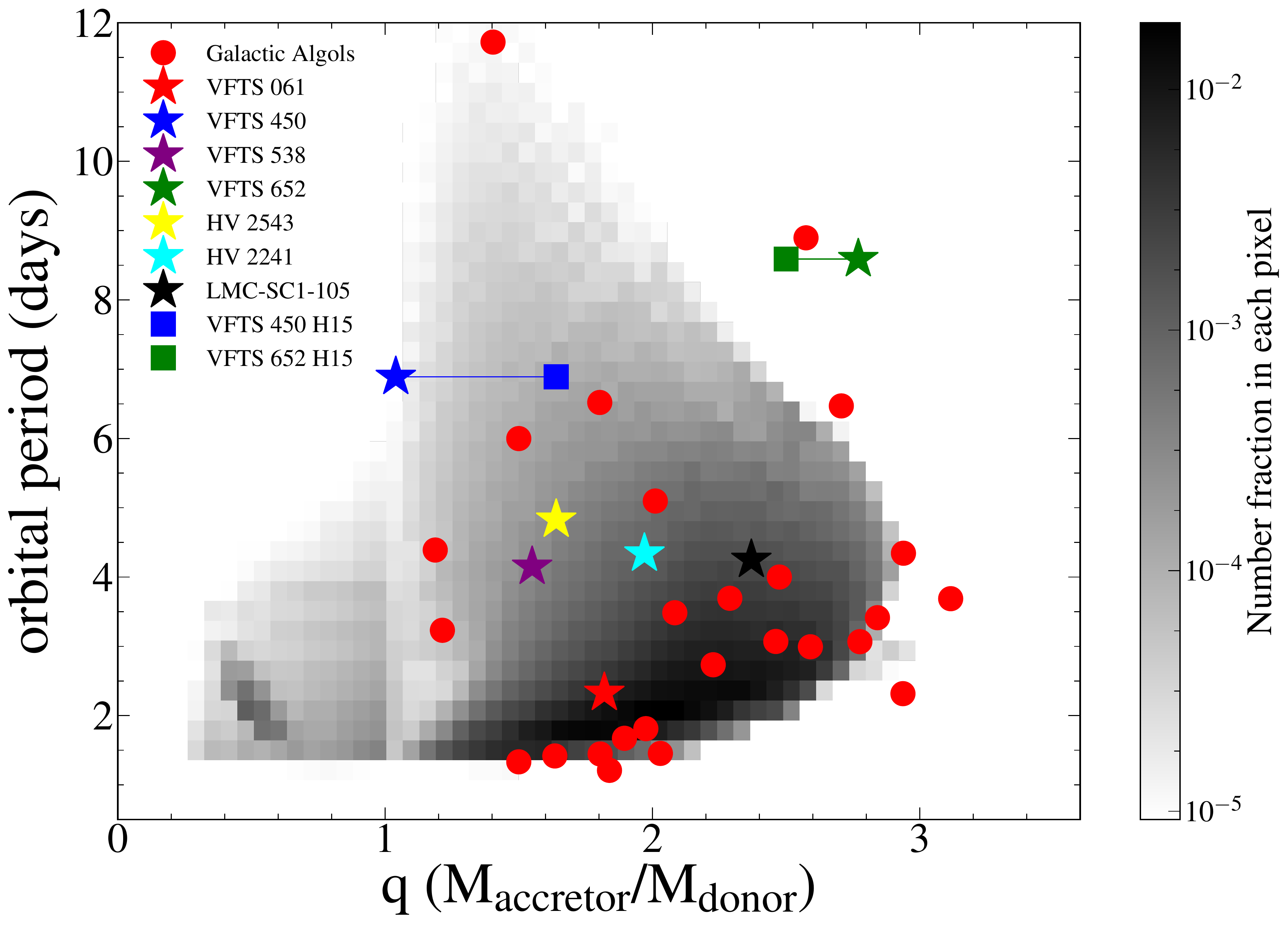}
\includegraphics[width=0.98\hsize]{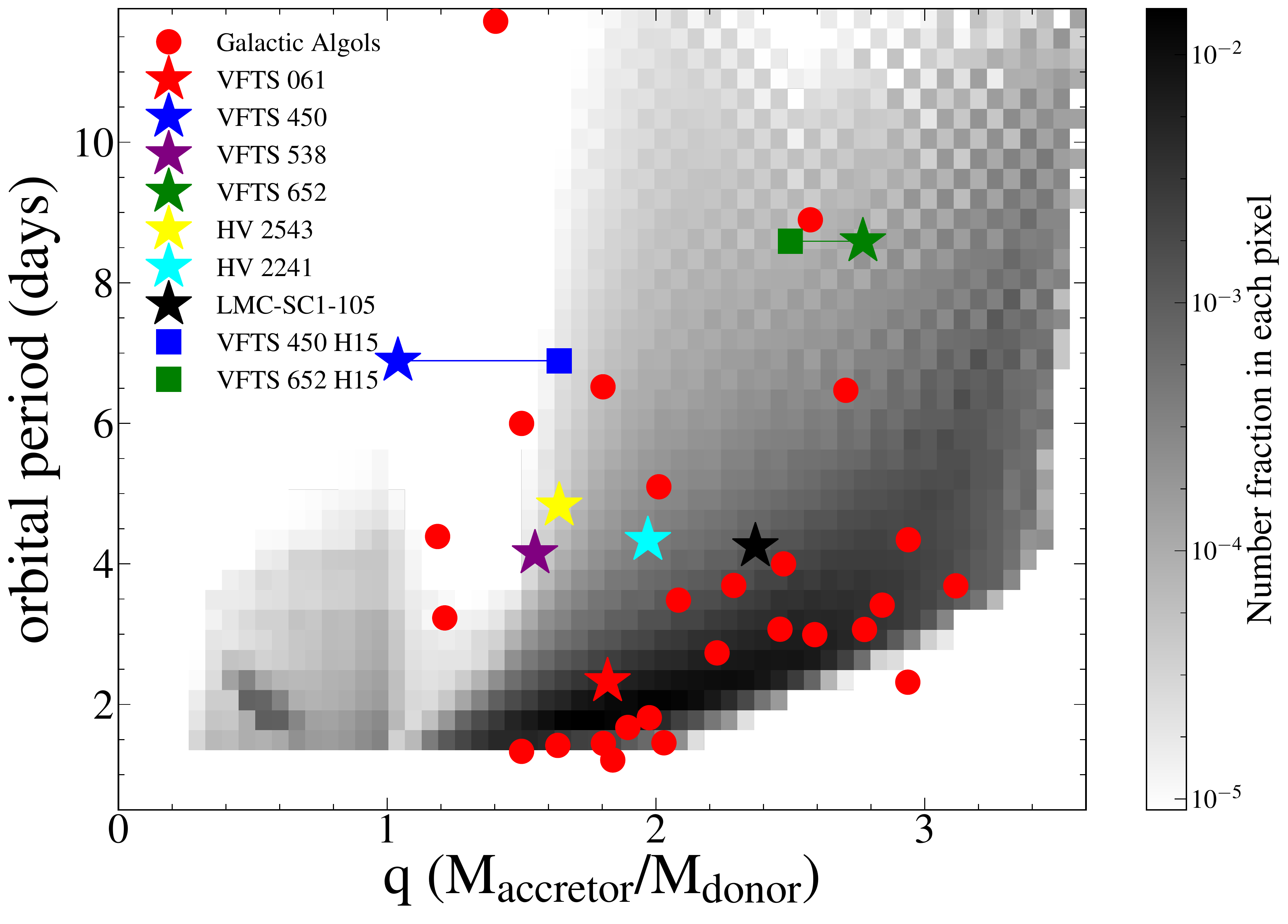}
\includegraphics[width=0.98\hsize]{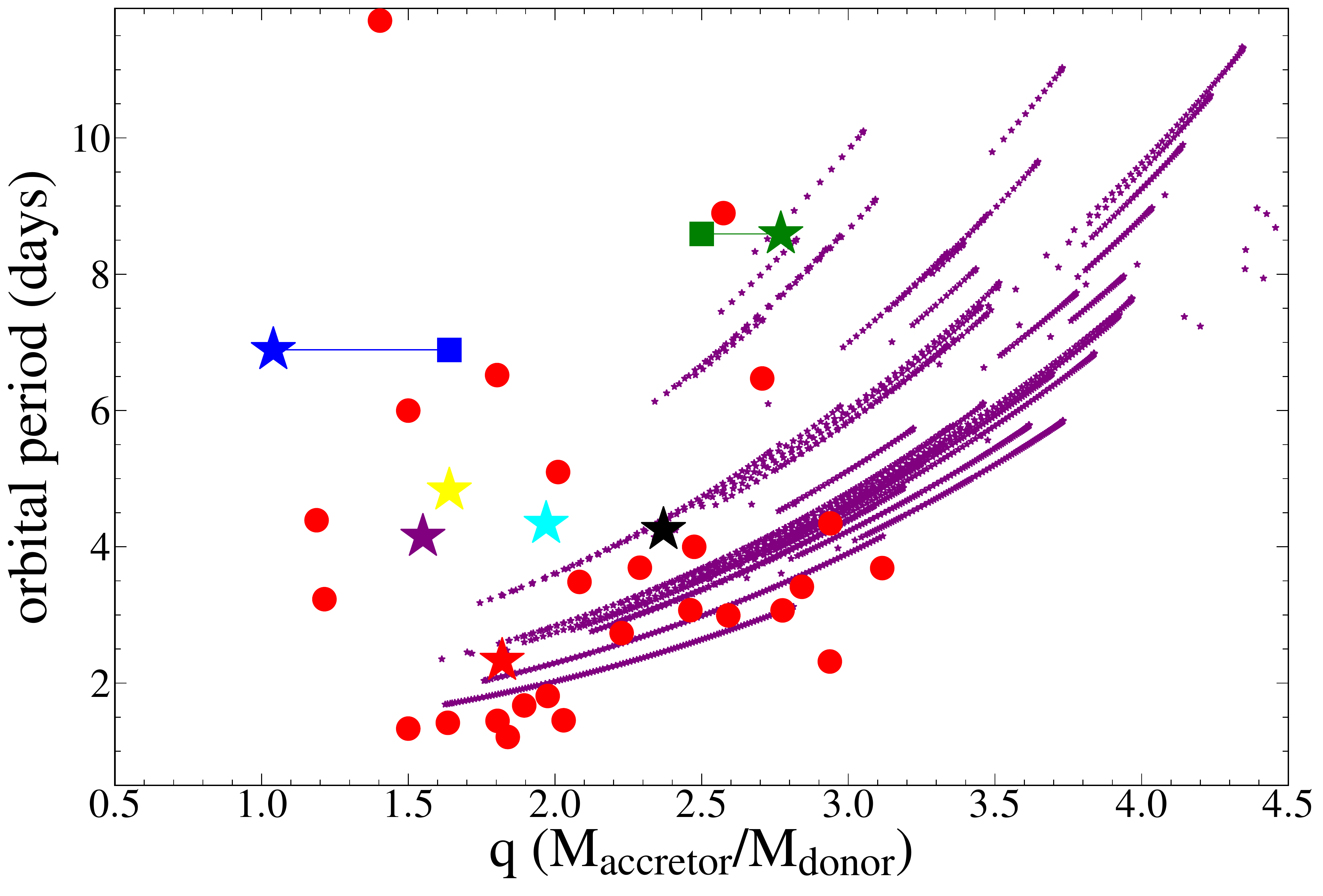}
\caption{\textit{Top panel:} Probability distribution of orbital 
periods and mass ratios of our synthetic population of semi-detached 
binaries. For the grey scale, see description in Fig.\,\ref{hessHe}.
\textit{Middle panel:} As the top panel, but with the retrospect 
assumption of fully conservative mass transfer (see text).
\textit{Bottom panel:} The evolution of orbital period versus mass 
ratio during the semi-detached phase of the 41 conservative Case\,A 
models of \citet{wellstein2001}, with primary masses of 12$\mso$, 
16$\mso$ and 25$\mso$ is marked by small purple stars. The small 
purple stars are placed with a time difference of 50\,000\,yrs. In 
all three plots, the different coloured star symbols denote the 
position of the semi-detached systems in the LMC (Table \ref{table}), 
and squares denote the binary parameters of VFTS 450 and 652 derived 
by \cite{howarth2015}. The colour coding is as in Fig.\,\ref{observed_period}. 
Red circles denote the parameters for the Galactic Algol systems.}
\label{hessSD}
\end{figure}

Figure\,\ref{hessSD} (top panel) shows the probability distribution of 
the predicted orbital period and mass ratio of our Algol models 
simultaneously. We also estimate the distribution of Algol binaries 
from our models in the $P-q$-plane if, in retrospect, mass transfer had 
been assumed to be fully conservative (cf. Sect.\,\ref{CaseA_example}; 
middle panel), and in the conservative models \citep[of][bottom panel]{wellstein2001}. 
For our \textit{pseudo} conservative approximation (middle panel), we 
assume that the binary will merge during Case\,A mass transfer if the 
mass transfer rate from the donor is greater than the mass divided by 
the thermal timescale of the accretor ($M_{\rm acc}/\tau_{\rm KH,acc}$). 

We see that in all the three panels, there is a sharp borderline leading 
to an absence of predicted high mass ratio models at the shortest orbital 
periods. This occurs since the shortest-period models in our grid are 
nearly conservative, and the retained mass and angular momentum during the 
mass transfer leads to a simultaneous increase of mass ratio and orbital 
period. Figure\,\ref{hessSD} shows further that for orbital periods above 
$\sim 4\,$d, our models predict a confinement of Algols to smaller mass 
ratios for larger periods (top panel). This occurs due to the fact that 
tides becomes weaker for larger periods, which leads to a smaller mass 
transfer efficiency (cf., Sect.\,\ref{Case_A}). In the conservative case 
(middle and lower panel of Fig.\,\ref{hessSD}), the opposite trend is seen. 
 
We note that our \textit{pseudo} conservative approximation predicts 
a maximum mass ratio of $\sim$3.6 while the maximum mass ratio from 
self-consistent conservative models can go above 4. This is primarily 
because the conservative models of \citep{wellstein2001} do not 
include overshooting while our models do. This leads to the availability 
of a larger envelope mass of the donor that can be transferred to the 
accretor in the self-consistent conservative models and a lighter 
stripped donor star. This leads to a  higher mass ratio for the 
self-consistent models towards the end of the slow Case\,A mass transfer. 

In a large unbiased observational sample, we expect to find 
$\sim$90\% of the semi-detached systems in the LMC in the 
orbital period range 1.5-5\,d and with mass ratios between one 
and three. About 75\% of the observed Algol 
binaries do fall into these ranges. Of the remaining ones,
we find four Galactic systems just below our short period cut-off
of 1.4\,d, which may have undergone a short-lived contact phase 
\citep{Menon2021}. However, while this means that the bulk
of the observed systems are well reproduced by our models (top panel), 
we see that this is also true for the conservative models (middle panel). 

A preference between the two sampled mass transfer efficiency assumptions
must be based on those Algol systems which are outliers.
Of these, V454\,Cyg ($P,q$=2.3\,d,2.93) is missed in both cases.
However, a handful of systems at the high mass ratio side,
in particular IZ\,Per ($P,q$=3.7\,d,3.11), V356\,Sgr ($P,q$=8.9\,d,2.57),
and VFTS\,652 ($P,q$=8.6\,d,2.78) are missed by 
our calculations but are well covered by the conservative models.
On the other hand, several low-$q$ systems are better reproduced 
by our non-conservative models, such as HH\,Car (3.2\,d,1.21), 29 CMa (4.39\,d,1.20), 
QZ\,Car (6.0\,d,1.50), VFTS\,450 (6.9\,d,1.04), and AQ\,Cas (11.7\,d,1.41).

An obvious cure to these discrepancies would be to assume that 
mass transfer is more efficient in the first group (at high mass ratio),
while retaining the low mass transfer efficiency otherwise. Notably,
recovering HH\,Car and VFTS\,450 needs a very non-conservative
evolution. We note that while one could tune the mass transfer efficiency
directly in binary model calculations, assuming more efficient tidal coupling
would likely serve the same purpose in our models, as it would diminish
the spin-up of the accretor and thus allow for more accretion. 

Recently, \citet{Justesen2021} found evidence for a higher efficiency 
of tidal circularization in binaries with temperatures between 6250 
K and 10000 K. A higher efficiency of tidal coupling in the 4-10\,d 
orbital period range than implemented in our models would lead to a 
higher mass transfer efficiency in this orbital period range, 
which could help reproduce the high mass ratio outliers in Fig.\,\ref{hessSD}
(top panel). 

We note from Fig.\,\ref{eta_fast_slow_ab} that the transition from 
conservative to non-conservative mass transfer in our Case\,A models 
is a function of all three initial binary parameters, the donor mass, 
orbital period and mass ratio. The self-consistently generated boundary 
line between conservative and non-conservative mass transfer is also 
different for different phases of the mass transfer (i.e. fast Case\,A, 
slow Case\,A and Case\,AB). These boundary lines essentially denote the 
orbital period (for binaries of a certain initial donor mass and mass 
ratio) at which tidal forces are unable to halt the spin-up of our 
accretor models. 

The above self-consistently derived mass transfer efficiency from 
our models can be easily implemented as an orbital period dependent 
mass transfer efficiency prescription in rapid binary evolution 
models. Then, quick tests on the extent to which the strength of 
tidal interactions needs to be increased can be gauged so as to 
reproduce some of the highest mass ratio Algols. This can be done 
by varying the boundary between efficient and inefficient mass 
transfer in the rapid codes, and looking at the maximum mass ratios 
attained by the models during the Algol phase. We also note that a 
completely non-conservative mass transfer at all orbital periods 
would not fit the observations as then we would underpredict the 
mass ratios of the very short period (1-3\,d) semi-detached models 
that undergo fairly conservative mass transfer in our original 
model grid. 

When looking at the mass dependence of the predictions (Fig.\,\ref{hessSD_mass_dependence}), for the most massive binary Algol 
systems we find that at the highest 
considered accretor masses (above 30\,M$_{\odot}$) the peak in 
the mass ratio distribution is much closer to one, and the orbital period 
distribution predicts a considerable fraction of Algol binaries 
above 5\,d. Indeed, Fig.\,\ref{hessSD_mass_dependence} does not
reveal any problem with our mass transfer scheme, where most of the 
observed Algols rather seem to be restricted to the lowest considered 
accretor mass interval (less than 20\,M$_{\odot}$).

\subsection{Surface abundances}

\label{sec: surf_abun_obs}

The four Roche lobe filling donor stars of the TMBM sample 
which we assume to be in the slow Case\,A mass transfer phase 
have observed surface nitrogen abundances consistent with CNO 
cycle equilibrium (Fig.\,\ref{hessHe}). These systems also 
show surface helium enrichment, signifying that their 
hydrogen-helium gradient region is exposed to the surface. 
We have shown (Fig.\,\ref{hessN_survivors}) that the models 
that survive the Case\,A mass transfer phase show high surface 
nitrogen enrichment. We conclude that these systems will 
likely survive the Case\,A mass transfer without merging.

On the other hand, there are two more VFTS systems (VFTS\,094 
and VFTS\,176) in the semi-detached phase which we did not 
include in our analysis (cf., Sect.\,\ref{obs_LMC}) that do not 
show surface helium and nitrogen enrichment \citep{mahy2019a}. 
These systems have very short orbital periods \citep{mahy2019b}. 
As we have shown in Sect.\,\ref{sec: surf_abun_models}, a very 
mild surface nitrogen mass fraction enhancement in our donor 
models occurs only in very short period model systems that 
undergo inverse mass transfer and subsequently a contact evolution 
before merging during the slow Case\,A phase. Hence, these two 
systems are likely not following the classical Case\,A evolution, 
and may be progenitors of binary mergers. 

In our models, a surface nitrogen enrichment above 5 is usually 
associated with an increase in surface helium mass fraction, as 
the slow Case\,A mass transfer exposes the hydrogen-helium 
gradient region. But this is not seen in VFTS 538, where the 
donor is consistent with no helium enhancement, while its nitrogen
enrichment is seemingly even above the CNO equilibrium value. 
On the other hand, VFTS 061 has a short orbital period and shows 
high nitrogen enrichment in both components. This is consistent 
with the region of our parameter space that predicts high surface 
nitrogen mass fraction enhancement in both donors and accretors 
(Fig.\,\ref{hessHe}), which occurs in our most conservative models.

Carbon deficiencies have also been observed in the more massive 
components of Algol binaries \citep{tomkin1993}, signifying that 
CNO cycled material has been transferred from the currently lower 
mass star. A high surface enrichment is also observed in several 
Galactic Algol binaries. \citet{mahy2011} find that the Roche 
lobe filling component of the semi-detached system LZ Cep shows 
a strong helium and nitrogen mass fraction enhancement. HD\,149404 
\citep{raucq2016} and XZ Cep \citep{martins2017} contain Roche lobe 
filling stars with surface abundances close to CNO equilibrium.

Overall, these observations reflect the range of enhancements 
we find in our models. While our donor models slowly increase 
their surface nitrogen abundance during the slow Case\,A evolution 
(see Figs.\,\ref{hessN_survivors} and \ref{hessHeN_10-20-30-40}), 
they can reach values near CNO equilibrium during the second half 
of Case\,A mass transfer (cf., Fig.\,\ref{CaseA_example}; see also 
\citealp{wang2017}). A statistically significant Algol sample is 
needed to test whether additional physics assumptions are required, 
such as tidally induced mixing in close binaries \citep{hastings2020}, 
to understand their surface abundances. 

\subsection{Synchronization}

For eight observed Algol binaries (Table \ref{table_velocities}), 
the projected rotational velocities and inclination angles are 
available in the literature. For those, we determine the deviation 
of both components from synchronous rotation. We obtain the spin 
angular velocity (v$_{\rm rot,a/d}$) from the measured projected 
rotational velocity (v$_{\rm rot,a/d}$ sin $i$), the stellar radius 
and the inclination angle $i$, while the orbital angular velocity 
is derived from orbital period and orbital separation. 

We find that the spins of three of the Roche lobe filling donors, 
VFTS\,061, VFTS\,652 and AQ\,Cas, to be synchronised with the 
orbit to within 10\%. The remaining five donor stars seem to 
rotate slower than synchronous, by 15\dots50\%. In our models, 
the donors do not deviate by more than 4\% from synchronous rotation 
(Fig.\,\ref{sync_donor}, left panel). The largest deviations are 
found for Galactic Algols (most extreme for LZ\,Cep) for which 
the derived radius my depend on the adopted distance. The resolution 
of this discrepancy needs a deeper investigation than is done here. 

The vertical dashes on the top in the right panel of Fig.\,\ref{sync_donor} 
denote the asynchronicities of the observed accretors (Table 
\ref{table_velocities}). All but one (LZ\,Cep) rotate super-synchronously, 
with an average value of 2.4 times synchronous rotation, and a maximum 
value of 5. Also a trend of stronger super-synchronicity with larger orbital period 
is apparent from the data (Table \ref{table_velocities}), which could imply 
that in closer systems, the accretion-induced spin-up is reduced by tidal interaction. 
Notably, the observed range of super-synchronicities, and its trend with 
orbital period, appears to be well reproduced by our models (Fig.\,\ref{sync_donor}, 
right panel). We note, however, that \cite{Dervisoglu2010} found in Algols of 
much lower mass, that the accretor rotates only at 10-40\% of their 
critical velocity, even though their orbital periods exceed 5\,d. 

\subsection{Overluminosity}
\label{ovl}

\begin{figure}
\centering
\includegraphics[width=\hsize]{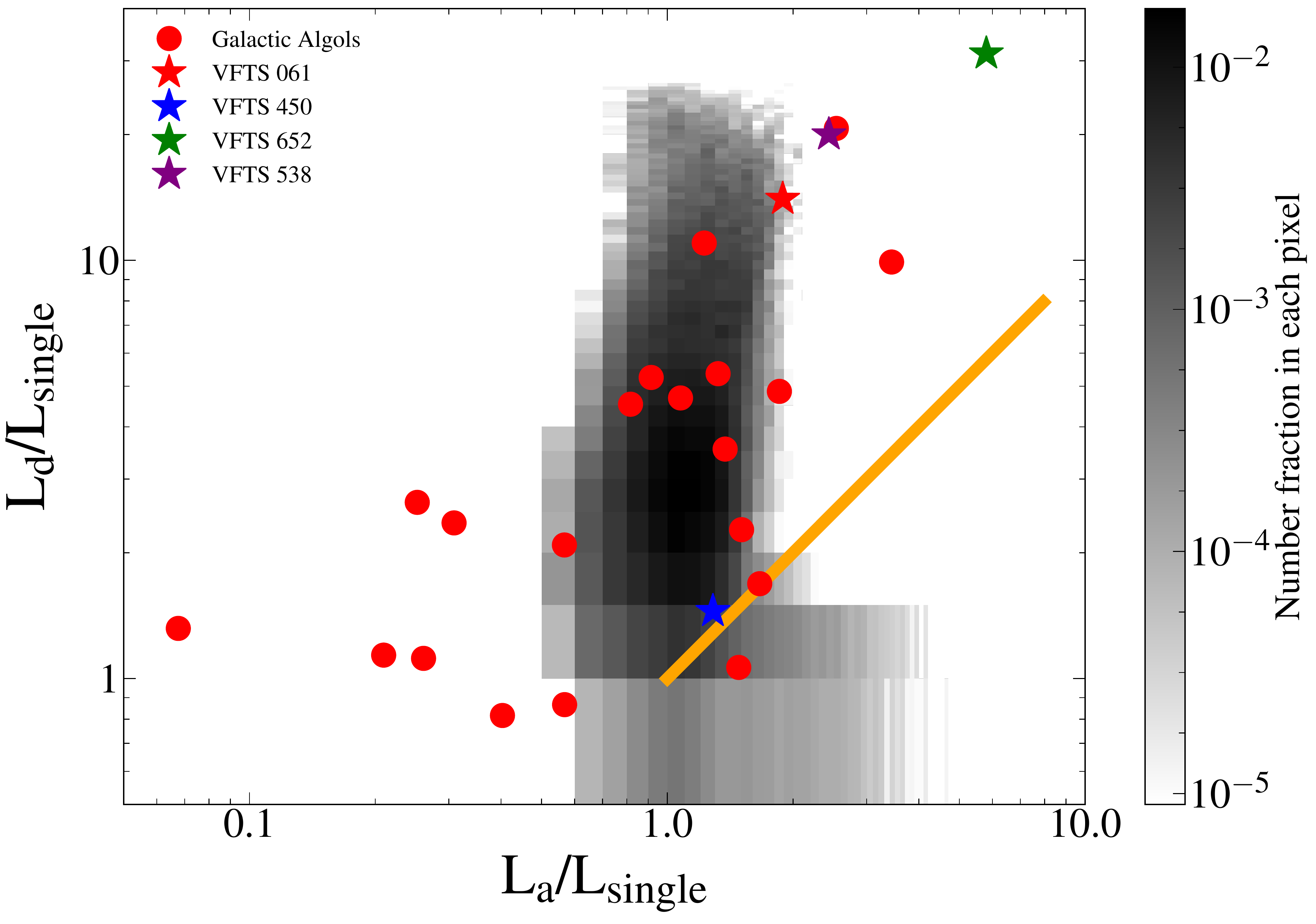}
\caption{Probability distribution of the ratios of the luminosities
of the Algol donor and accretor models over the luminosities of single star models
of the same mass and effective temperature (grey-scale, cf., description in Fig.\,\ref{hessHe}). Red circles and coloured star symbols denote the corresponding parameters for the Galactic and LMC Algol systems, respectively (see Table \ref{table_Lov}). The orange line shows where the luminosity ratios of donor and accretor are equal. }
\label{hessOVL}
\end{figure}

We compare the luminosity of the individual stellar components 
of the observed semi-detached binaries in the LMC and Galaxy to 
the luminosity of a single star that has the same mass and 
effective temperature. For this, we use the 
single star models of \citet{brott2011} with 
initial rotational velocity of $\sim$200\,km/s. We create 
50,000 points with random masses between 
3-80\,$M_{\odot}$ and random ages between zero and terminal 
age main sequence. Through interpolation we 
assign a luminosity and effective temperature to each of the 50,000 
points. Finally, we use the mass and effective 
temperature of an observed Algol binary component to find the 
nearest point on the stellar mass-effective temperature plane 
and note its luminosity. The ratio of the observed luminosity
to this single star luminosity defines our overluminosity.

Table \ref{table_Lov} lists 
the luminosities of the observed semi-detached binaries and the 
derived overluminosities; see also Fig.\,\ref{hessOVL}. 
We find that the observed mass donors 
are generally significantly overluminous for their mass.
Exceptions are AI\,Cru and $\delta$\,Pic, but their 
underluminosity is so small (10 and 20\%, respectively)
that an overluminosity can not be excluded. 
On average, the donors are overluminous by a factor of seven,
with the record holder (VFTS\,652) showing a factor of\,30.

Such large overlumnosities are expected for stars whose average
mean molecular weight $\mu$ is much larger than in single stars, 
since for chemical homogeneous stellar models of mass $M$, their
luminosity $L$ behaves as $L \sim M^{\alpha} \mu^{\beta}$ 
\citep{Kippenhahn1990}, where, in the mass range $10\dots30\,$M$_{\odot}$, 
the exponent $\beta$ is $5\dots 2.5$ \citep{grafener2011,Kohler2015}. 
As the mean molecular weight in the stellar core increases by a factor of 
$\sim 2.2$ during hydrogen burning, and as the donor stars may be 
almost entirely stripped of their hydrogen envelope, we expect 
overluminosities of up to a factor of 50 for the donor stars 
\citep{wellstein2001}. The observed large overluminosities 
are therefore a direct confirmation of the loss of a large fraction 
of the envelope of the Roche-lobe filling components of the Algol 
systems. 

For the accretors, we expect a much smaller effect, since 
most of the transferred matter from the donor is comprised of its
unenriched envelope. The accretor could be slightly underluminous
in case it does not undergo complete rejuvenation \citep{braun1995}, or 
it could be slightly overluminous due to
the accretion of some helium-enriched material, but both effects
are expected to be small. 

Figure\,\ref{hessOVL} shows the distribution of the overluminosities 
of our Algol model components in comparison with the observed values. 
We see that the accretors cluster around an overluminosity factor of 
one. However, we find several stars with significantly smaller and 
larger values. Here, we can only speculate about the reasons for this 
mismatch. One interesting feature is that the smallest overluminosity 
factors of the accretors occur in systems which also have small 
overluminosities of their donors, and the most overluminous accretors 
are accompanied by the most overluminous donors. The cloud of points 
is stretched in the direction parallel to the yellow line, which 
indicates equal donor and accretor mass. Potentially, any error in 
the distance would move observed points in this direction. While such 
errors could well affect the location of the Galactic Algols in this 
diagram, this is unlikely for the VFTS binaries in the LMC. On the 
other hand we note that most of the observational data for the Galactic 
Algols has been obtained and analysed several decades ago\footnote{For 
example, \citet{McSwain2008} note on the most extreme outlier in 
Fig.\,\ref{hessOVL}, BF\,Cen, that `Even the eclipsing double-lined 
spectroscopic binary BF Centauri (=HD 100915), a member of NGC 3766, 
has been largely neglected by modern spectroscopic observations.'}. 
As it stands, the average of the observationally derived overluminosities 
of the accretors does agree with our model prediction, but the observed 
distribution is much broader. 

\begin{table*}[t]
\centering
\caption{Overluminosity of observed Algol binaries in the Galaxy and the LMC, whenever available. }
\label{table_Lov}
\begin{tabular}{l r r r r r r r r r r}
\hline
\hline
Name & Period & $M_{\rm a}$\;\;\;\;\;\;\;\;\, & $M_{\rm d}$\;\;\;\;\;\;\;\;\, & log $L_{\rm a}$ & log $L_{\rm d}$ &  $L_{\rm a}/L_{\rm single}$ & $L_{\rm d}/L_{\rm single}$ & log ($L_{\rm a}$/$M_{\rm a}$) & log ($L_{\rm d}$/$M_{\rm d}$) & Ref\\    
     & days & M$_{\odot}$\;\;\;\;\;\;\;\;\, & M$_{\odot}$\;\;\;\;\;\;\;\;\, &  &  &  &  &  &  & \\    
\hline
TT Aur  & 1.33 & 8.10\;\;\;\;\;\;\;\;\, & 5.40\;\;\;\;\;\;\;\;\, & 3.72 & 3.24 & 1.7 & 1.7 & 2.81 & 2.51 & [1]  \\
$\mu_{1}$ Sco & 1.44 &  8.30$\pm$1.00 & 4.60$\pm$1.00 & 3.66 & 3.20 & 1.4 & 3.5 & 2.74 & 2.54 & [2] \\
BF Cen   & 3.69 &  8.70\;\;\;\;\;\;\;\;\, & 3.80\;\;\;\;\;\;\;\;\, & 2.82 & 2.55 & 0.06 & 1.3 & 1.88 & 1.97 & [1]  \\
IZ Per   & 3.68 &  9.97$\pm$0.55 & 3.20$\pm$0.17 & 3.85 & 2.57 & 0.5 & 2.3 & 2.85 & 2.06 & [3] \\
AI Cru   & 1.41 & 10.30$\pm$0.20 & 6.30$\pm$0.10 & 3.88 & 3.24 & 0.6  & 0.9 & 2.87 & 2.44 & [4] \\
MP Cen   & 2.99 & 11.40$\pm$0.40 & 4.40$\pm$0.20 & 3.80 & 3.00 & 0.25 & 2.63 & 2.74 & 2.35 & [5]  \\
IU Aur   & 1.81 & 11.99$\pm$0.08 & 6.07$\pm$0.04 & 4.64 & 4.22 & 3.4 & 9.9 & 3.56 & 3.43 & [6]  \\
V356 Sgr & 8.89 & 12.10\;\;\;\;\;\;\;\;\, & 4.70\;\;\;\;\;\;\;\;\, & 3.81 & 3.02 & 0.3 & 2.3 & 2.73 & 2.35 & [1]  \\
V Pup    & 1.45 & 14.00$\pm$0.50 & 7.30$\pm$0.30 & 4.28 & 3.97 & 1.7  & 8.1 & 3.13 & 3.10 & [7] \\
$\delta$ Pic  & 1.67 & 16.30\;\;\;\;\;\;\;\;\, & 8.60\;\;\;\;\;\;\;\;\, & 4.32 & 3.68 & 0.4 & 0.8 & 3.10 & 2.74 & [8] \\
HH Car   & 3.23 & 17.00\;\;\;\;\;\;\;\;\, &14.00\;\;\;\;\;\;\;\;\, & 4.73 & 4.91 & 1.8 & 4.8 & 3.50 & 3.76 & [1]  \\
V337 Aql & 2.73 & 17.44$\pm$0.31 & 7.83$\pm$0.18 & 4.73 & 4.20 & 1.1  & 4.7 & 3.49 & 3.30 & [9]  \\
AQ Cas   &11.70 & 17.63$\pm$0.91 &12.50$\pm$0.81 & 4.94 & 4.59 & 1.48  & 1.1 & 3.69 & 3.49 & [10]  \\
XZ Cep   & 5.09 & 18.70$\pm$1.30 & 9.30$\pm$0.50 & 5.05 & 4.79 & 1.3  & 11 & 3.78 & 3.82 & [11]  \\
AB Cru   & 3.41 & 19.75$\pm$1.04 & 6.95$\pm$0.65 & 5.21 & 4.58 & 2.5  & 20.6 & 3.91 & 3.74 & [12] \\
V448 Cyg & 6.51 & 24.70$\pm$0.70 &13.70$\pm$0.70 & 4.63 & 4.57 & 0.3 & 1.1 & 3.23 & 3.43 & [13] \\
QZ Car   & 5.99 & 30.00$\pm$3.00 &20.00$\pm$3.00 & 5.18 & 5.46 & 0.9  & 5.2 & 3.70 & 4.16 & [14]  \\
\hline
VFTS 061 & 2.33 & 16.30$\pm$1.40 & 8.70$\pm$0.60 & 4.77 & 4.75 & 1.9  & 14 & 3.55 & 3.81 & [15]  \\
VFTS 652 & 8.59 & 18.10$\pm$3.90 & 6.50$\pm$1.10 & 5.35 & 4.92 & 5.8  & 31 & 4.09 & 4.10 & [15]  \\
VFTS 538 & 4.15 & 18.30$\pm$1.90 & 11.80$\pm$1.40 & 4.96 & 5.31 & 2.4 & 20 & 3.70 & 4.24 & [15]  \\
VFTS 450 & 6.89 & 29.00$\pm$4.10 & 27.80$\pm$3.90 & 5.30 & 5.46 & 1.3 & 1.5 & 3.84 & 4.01 & [15]  \\

\hline
\end{tabular}
\tablebib{
(1) \citet{surkova2004}; (2) \citet{budding2015}; (3) \citet{hilditch2007}; (4) \citet{bell1987}; 
(5) \citet{terrell2005}; (6) \citet{surina2009}; (7) \citet{stickland1998};
(8) \citet{evans1974}; (9) \citet{tuysuz2014}; (10) \citet{ibanoglu2013}; (11) \citet{martins2017}; 
(12) \citet{lorenz1994}; (13) \citet{djurasevi2009}; (14) \citet{walker2017}; (15) \citet{mahy2019b,mahy2019a}.
\tablefoot{We give here the luminosity of the accretors ($L_{\rm a}$) and donors ($L_{\rm d}$) of massive semi-detached double-lined eclipsing binaries in the Galaxy and the LMC, and derive the ratio of their respective luminosity to that of single star models ($L_{\rm single}$) of the same mass and effective temperature. Masses and luminosities are in Solar units.}
}
\end{table*}

\section{Comparison with earlier work}
\label{earlier_work}

\subsection{Properties before Case\,AB mass transfer}

Previous studies on the Case\,A mass transfer phase were focussed 
on low to intermediate mass Algol binaries having masses 
0.1\dots8 $M_{\odot}$ \citep[][]{mennekens2017,negu2018}.
\cite{mennekens2017} compared the observed Algol systems 
having B-type companions with binary stellar evolution 
models in the solar neighbourhood. We note that while 
their models reached up to 17 $M_{\odot}$ for the initial 
donor mass, the observed binaries that they compare their 
models to reached a maximum mass of $\sim$8 $M_{\odot}$.
They found evidence for non-conservative mass transfer. 

In the low to intermediate star mass range, their models 
predict a peak in orbital periods of Algols between 5-10\,d, 
and mass ratios (inverting their definition of mass ratio 
to match with ours) around 2. While their predicted mass 
ratio distribution matches with ours, we note that the most 
probable orbital periods are higher in their predictions. 
At the same time, the orbital period distribution of the 
observed Algol binaries, that they compare their model 
predictions to, has a peak between 2.5-4\,d, which is 
closer to our predicted distribution of orbital periods. 
They conclude that the mass transfer in the intermediate-mass 
Algol binaries is non-conservative \citep[see also][]{VanRensbergen2010}. 

Evidence of low mass transfer efficiency was found in the 
study by \cite{petrovic2005}, who compare observations of 
three short-period WR+O binaries with their binary models. 
Investigation into the mass transfer efficiency 
(Fig.\,\ref{mass_trans_eff_fast_period}) of our models also 
reveals that models that do not merge during the main sequence 
have typically low mass transfer efficiencies and, as an 
outcome of how our mass transfer is calculated, decreases 
with increasing orbital period. However, we note that we 
predict the shortest period observed Algol binaries might 
have undergone a largely conservative mass transfer and is 
destined to eventually merge later during the slow Case\,A 
mass transfer phase. 

Evidence of an orbital period dependent mass transfer efficiency 
was found by \citet{selma2007}, who compared observed massive 
Algol binaries in the Small Magellanic Cloud to detailed binary 
evolution models. They found that the mass transfer efficiency 
of their models, that can reproduce the observations, has to 
decrease with increasing orbital period. We have shown that our 
accretor spin-up dependent mass transfer efficiency does reproduce 
this orbital period dependency naturally (Fig.\,\ref{mass_trans_eff_fast_period}), 
giving additional credibility to our modelling of the mass transfer phase. 

\cite{selma2014} studied the effects of a binary population on a 
synthetic massive star population assuming a binary fraction of 
0.7. They assumed that, in their standard simulation, binaries 
with initial mass ratios less than 0.65 merge. They found that 
3\% of their total population of models to be binaries in the 
semi-detached phase. We also predicted the same fraction of such 
systems from our binary model grid, which can be said to 
be well in harmony with the former study, when the uncertainties 
related to mass transfer efficiency, angular momentum loss are 
taken into consideration. In our models, the dividing line (in 
initial mass ratio) between stable and unstable mass transfer 
depends on the initial donor mass (c.f. Fig.\,\ref{1.350_summary}, 
upper and lower panels of \ref{1.100_summary}), with more low initial 
mass ratio binaries undergoing stable Case\,A mass transfer at 
higher initial donor masses. By inspection, we found that lower 
initial mass ratio models spend less time in the slow Case\,A 
mass transfer phase, such that the overall prediction of the 
properties of Algol systems are not significantly affected by 
the smaller initial mass ratio models that also undergo stable 
mass transfer in our binary grid.

The predicted velocity semi-amplitude of the semi-detached models 
in \citet[][fig.\,2]{selma2014} matches well with the orbital velocity 
distribution of the accretors in our model grid (Fig.\,\ref{hessVorb}). 
However, the orbital velocities of the donor are considerably higher 
on average than predicted by the former study. This is in part be due 
to the fact that \cite{selma2014} accounted for the random 3D orientation 
of the plane of the orbit, while we only report the absolute value of 
the orbital velocity. 

However, incorporating this randomness in orientation will also reduce 
the average orbital velocity of the accretors and the resolution of this 
conflict requires a deeper investigation than we do here. We also note 
that the calculation of this phase in \cite{selma2014} was done using a 
rapid binary stellar evolution code, which does not calculate in detail 
through the mass transfer phase but rather uses fitting 
recipes \citep{tout1997,hurley2002} to simulate the properties of the 
stars during and after the mass transfer phase. 

The general properties of our models (Sect.\,\ref{general_props}) are 
in agreement with the binary models of \citet{pols1994} who find that 
their models having initial donor mass of 16\,M$_{\odot}$ and orbital 
periods below 1.6\,d enter into contact during the slow Case\,A mass 
transfer phase (c.f. Fig.\,\ref{1.100_summary}). Moreover, as in \citet{pols1994}, 
we also see that our models having low orbital periods and low mass ratios 
enter into a thermal timescale contact during the fast Case\,A mass transfer 
phase (compare hatching in bottom panel of Fig.\,\ref{1.100_summary} with 
black frames around coloured squares in right-middle panel of Fig.\,\ref{eta_fast_slow_ab}). 

\subsection{Properties after Case\,AB mass transfer}

Recently, \cite{langer2020} investigated the properties of black hole+OB star 
binaries in the LMC using the same grid of binary evolution models 
as is used in our work. Comparing their predictions of the mass ratio and 
orbital period of black hole+OB binaries arising from Case\,A mass transfer 
(figs. 4 \& 6), we find that our prediction of the most probable mass 
ratios and orbital periods (Fig.\,\ref{obs_period_CaseAB}) after the 
Case\,AB mass transfer phase are very similar to their predicted 
distribution. We expect this to be the case because the models do not 
go through any more mass transfer phases, before the collapse of the 
initially more massive star, that can significantly alter the binary 
properties \citep[see however][]{laplace2020}. Likewise, predictions of 
surface nitrogen mass fraction enhancement and rotational velocity of 
the mass accretor after the Case\,AB mass transfer (Fig.\,\ref{Nenh_vrot_donors_CaseAB}) 
are very similar to the corresponding properties of the OB star companions 
to the BHs in \citet[][fig.\,8\,\&\,9]{langer2020}.

\section{Conclusions}
\label{conclusions}

Semi-detached binaries pose a crucial test for massive binary 
evolution models. On one hand, their evolution is typically 
simpler than that of the shortest period binaries, whose 
components may swap mass back-and-forth and evolve through 
long-lived contact stages \citep{pablo2016,Menon2021}. On the 
other hand, longer-period binaries undergo thermal timescale 
mass transfer after the donor star has exhausted hydrogen in 
its core, after which their binary nature remains elusive 
\citep{selma2014}.

The nuclear timescale semi-detached phase leads to the prediction 
that about 3\% of all massive binaries in the LMC, with initial 
donor masses 10\dots40\,M$_{\odot}$ and orbital periods above 
1.4\,d, should be found in this stage, which corresponds to a 
significant population in any star-forming galaxy. They are easily 
identified, since they are photometrically variable and show large 
radial velocity variations. While they undergo nuclear-timescale 
mass transfer currently, they underwent a phase of thermal timescale 
mass transfer previously. It is the latter which leads to the most 
drastic changes of the binary properties. This thermal timescale 
mass transfer is crucial for all mass transferring binaries, and 
the semi-detached binaries allow us to examine its consequences.

Based on our large grid of $\sim 10\,000$ detailed binary evolution 
models \citep{pablothesis}, we predict the distribution functions 
of the orbital properties of massive semi-detached binaries (orbital 
period, period derivative, orbital velocities, mass ratio) as well 
as the surface properties of both stellar components (luminosity, 
temperature, chemical composition, spin). A comparison with the 
available observations evidently confirms the classical Algol scenario 
\citep{Crawford1955,Kopal1971} for most of the observed massive 
semi-detached systems, most spectacularly by the observed large 
overluminosities of the donor stars by up to a factor of 30 
(Fig.\,\ref{hessOVL}). These, as most other observables, are found 
in fair agreement with the model predictions. 

Some necessary corrections of our models are also signified by the 
observations. The observed period and mass ratio distribution demonstrates 
that some semi-detached binaries did undergo nearly conservative mass 
transfer, while others can only be explained assuming very non-conservative 
evolution (Fig.\,\ref{hessSD}). This empirical range of accretion efficiencies 
is well reproduced by our models (cf., Appendix\,\ref{appendix_summary_models}). 
But a more efficient mass accretion in the lowest considered mass range 
($10\dots 20$M$_{\odot}$; Fig.\,\ref{hessSD_mass_dependence}), perhaps 
mediated by a stronger tidal coupling than the one adopted here, could 
help to understand the Algol systems with the highest mass ratios.
New binary models are required to explore this. 

However, the impetus from the current models can not yet be fully 
harvested, since the number of well observed massive Algol systems 
is small. For the LMC, large progress has been made by the TMBM 
\citep{almeida2017} and BBC \citep{Villasenor2021} surveys. Through 
detailed analyses, \citet{mahy2019b,mahy2019a} derived the most 
relevant observational constraints for our models for four TMBM 
targets. However, in agreement with our predictions, the Algol 
binaries comprise only a few percent of the targets in these surveys. 
Of the putative 90 LMC Algols above $10\,$M$_{\odot}$, we have basic 
data for seven, and highest quality analyses for four. While we know 
more massive Algols in the Milky Way, the data here is scarce, very 
heterogeneous and often without error estimates. This calls for a 
high quality multi-epoch survey dedicated to massive Algol binaries. 

When confronted with the results of such a survey, we expect more 
crisp constraints on the input physics of our models (see App.\,\ref{uncertainties}). 
In addition, model grids for solar and sub-LMC metallicity would form a big step 
forward to understand the impact of metallicity on the mass transfer process. 
While parameter variations are difficult for model grids as large as the 
one analysed here, the best studied Algol binaries could be attempted 
to be reproduced by tailored models where the input physics could be 
varied. As semi-detached massive binaries may be progenitors of 
magnetic main sequence stars \citep{schneider2016,schneider2019,takahashi2021}, 
hydrogen-poor supernovae \citep{Yoon2010,Dessart2020,Stanway2020}, 
neutron star and black hole binaries \citep{Tauris2006,VanBever2000,langer2020}, 
and double-compact binary mergers \citep{Eldridge2016,kruckow2018}, 
pursuing these routes promises to be fruitful.

\begin{acknowledgements}

We thank the anonymous referee, Dany Vanbeveren, Jan J. Eldridge and Mario Spera for helpful comments that refined the submitted version of the manuscript. SdM and AM acknowledge partial financial support from the European Union’s Horizon 2020 research and innovation program  (ERC, Grant agreement No. 715063), and from the Netherlands Organization for Scientific Research (NWO) as part of the Vidi research program BinWaves with project number 639.042.728. PM acknowledges support from the FWO junior postdoctoral fellowship No. 12ZY520N. AM was also supported by the Alexander von Humboldt foundation. LM thanks the European Space Agency (ESA) and the Belgian Federal Science Policy Office (BELSPO) for their support in the framework of the PRODEX Program. This project has received funding from the European Research Council under European Union's Horizon 2020 research programme (grant agreement No 772225: MULTIPLES). This research has made use of NASA’s Astrophysics Data System and the VizieR catalogue access tool, CDS, Strasbourg, France.

\end{acknowledgements}

\bibliographystyle{aa}
\bibliography{caseA}

\begin{appendix}

\newpage

\section{Discussion and uncertainties}
\label{uncertainties}
\subsection{Envelope inflation}

\cite{sanyal2015} investigated the single star models of \cite{brott2011} 
at LMC metallicity and found that models exceeding $\sim$40 $M_{\odot}$ 
reach their Eddington limit inside the stellar envelope during core 
hydrogen burning. This leads to a large expansion of the stellar 
envelope, which makes the star grow to red supergiant proportions 
during core hydrogen burning. Our models, with very similar physics 
assumptions as those of \cite{brott2011}, are also expected to show 
similar behaviour. Indeed, for the next higher initial donor mass 
to be considered (44.7 $M_{\odot}$), the envelope of the donor 
inflates and leads to unstable mass transfer which MESA is unable 
to calculate through. In this sense, the upper limit on the initial 
donor mass in our grid is a computational result. For a more 
comprehensive discussion of this topic, we refer the interested 
readers to sect. 4.1 of \cite{langer2020}. We discuss the parts 
relevant to our work.

The implication of inflation above $\sim$40 $M_{\odot}$ at the 
LMC metallicity is that all binary models having initial donor 
masses above $\sim$40 $M_{\odot}$ will undergo Case\,A 
mass transfer since the donors will expand to red supergiant 
proportions during core hydrogen burning. Such donors 
have inflated convective envelopes at the onset of Roche-lobe 
overflow \citep{sanyal2015}. The subsequent mass transfer due 
to Roche-lobe overflow from stars with convective envelopes is 
expected to be unstable as the radius of a star with convective 
envelope increases when mass is lost from the envelope \citep{quast2019,Ge2020}. 
\cite{langer2020} assume that, in the absence of detailed model 
calculations, mass transfer due to Roche-lobe overflow from an 
inflated mass donor will lead to an unstable mass transfer and 
the binary would merge.

Due to inflation and the inability of MESA to calculate through 
the mass transfer phase of such models, our models cannot predict 
the observable properties of binary systems in semi-detached 
configurations originating from initial donor masses greater than 
$\sim$40 $M_{\odot}$. In the TMBM sample of semi-detached systems, 
we find a system VFTS 094 in which both components have masses 
around $\sim$30 $M_{\odot}$. This might be an observed system 
which shows evidence of stable mass transfer during fast Case\,A 
when the mass of the donor was arguably greater than $\sim$40 
$M_{\odot}$. However, the error bars are too big to draw any 
conclusive arguments. The system VFTS 450 also shows that both 
components have high dynamical masses $\sim$30 $M_{\odot}$ and 
is a possible candidate to have undergone stable mass transfer 
above the cut-off mass for inflation to set in our models. 

We find another system LH 54-425 \citep{williams2008} in which the 
more massive component is estimated to have a mass of $\sim$47 
$\pm$ 2 $M_{\odot}$. The binary system R136-38 \citep{massey2002} 
is estimated to have a star with mass $\sim$56.9 $\pm$ 0.6 $M_{\odot}$. 
We note that the 04f + 06V eclipsing binary system Sk—67$^{\circ}$105 
is found to have one component which has mass greater than 40 
$M_{\odot}$, and both the components are nearly filling their 
Roche lobe. The existence of these observed high mass systems 
provides a motivation for a comprehensive investigation of the 
mass transfer phase in mass donors with inflated envelopes. 

However, we note that inflation at LMC metallicity sets in 
at 40 $M_{\odot}$ during the late stage of core hydrogen burning 
and the estimated age of LH 54-425 and R136 is 1.5-2 Myrs. Hence, 
these systems might not have inflated yet. On the other hand, 
the authors that studied Sk—67$^{\circ}$105 reported that it is 
in a very distorted configuration and is likely that its components 
are undergoing very strong interactions. 

If inflated stars do undergo stable mass transfer in nature, then 
the orbital period distribution of our semi-detached models can 
be taken as skewed to lower periods. However, the initial mass 
function and the empirical determination of the distribution of 
orbital periods in binaries are heavily weighted towards lower 
masses and orbital periods. Moreover, the time spent in Case\,A 
mass transfer phase decreases with the increase in the initial 
orbital period of the binary. Hence, we do not expect a significant 
impact on the predicted orbital period distributions as a whole. 
However, the properties of very massive binaries (> $\sim$30 $M_{\odot}$) 
in the Algol configuration can be quite different than comparatively 
lower mass Algol binaries (see for eg. Fig.\,\ref{observed_q}).

\subsection{Mass transfer efficiency and stability of mass transfer}

Observations of binary systems which have undergone a mass transfer 
phase indicate that the mass transfer efficiency varies from binary 
to binary. While some binaries require models to have low mass transfer 
efficiency \citep{langer2003}, others indicate a need for higher mass 
transfer efficiency \citep{wellstein1999}. It has also been argued 
that binaries with lower mass ratio experience lower mass transfer 
efficiency \citep{petrovic2005}. \citet{selma2007} also find hints 
for a low mass transfer efficiency for higher period binaries. However, 
a recent study by \citet{deschamps2015} reported that direct observational 
imprints of mass loss due to non-conservative mass transfer might not 
be visible in Algol systems. 

Our mass transfer model based on the principle of decreasing 
mass transfer efficiency with increasing rotation of the 
mass accreting star does in theory take into consideration 
these variations and we do see a trend of decreasing mass transfer 
efficiency with increasing period and decreasing mass ratios. 
This model needs excess mass to be removed from the binary so 
that the mass gainer does not exceed critical rotation. We 
model this by requiring that the combined photon energy from 
both the stars in the binary is larger than the gravitational 
energy needed to remove the excess unaccretable mass due to 
the accretor reaching critical rotation. If this condition fails,
the evolution of the model is stopped and assumed to merge. 
However, as already demonstrated, our mass transfer scheme 
is able to reproduce the observed population of Algol binaries 
in the Galaxy and LMC. 

Our criterion to determine the stability of mass transfer 
is simple. The dividing line between stable and unstable 
mass transfer in our models is a function of the initial 
orbital period and mass ratio for a particular initial donor 
mass of the binary. For a discussion, we direct the interested 
reader to fig.\,2 of \cite{langer2020}. 

\cite{wang2020} has shown that a large fraction of the Be 
star population in NGC 330 \citep{milone2018} is well 
represented by binary evolution models. To correctly predict 
their numbers in comparison to the number of mergers, however, 
they posit that the boundary between stable and unstable 
mass transfer has to be relaxed such that there are more 
models that can undergo stable mass transfer. Doing such 
would increase the number of Case\,A mass transfer systems 
with lower initial mass ratios. Inspection of our model grid 
shows that the time spent in the semi-detached phase 
decreases with lower mass ratios, such that models having 
initial mass ratio less than 0.5 spend less than 10\% of 
their main sequence lifetime in the semi-detached phase. Hence, 
relaxing the criterion for stable mass transfer will not 
significantly affect the results we derive in this work.

\subsection{Semiconvection}

We are aware that recent studies on single star models 
\citep{schootemeijer2019} argue for a higher efficiency 
of semiconvection in stellar models. We note that a 
higher semiconvective efficiency than what is implemented 
in our models will make the mass gaining stars rejuvenate 
\citep{braun1995} and develop a core-envelope structure 
similar to that of a single star \citep{braun1995}. The 
core mass and stellar radius of rejuvenated accretors 
will be larger and smaller, respectively, compared to 
our nonrejuvenated models \citep[compare model 47 and 48 of][]{wellstein2001}. 
This implies that our binary models enter into a contact 
configuration and can undergo L2 overflow (purple models 
in Fig.\,\ref{1.350_summary}) at longer orbital periods 
compared to models calculated with a higher semiconvective 
efficiency, due to the larger radii of our accretors 
\citep[compare Fig.\,\ref{1.350_summary} to Fig.\,A1 of][]{Menon2021}. 

Rejuvenated accretors also have a greater remaining core 
hydrogen burning lifetime than nonrejuvenated accretors. 
Accordingly, our binary models are more likely to undergo 
inverse mass transfer from a post main sequence accretor 
onto a main sequence donor (orange models in Fig.\,\ref{1.350_summary} 
than binary models calculated using 
a high semiconvective efficiency. After Case\,AB mass 
transfer, our nonrejuvenated accretors may live their 
core helium burning lifetime as a blue supergiant instead 
of a red supergiant \citep{philip1992}. However, the amount 
of increase in the core masses of mass gainers due to 
rejuvenation depends on the amount of mass that can be 
successfully accreted \citep{braun1995}. We showed that 
the mass transfer efficiency is low in most of our models 
that survive the Case\,A mass transfer phase 
(Fig.\,\ref{mass_trans_eff_fast_period}). 

From the above discussion, we can speculate that more 
short period binaries will contribute to our predicted 
Algol population for models with efficient semiconvection. 
As such, the peak in the predicted orbital period distribution 
may be shifted to a lower value, although we cannot quantify 
it. The maximum lifespan of the slow Case\,A phase is also 
expected to increase since fewer models will merge due to L2 
overflow during slow Case\,A mass transfer. This may increase 
our predicted number of Algol binaries in the LMC. As less 
binaries will undergo contact and inverse slow Case\,A mass 
transfer, the peak in the predicted surface abundance is 
also expected to increase to higher values, especially the 
nitrogen surface enhancement. Since shorter period binaries 
undergo more conservative Case\,A mass transfer in our mass 
transfer scheme, we expect more contribution from models that 
undergo conservative mass transfer to the population of 
WR/He+OB star binaries. 

\section{Additional figures for surface abundances during slow Case\,A}
\label{surface_abundance_additional_figures}

\begin{figure*}
{\centering
\includegraphics[width=0.5\hsize]{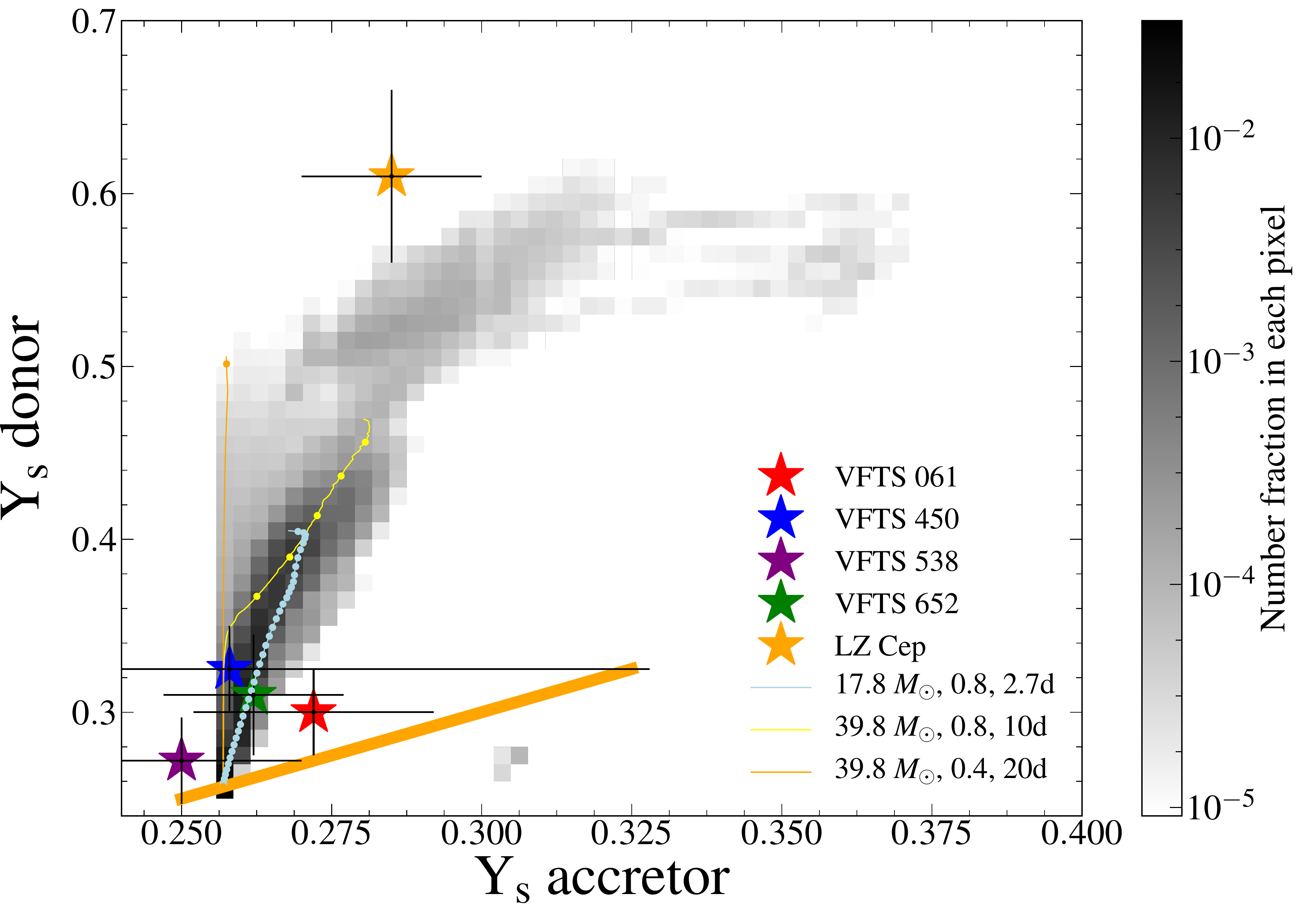}
\includegraphics[width=0.5\hsize]{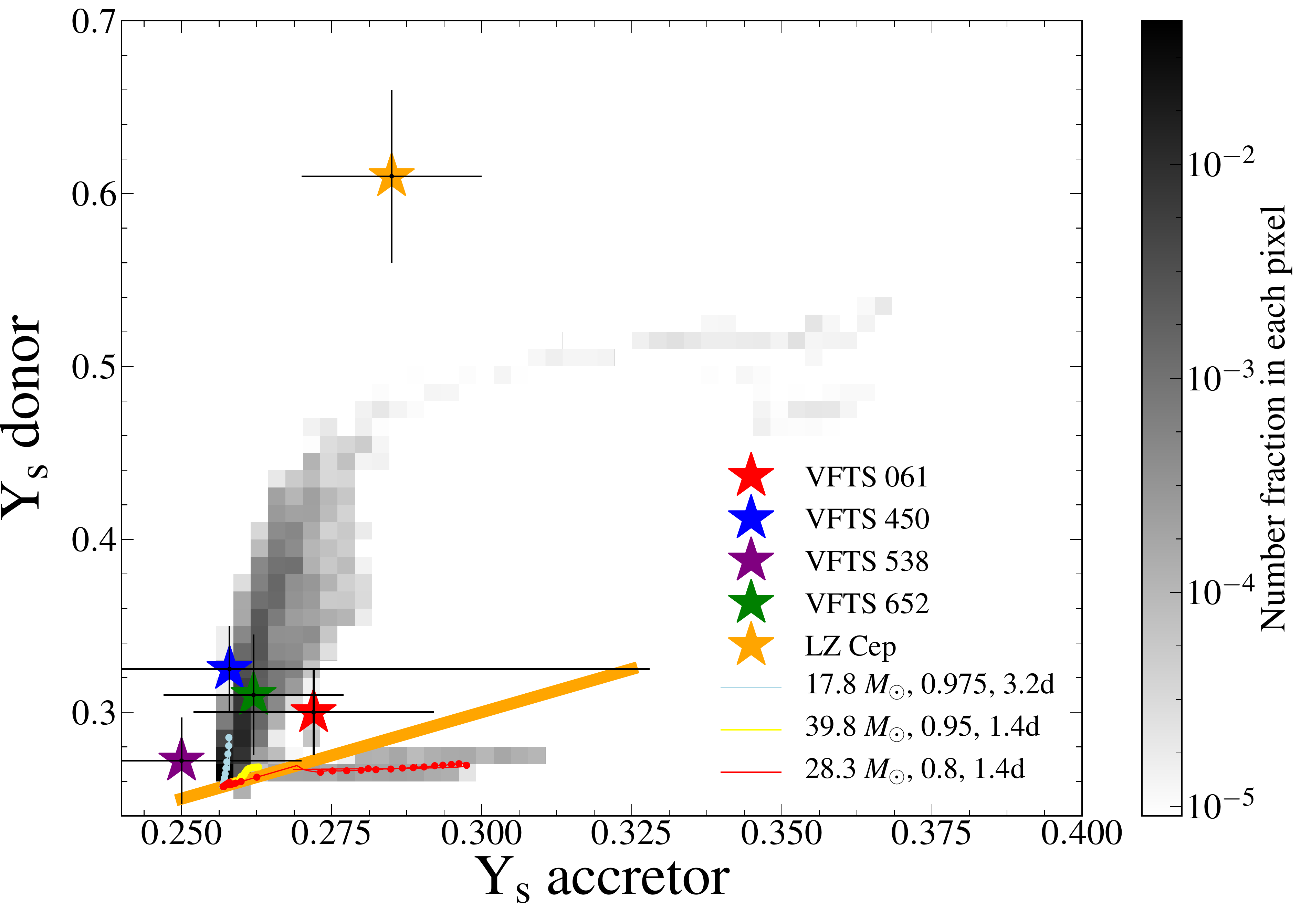}
}
{\centering
\includegraphics[width=0.5\hsize]{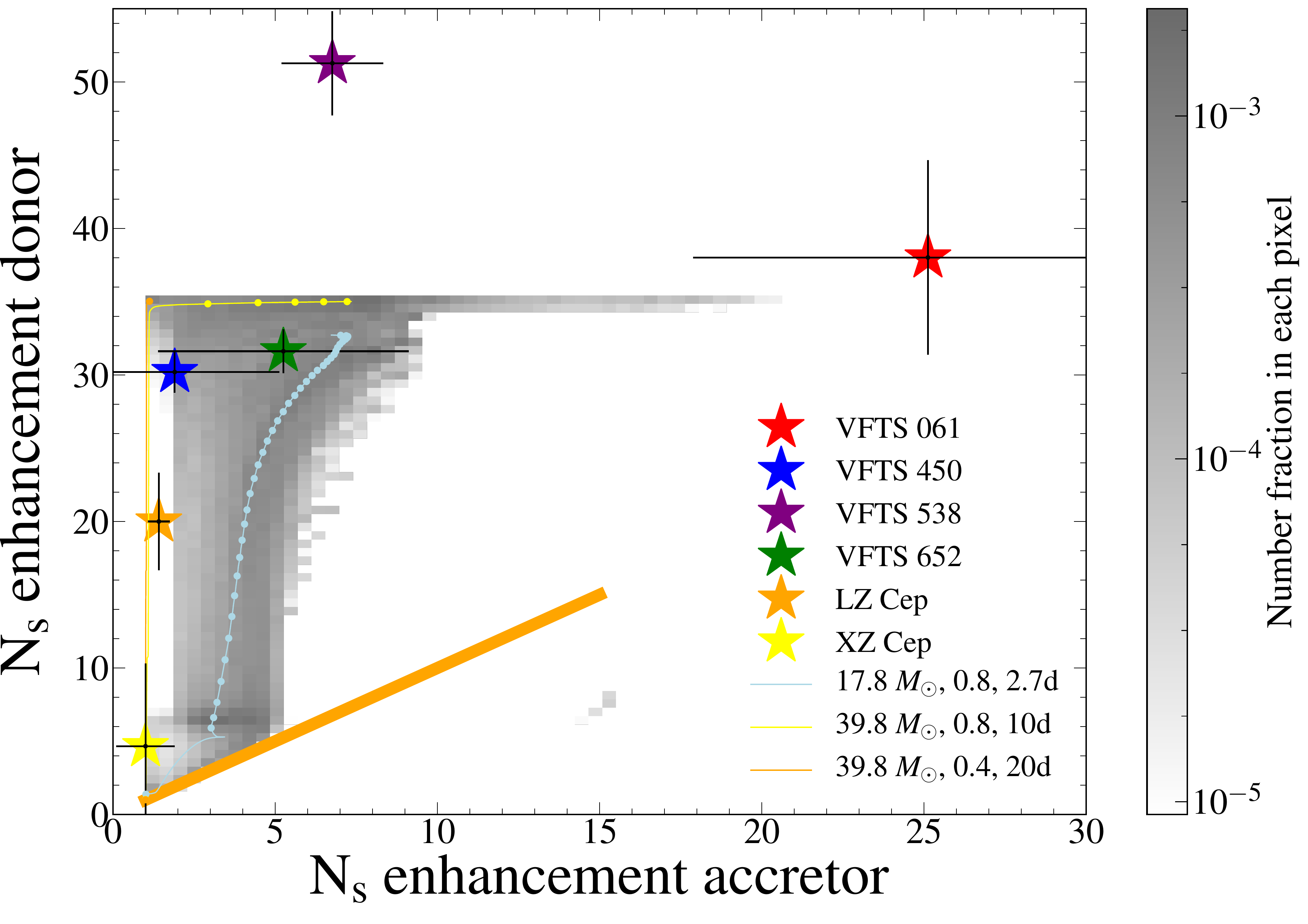}
\includegraphics[width=0.5\hsize]{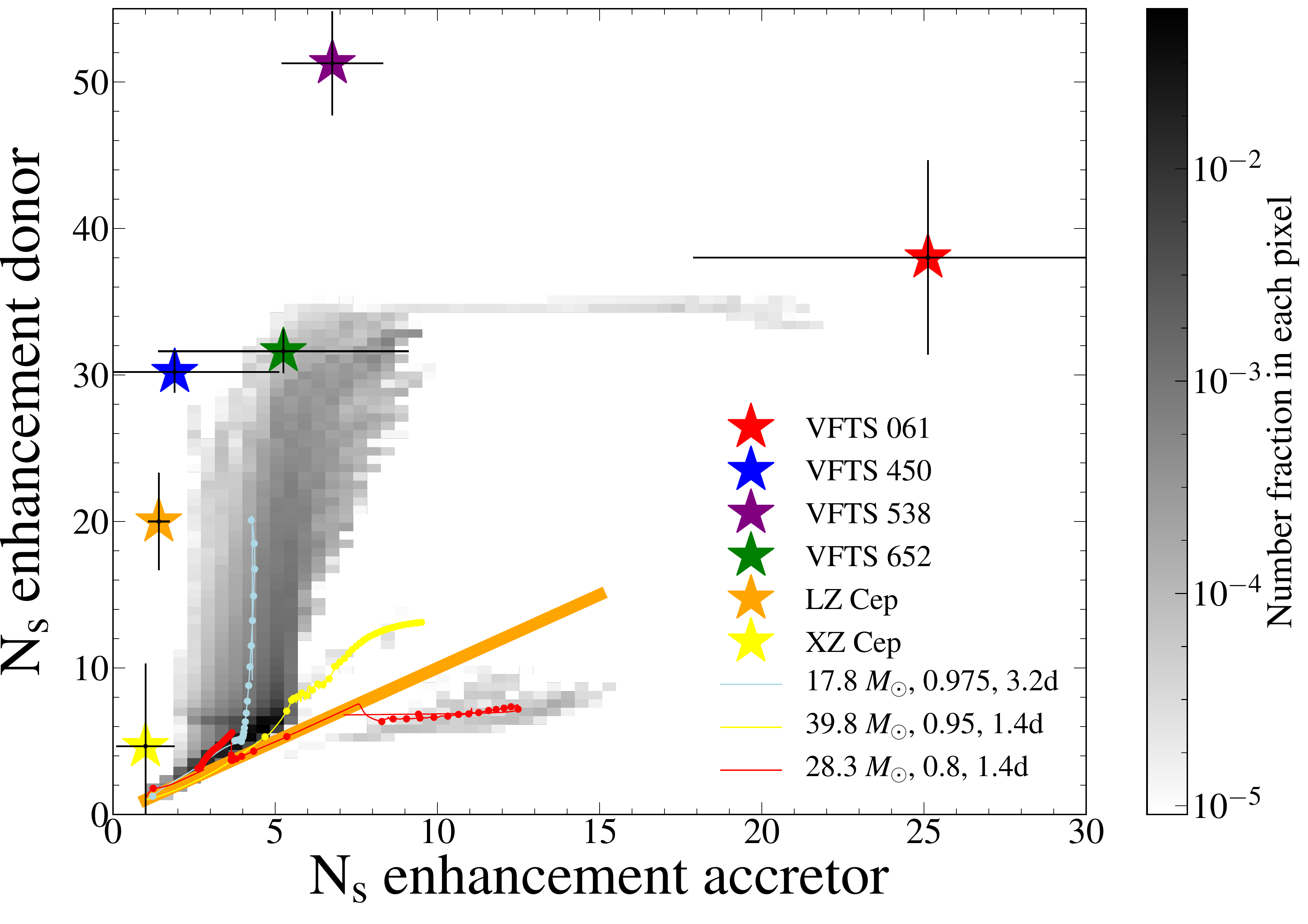}
}
\caption{Probability distribution of the surface helium mass fraction (top panels) and surface nitrogen mass fraction enhancement (bottom panels) of donor vs the accretor that is predicted to be observed in the semi-detached configuration of the Case\,A mass transfer phase. The left panels show the contribution to Fig.\,\ref{hessHe} from models that survive the Case\,A mass transfer phase while the right panels show the contribution to Fig.\,\ref{hessHe} from models that eventually merge during the slow Case\,A phase. The different coloured `stars' with error bars denote the position of the semi-detached systems of the TMBM survey \citep{mahy2019a} and Galaxy. The thick orange line indicates where the surface helium mass fraction or surface nitrogen enrichment of the donor and accretor is the same. The thin coloured lines show the evolution of surface abundances during the main sequence with initial parameters (donor mass, mass ratio, period) as given in the legend. The corresponding coloured dots are at 50000 years during the semi-detached phase. The grey-scale gives the probability fraction in each pixel. The total probability is normalised such that the integrated sum over the entire area is 1.}
\label{hessN_survivors}
\label{hessN_mergers}
\end{figure*}

\begin{figure*}
{\centering
\includegraphics[width=0.48\hsize]{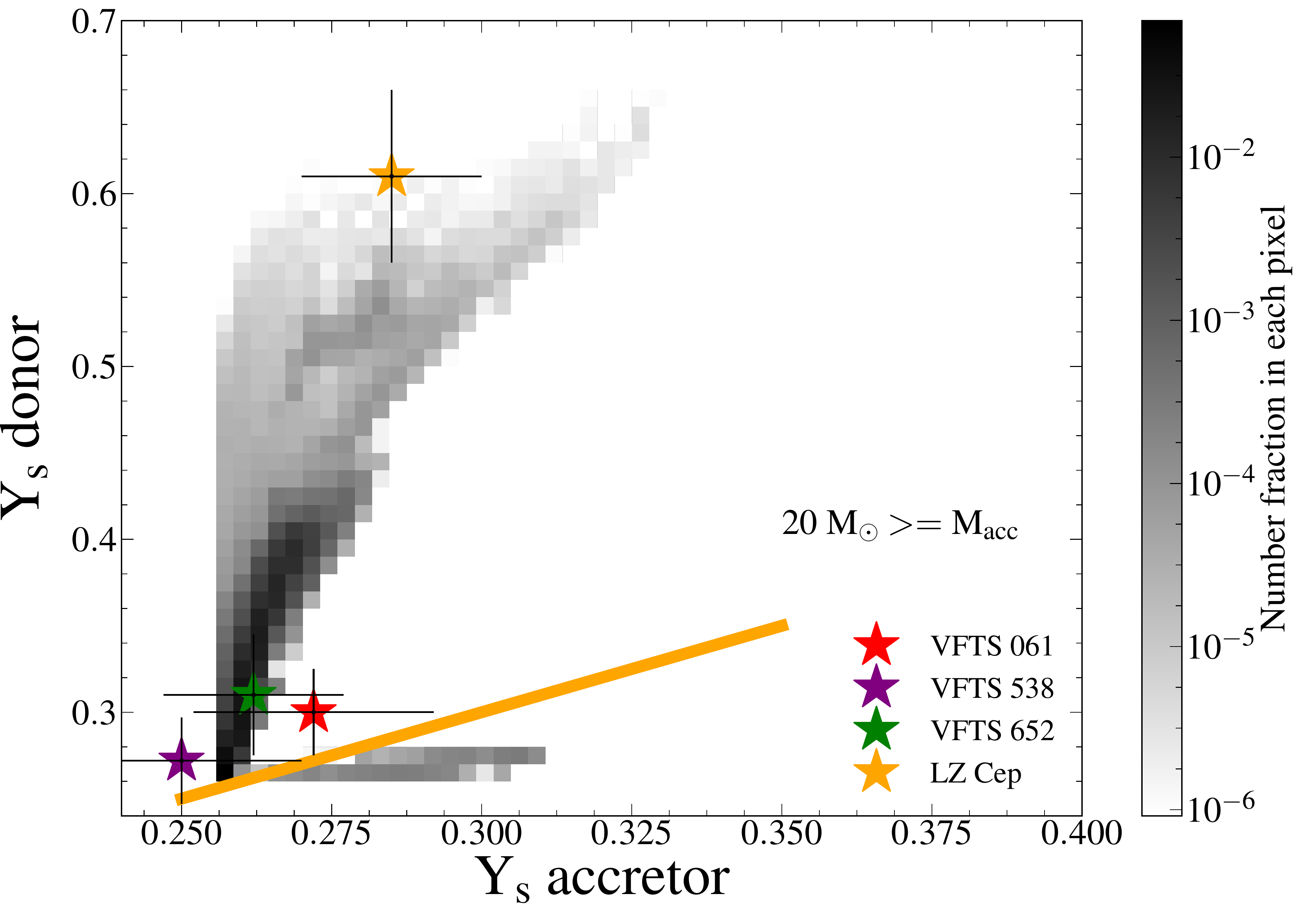}
\includegraphics[width=0.48\hsize]{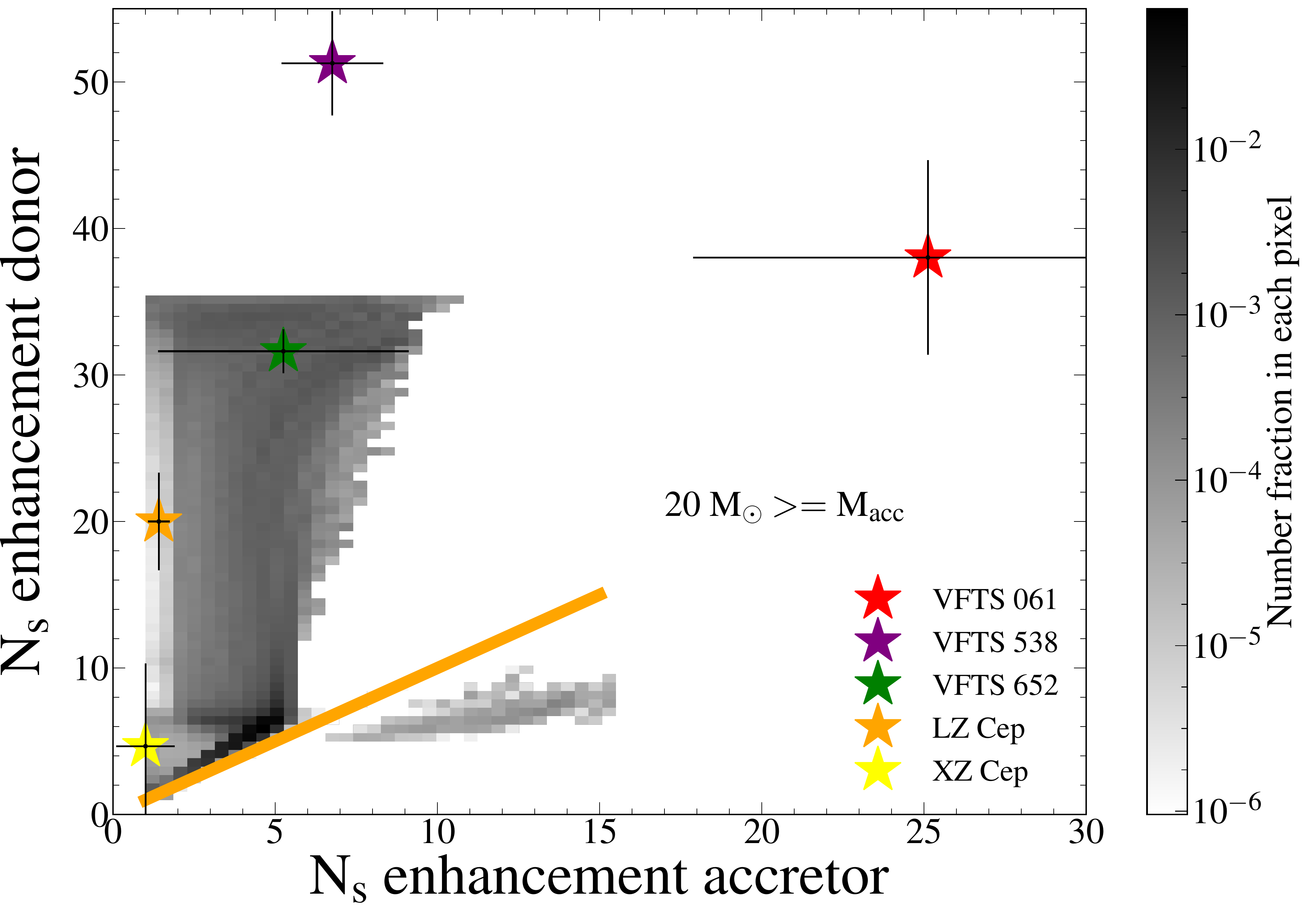}}
{\centering
\includegraphics[width=0.48\hsize]{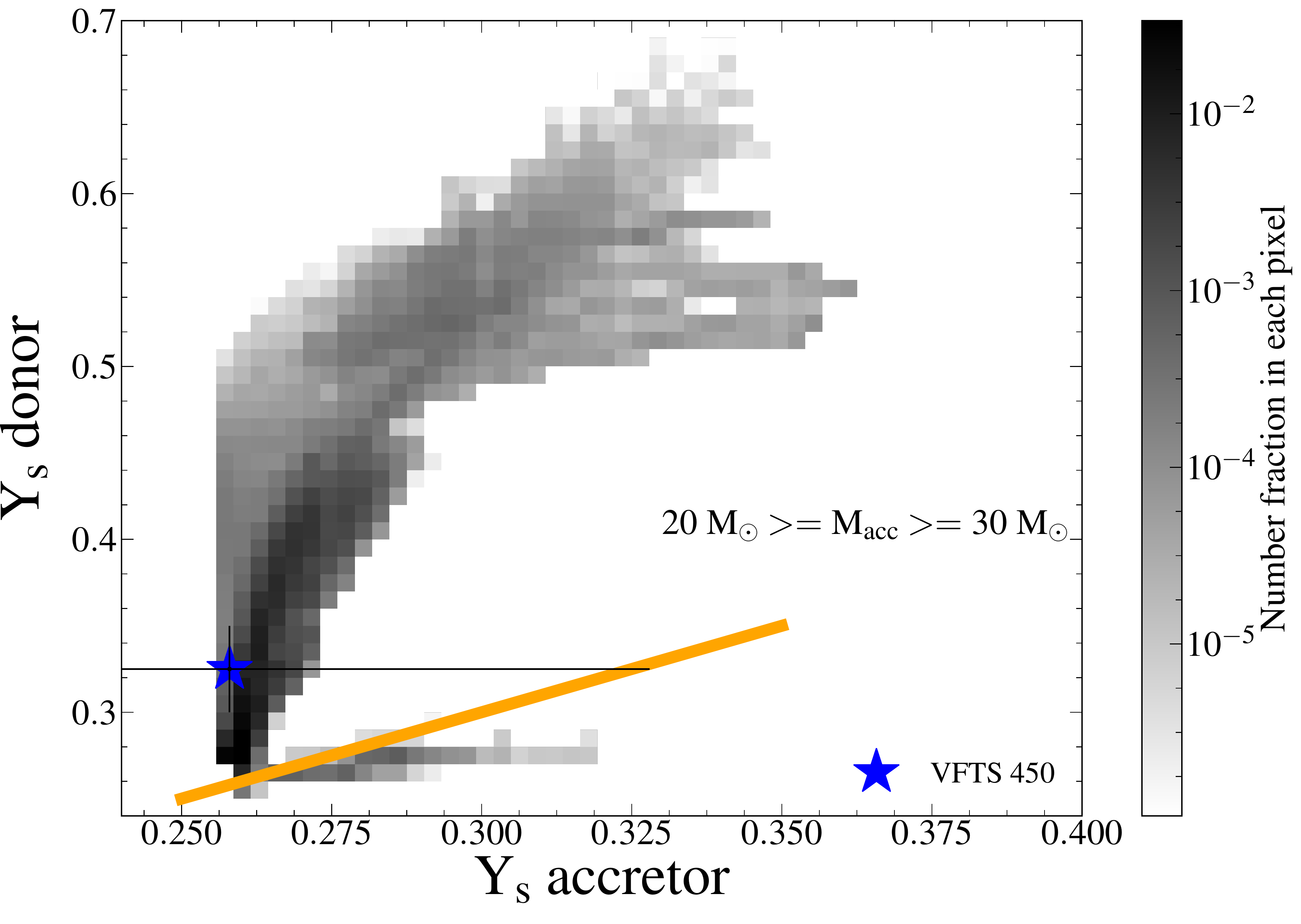}
\includegraphics[width=0.48\hsize]{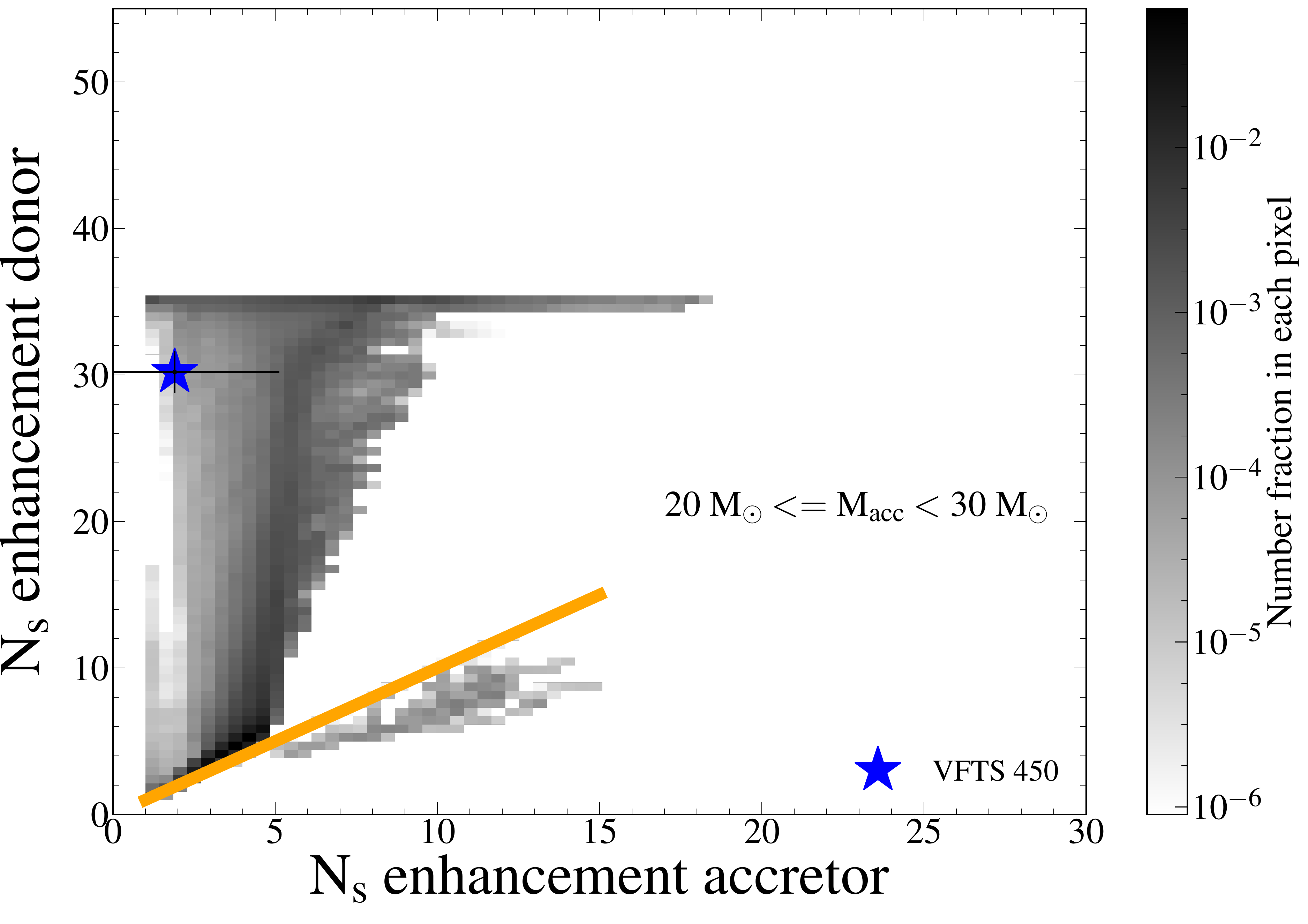}}
{\centering
\includegraphics[width=0.48\hsize]{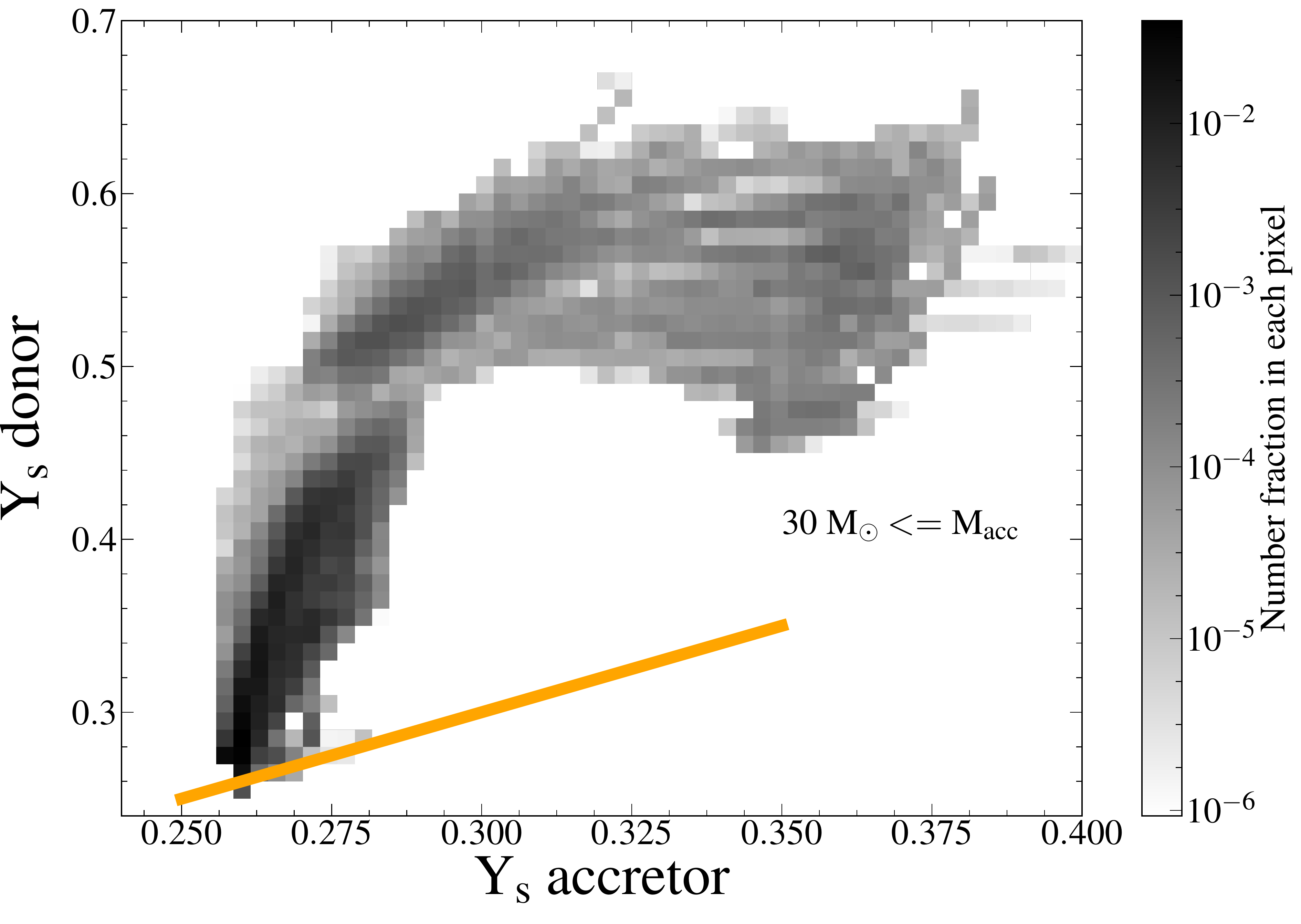}
\includegraphics[width=0.48\hsize]{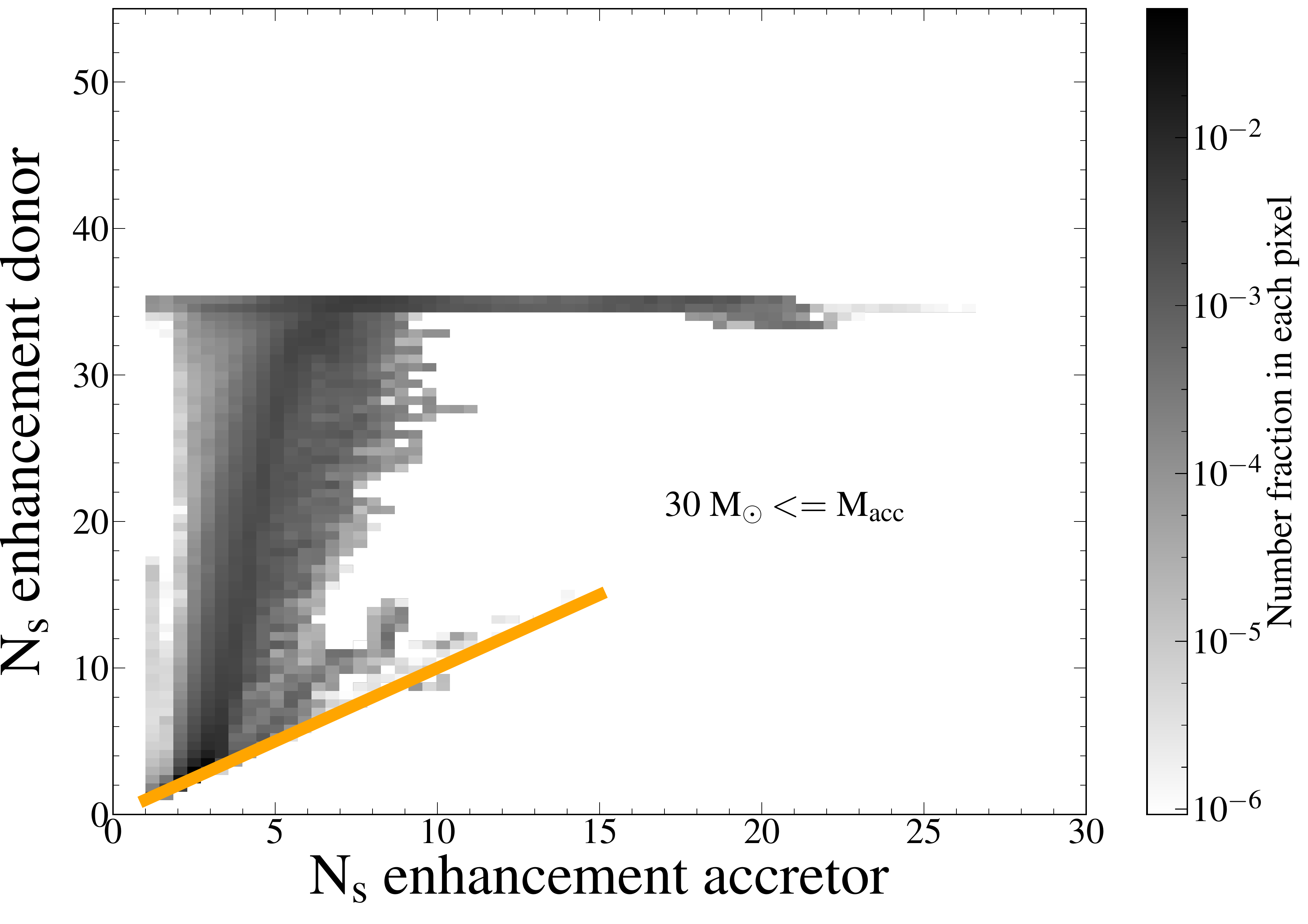}}
\caption{Probability distribution of the surface helium mass fraction (left three panels) and surface nitrogen mass fraction enhancement factors (right three panels) of the donor vs the accretor for different ranges of accretor masses during the semi-detached phase: below 20 $M_{\odot}$ (top panels), 20-30 $M_{\odot}$ (middle panels) and above 30 $M_{\odot}$ (bottom panels). The different coloured stars symbols with error bars denote the positions of the semi-detached systems from the TMBM survey \citep{mahy2019b,mahy2019a}. The orange line indicates where the donor and accretor abundances are the same. The grey-scale gives the probability fraction in each pixel. The total probability is normalised such that the sum over the entire area equals unity.}
\label{hessHeN_10-20-30-40}
\end{figure*}

Here, we show the prediction of surface helium mass fraction 
and surface nitrogen mass fraction enhancement for only the models that survive 
the Case\,A mass transfer phase (Fig.\,\ref{hessN_survivors}) or 
only for models that merge during the Case\,A mass transfer 
(Fig.\,\ref{hessN_survivors}). We 
also provide additional figures of the surface helium mass fraction 
and nitrogen mass fraction enhancement of donors and accretors during the slow 
Case\,A mass transfer phase, for different ranges of accretor 
masses of the binary during the semi-detached configuration 
(Fig.\,\ref{hessHeN_10-20-30-40}). 

\newpage

\section{Mass dependence of orbital period and mass ratio during slow Case\,A}

Here, we provide additional figures of the orbital period 
and mass ratio during the slow Case\,A mass transfer phase, 
for different ranges of accretor masses of the binary. 

\begin{figure}
\begin{minipage}{1.0\linewidth}
\includegraphics[width=\hsize]{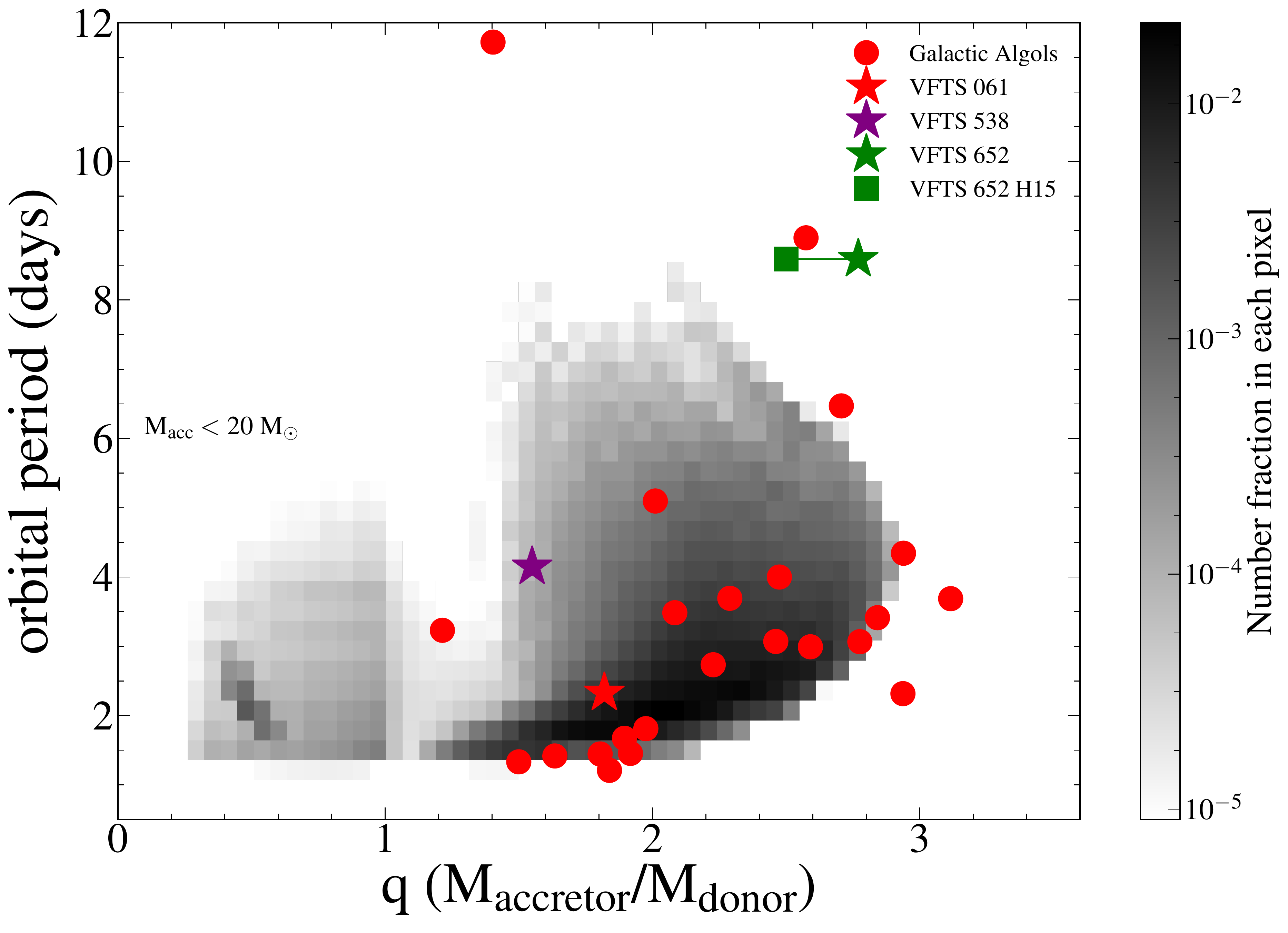}
\end{minipage}
\begin{minipage}{1.0\linewidth}
\includegraphics[width=\hsize]{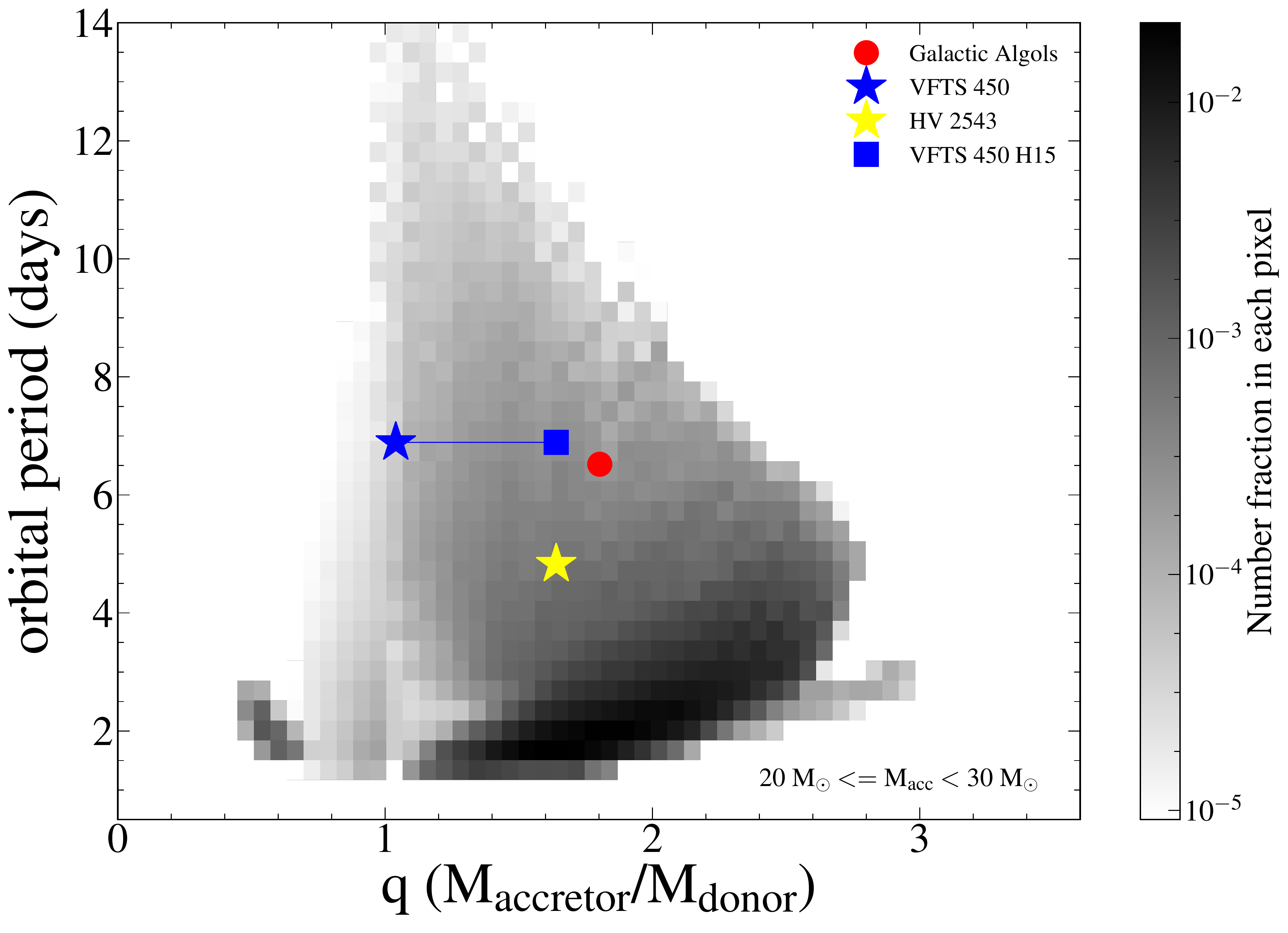}
\end{minipage}
\begin{minipage}{1.0\linewidth}
\includegraphics[width=\hsize]{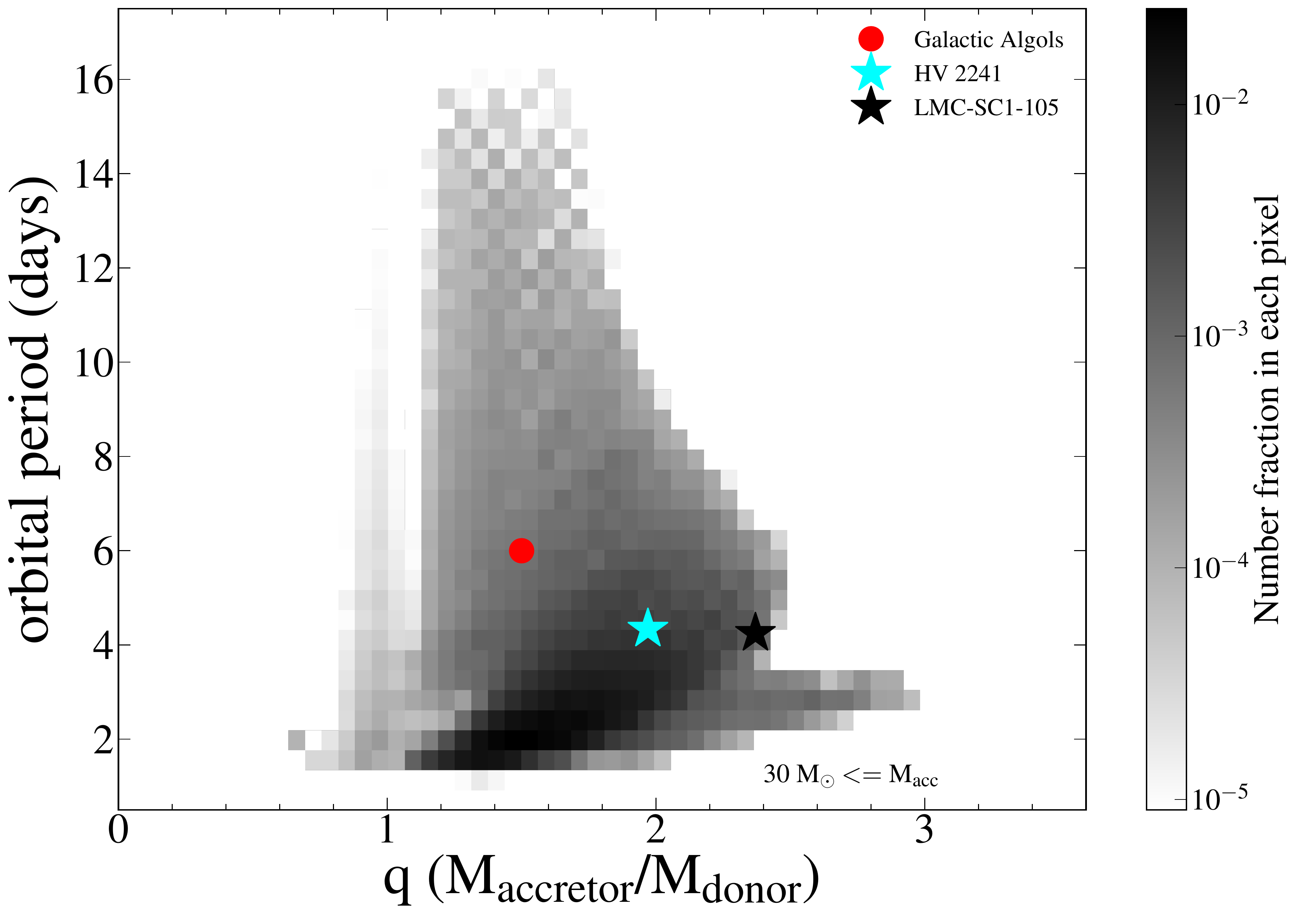}
\end{minipage}
\caption{Probability distribution of the orbital period and mass ratio that is predicted to be observed in the semi-detached configuration of the Case\,A mass transfer phase for models with accretor mass below 20 $M_{\odot}$ (top panel), 20-30 $M_{\odot}$ (middle panel) and above 30 $M_{\odot}$ (bottom panel). The grey-scale gives the probability fraction in each pixel. The total probability is normalised such that the sum over the entire area is 1.}
\label{hessSD_mass_dependence}
\end{figure}

\section{Orbital velocities}

\begin{figure}
\centering
\includegraphics[width=\hsize]{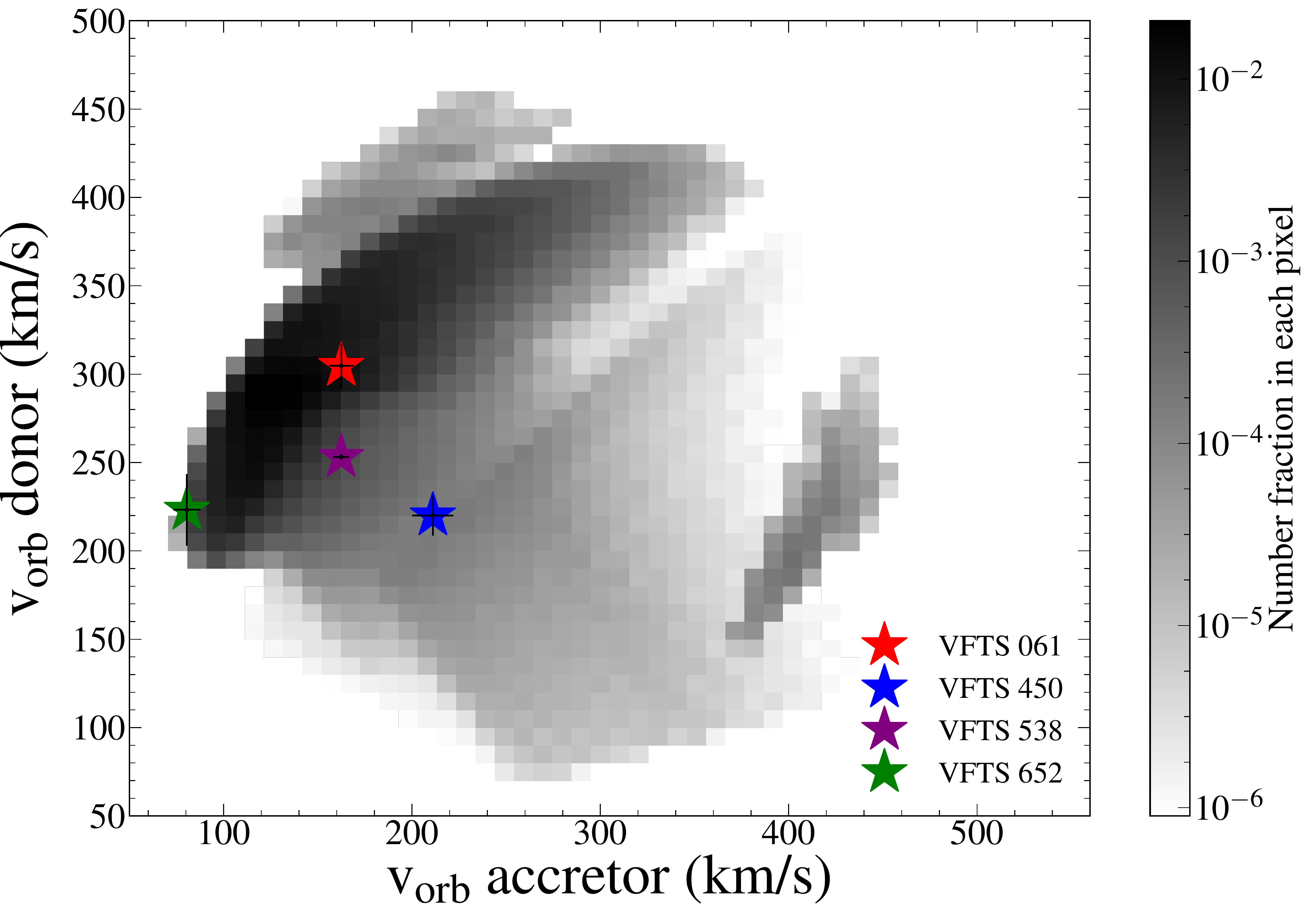}
\caption{Probability distribution of the orbital velocities of the donor and the accretor in the semi-detached configuration. The different coloured stars with error bars denote the position of the semi-detached systems of the TMBM survey \citep{almeida2017}. Grey-scale: See description in Fig.\,\ref{hessHe}.}
\label{hessVorb}
\end{figure}

Figure\,\ref{hessVorb} shows the probability distribution 
of the orbital velocities of the two components during 
the semi-detached phase. In the darker shaded regions, 
we see that the donor has a higher orbital velocity 
than the accretor. This indicates that the donor 
is currently the less massive component of the system.

\section{Orbital period derivatives}

\begin{figure}
\begin{minipage}{\linewidth}
\includegraphics[width=\hsize]{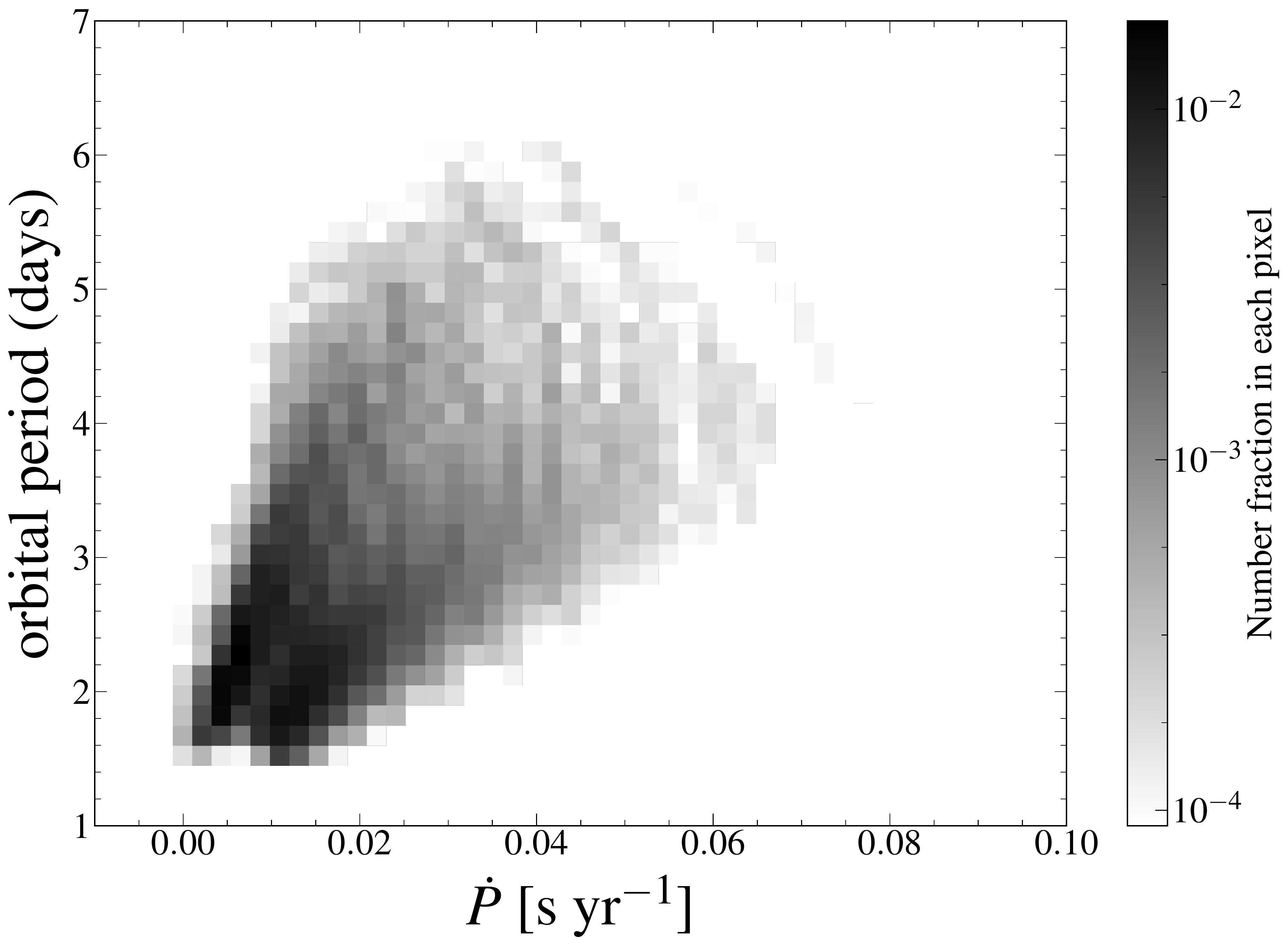}
\end{minipage}
\begin{minipage}{\linewidth}
\includegraphics[width=\hsize]{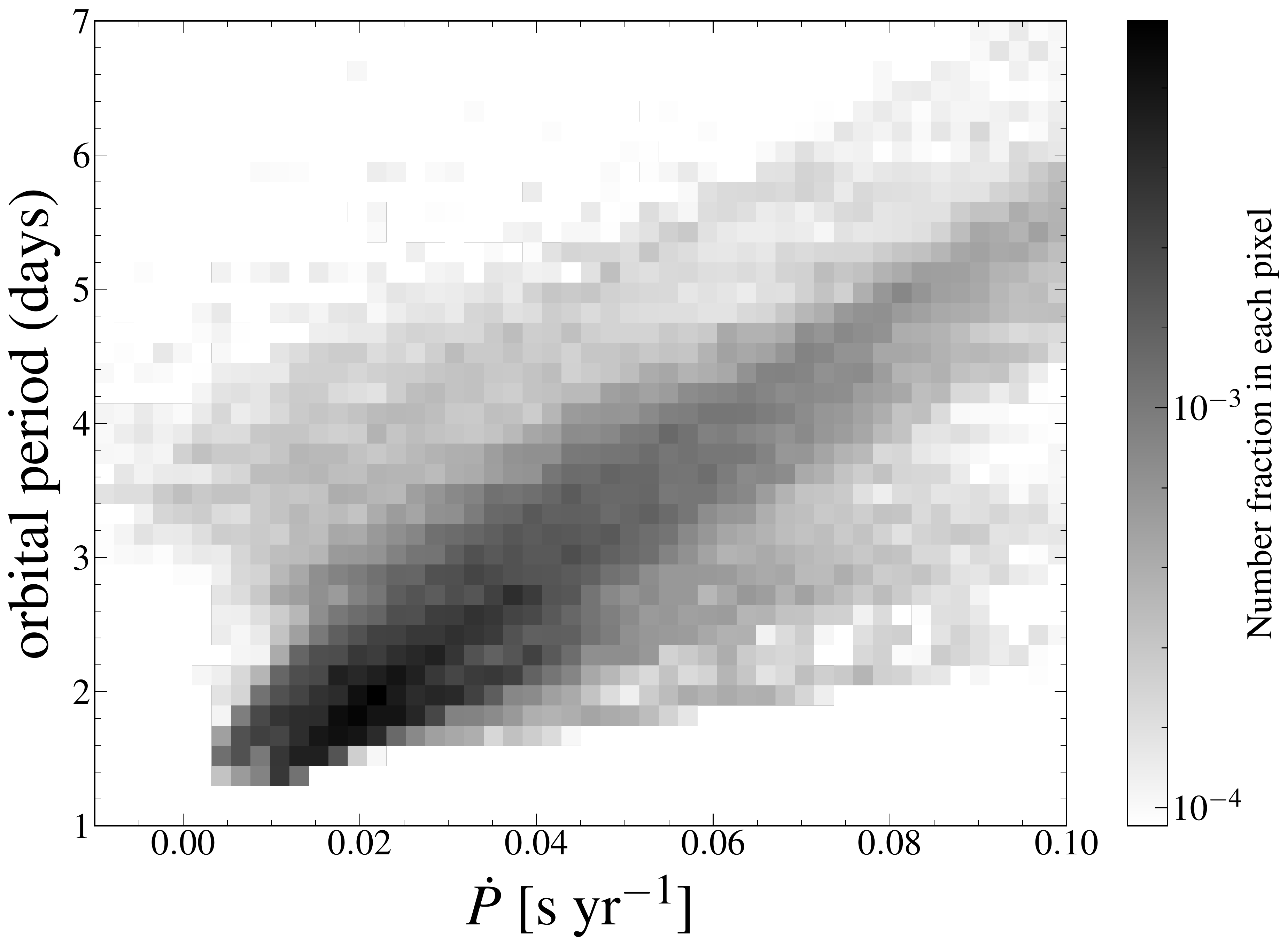}
\end{minipage}
\caption{Same as Fig.\,\ref{observed_Pdot}, but only for initial donor mass of 10-15 $M_{\odot}$ (top panel) and 30-40 $M_{\odot}$ (bottom panel). 
}
\label{hess_Pdot_diff_ranges}
\end{figure}

Figure\,\ref{hess_Pdot_diff_ranges} shows the orbital period 
derivatives during the semi-detached phase for different 
ranges of initial donor masses.

\section{Summary of the outcome of binary models for different initial donor masses in our grid}
\label{appendix_summary_models}
\begin{figure*}
\centering
\includegraphics[width=\hsize]{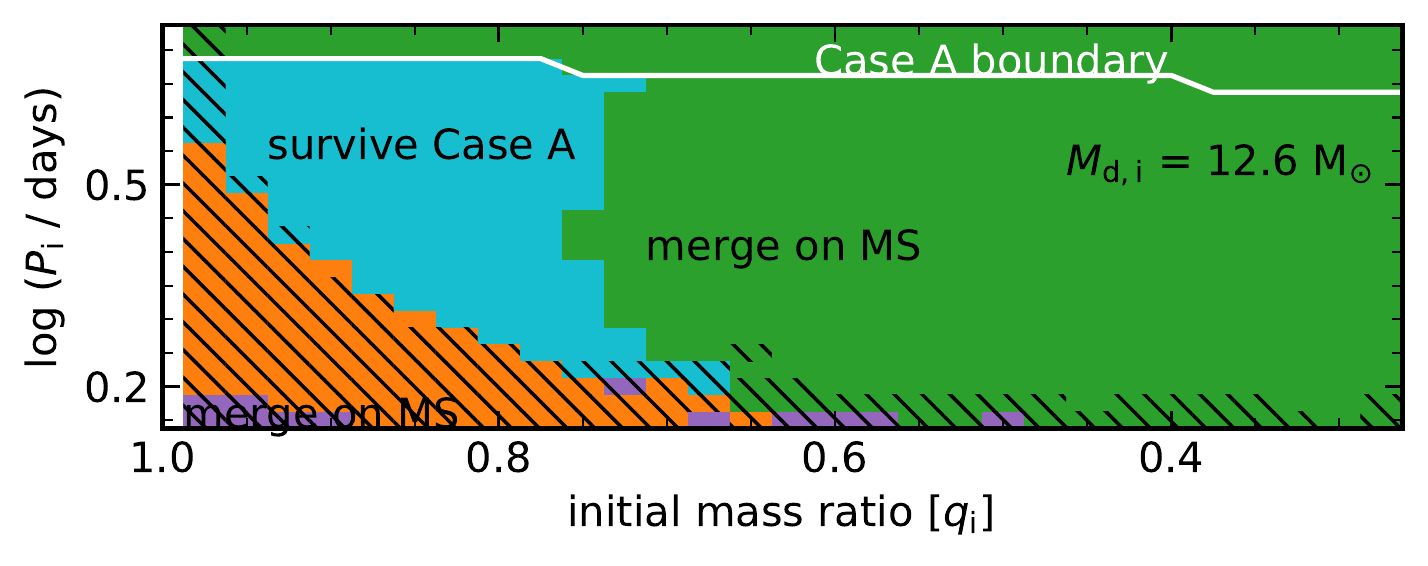}
\includegraphics[width=\hsize]{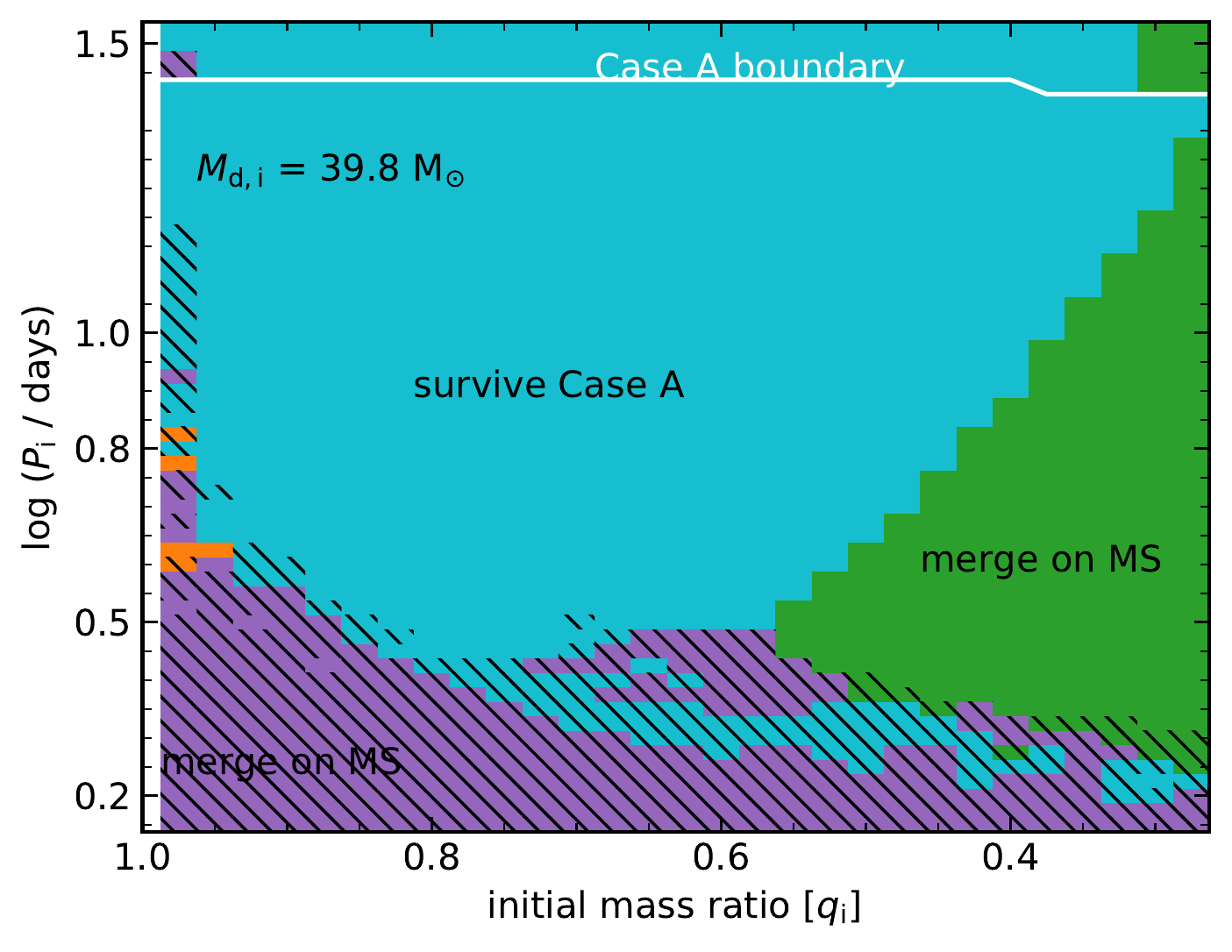}
\caption{Summary plot, same as Fig.\,\ref{1.350_summary}, but for an initial donor mass of $M \simeq 12.6\mso$ (top plot) and $M\simeq 39.8\mso$ (bottom plot).}
\label{1.100_summary}
\end{figure*}

We present the summary plots for different initial donor masses in our grid. Our work is mainly focussed on the systems that survive the Case\,A mass transfer phase (light blue models) or undergo an extended phase of interaction (orange and purple models) before merging on the main sequence (Fig.\ref{1.100_summary}).

\begin{figure*}
\centering{
\includegraphics[width=0.4\hsize]{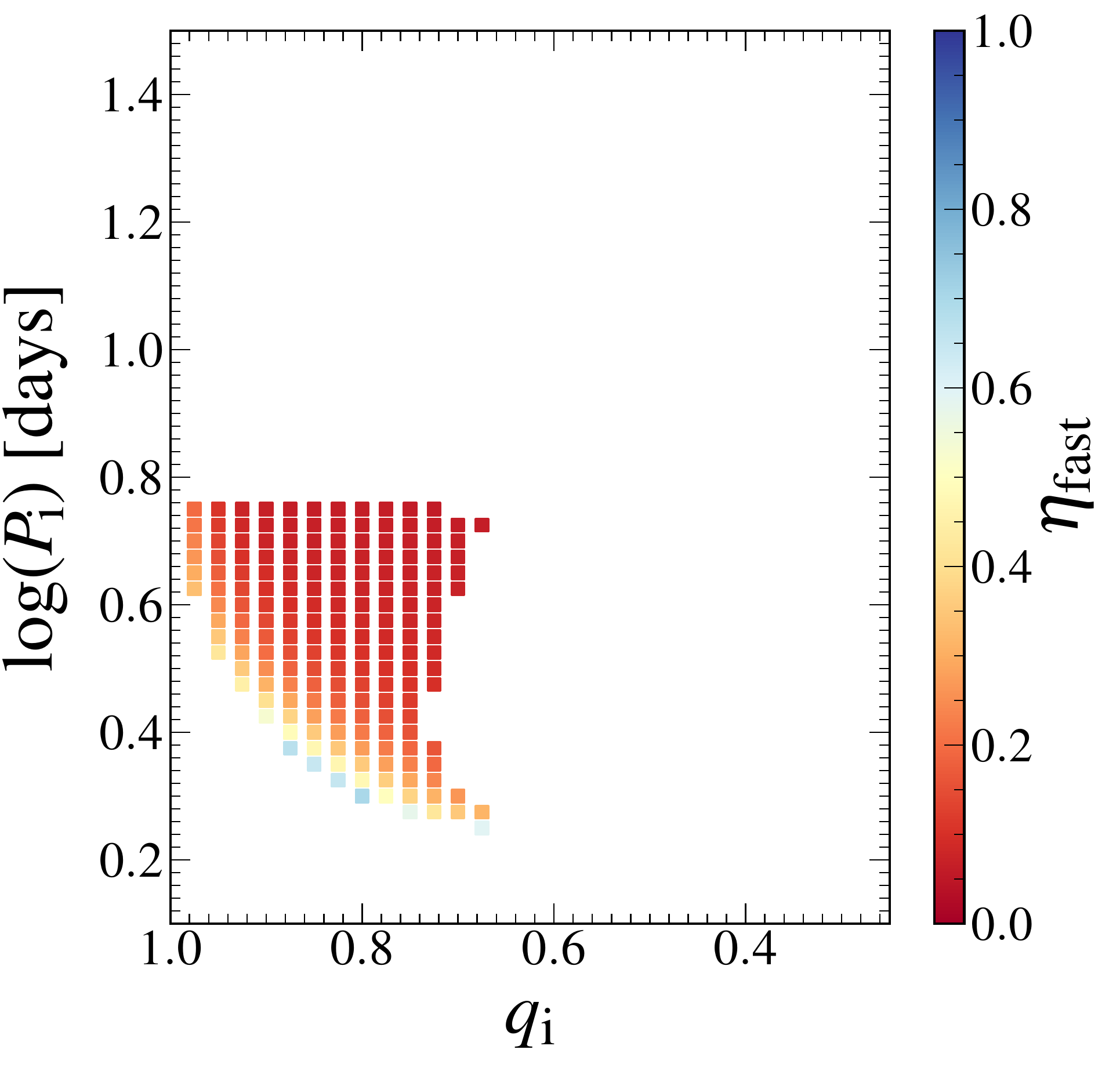} 
\includegraphics[width=0.4\hsize]{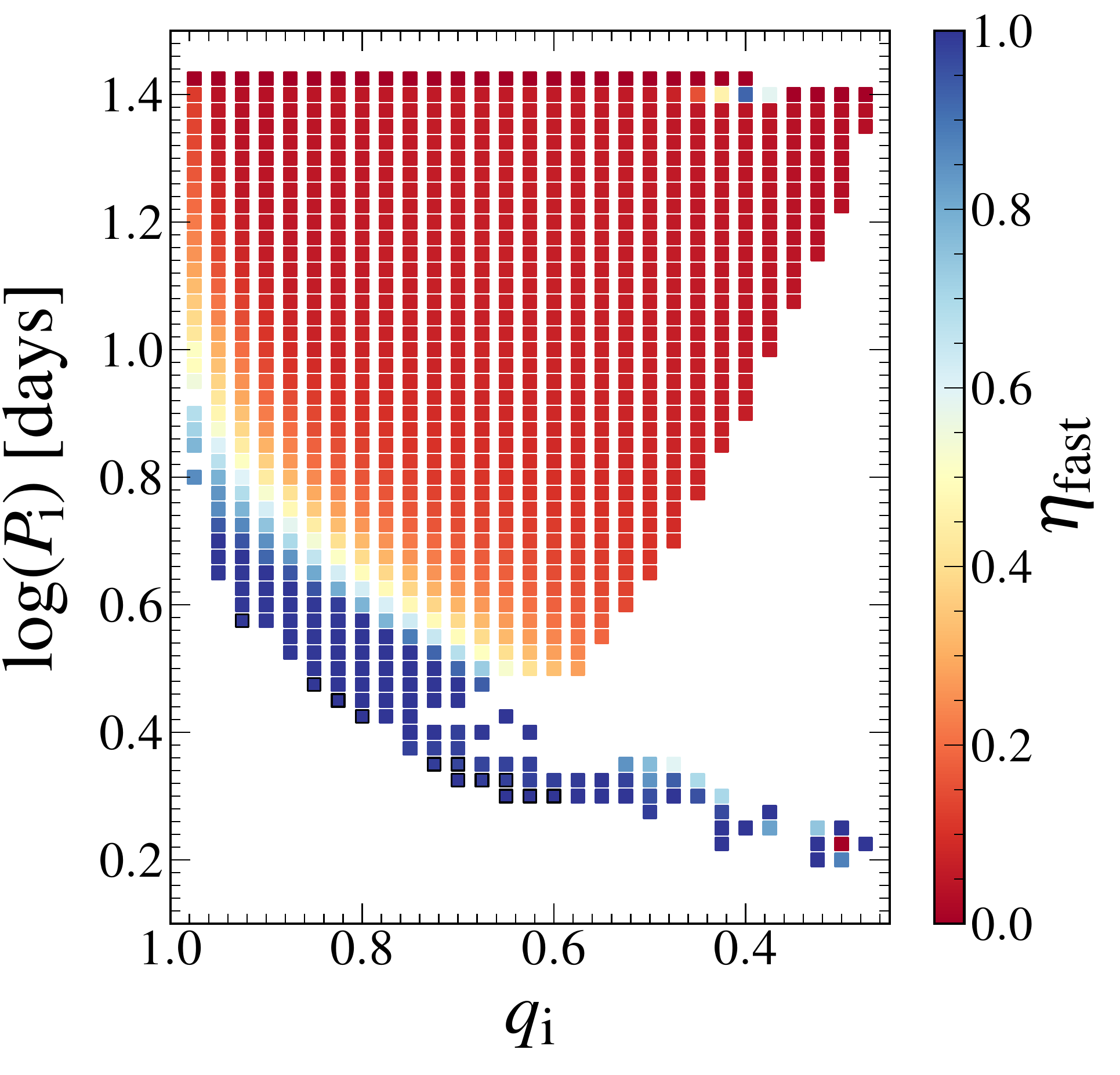}\hfill}
\centering{
\includegraphics[width=0.4\hsize]{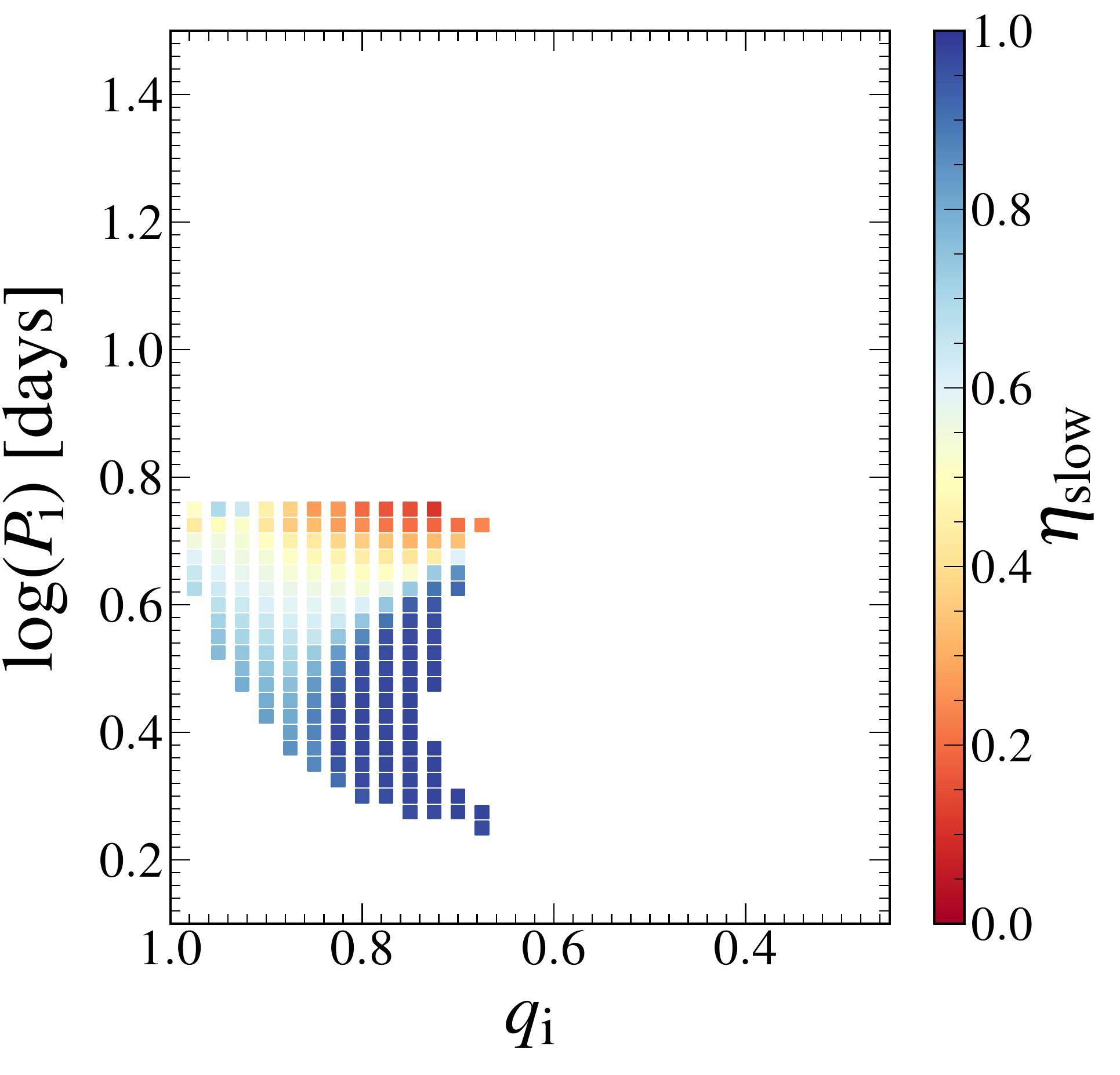}
\includegraphics[width=0.4\hsize]{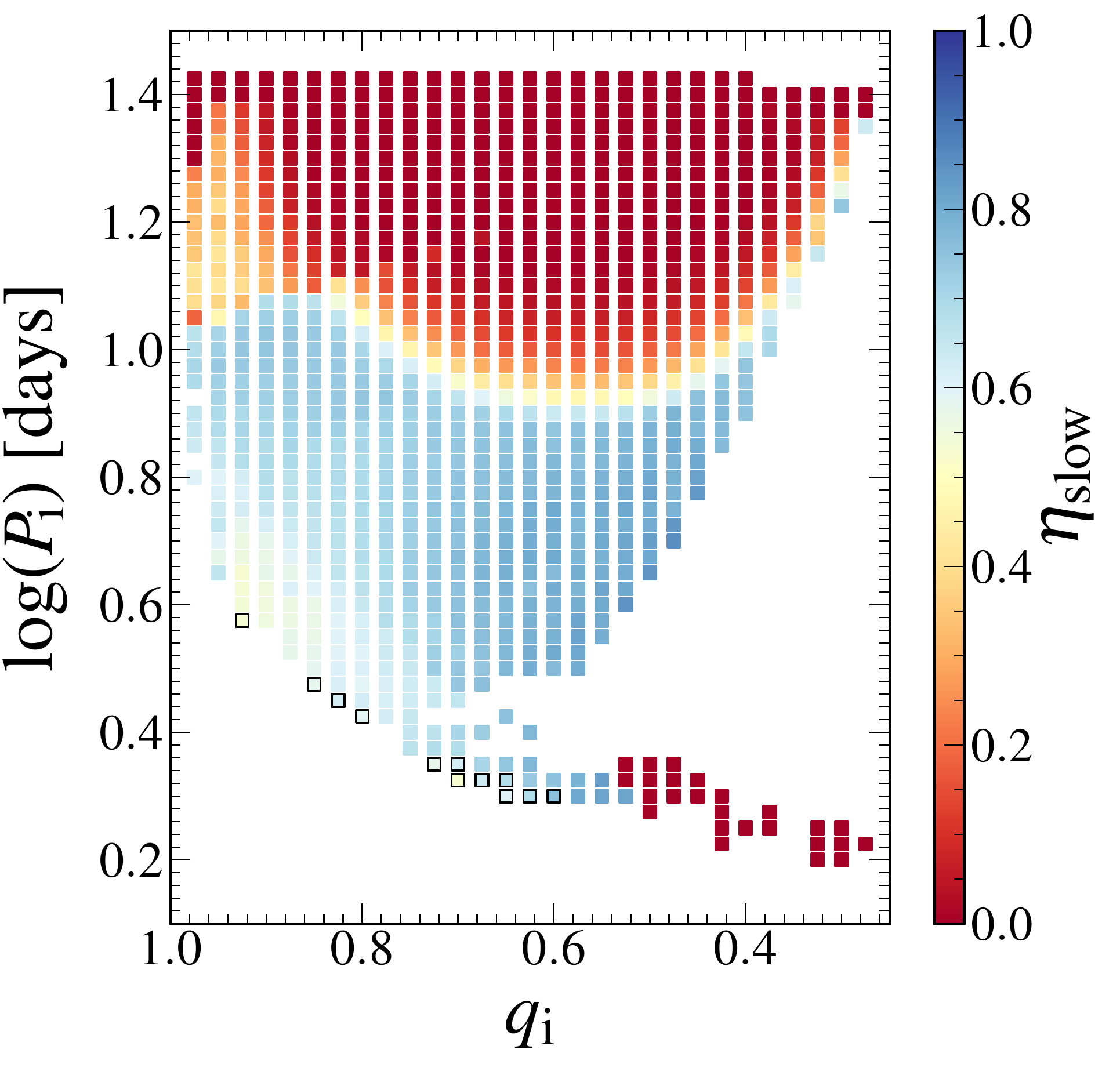}\hfill}
\centering{
\includegraphics[width=0.4\hsize]{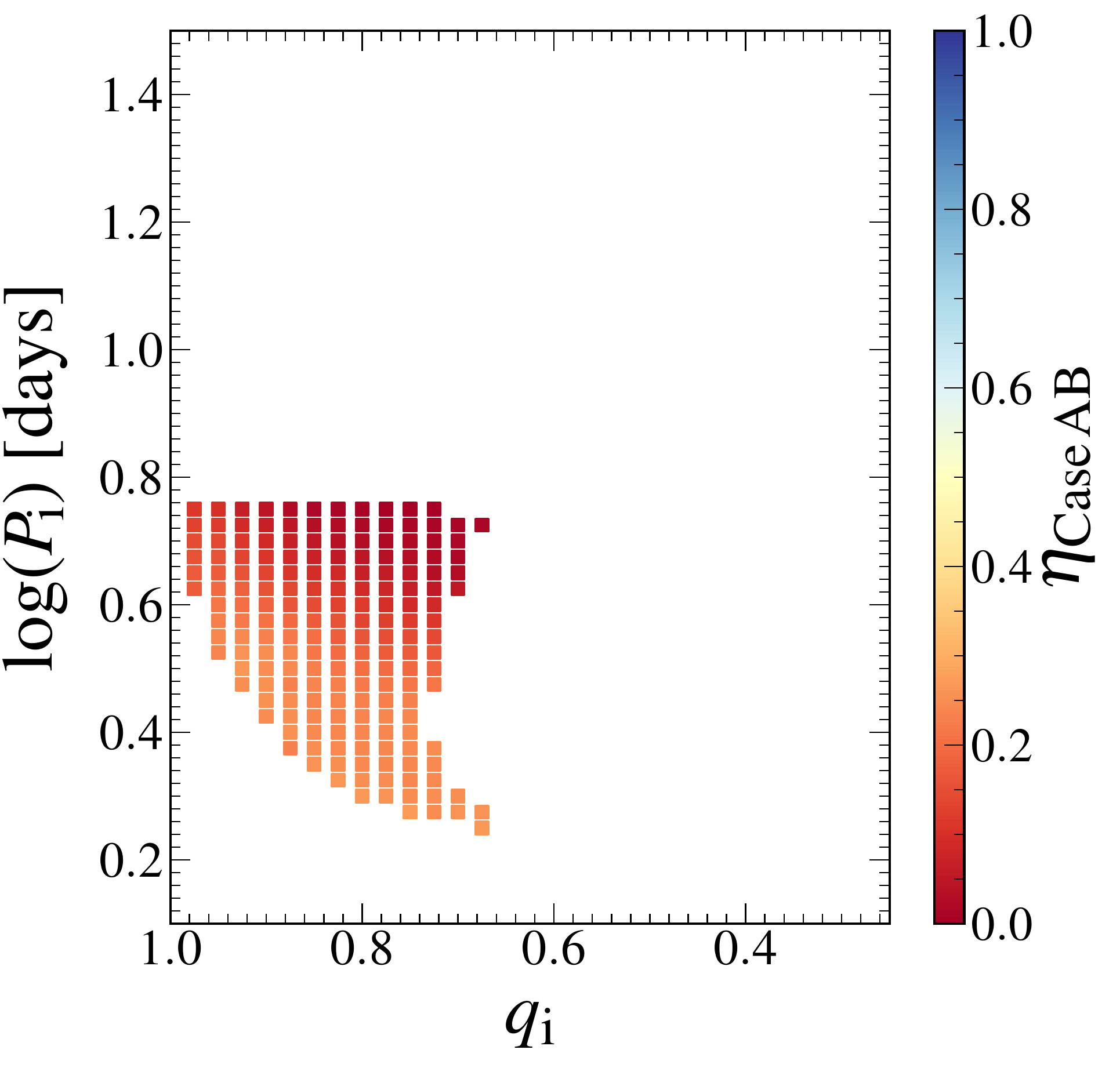}
\includegraphics[width=0.4\hsize]{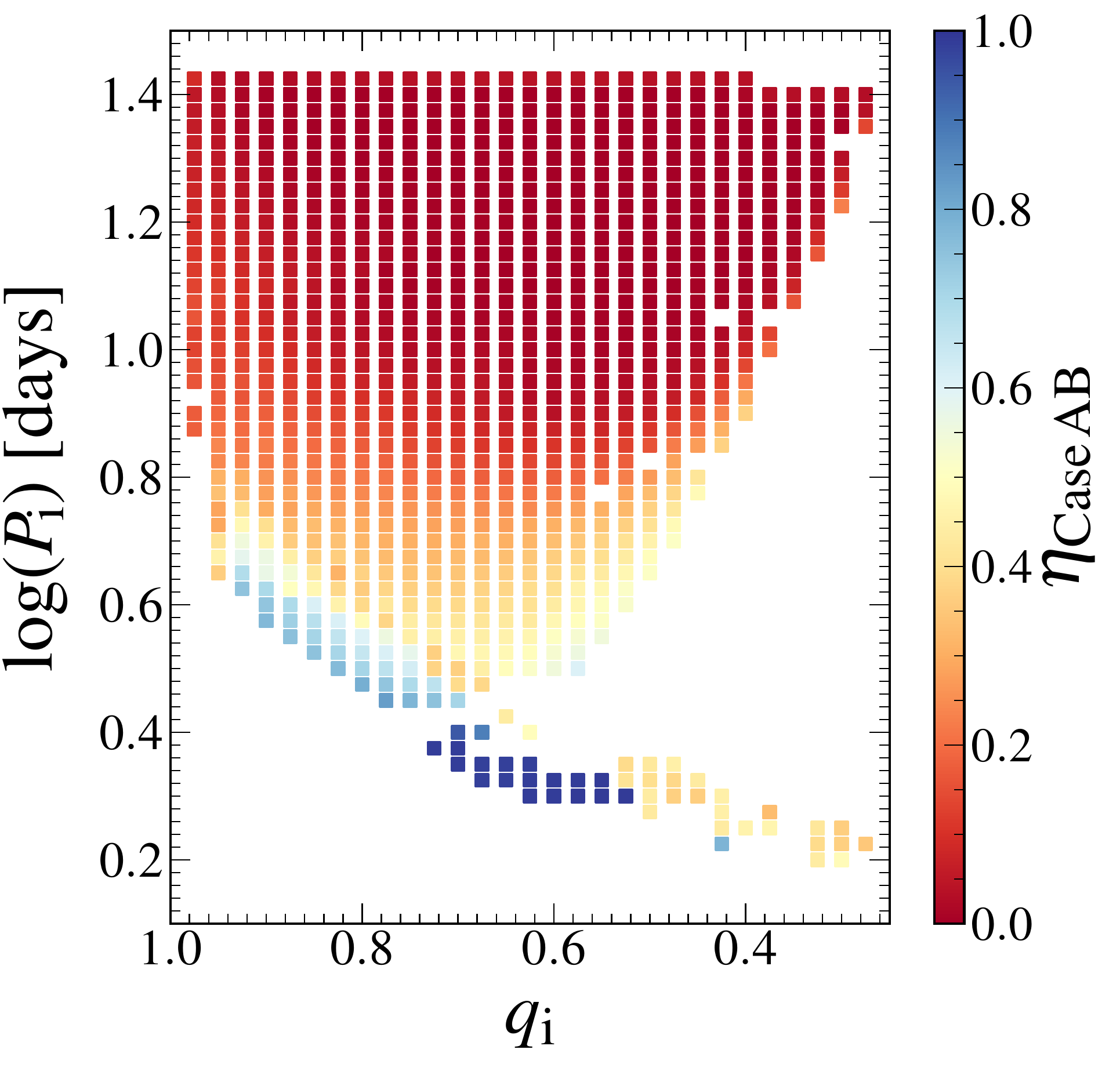}}
\caption{Mass transfer efficiency $\eta$ (colour coded) during fast 
(top panels) and slow (middle panels) Case\,A, and during
Case\,AB (bottom panels) as function of initial orbital period and 
initial mass ratio, for systems with initial donor masses of $\sim 16 \mso$
and $\sim 40\mso$ in the left and right panels, respectively. 
Each coloured square represents one binary evolution model, with the colour denoting the 
mass transfer efficiency. Black frames around coloured square indicate models 
that has undergone a contact phase lasting more than three thermal timescales of the mass donor.} 
\label{eta_fast_slow_ab}
\end{figure*}

\begin{figure*}
\centering{
\includegraphics[width=0.45\hsize]{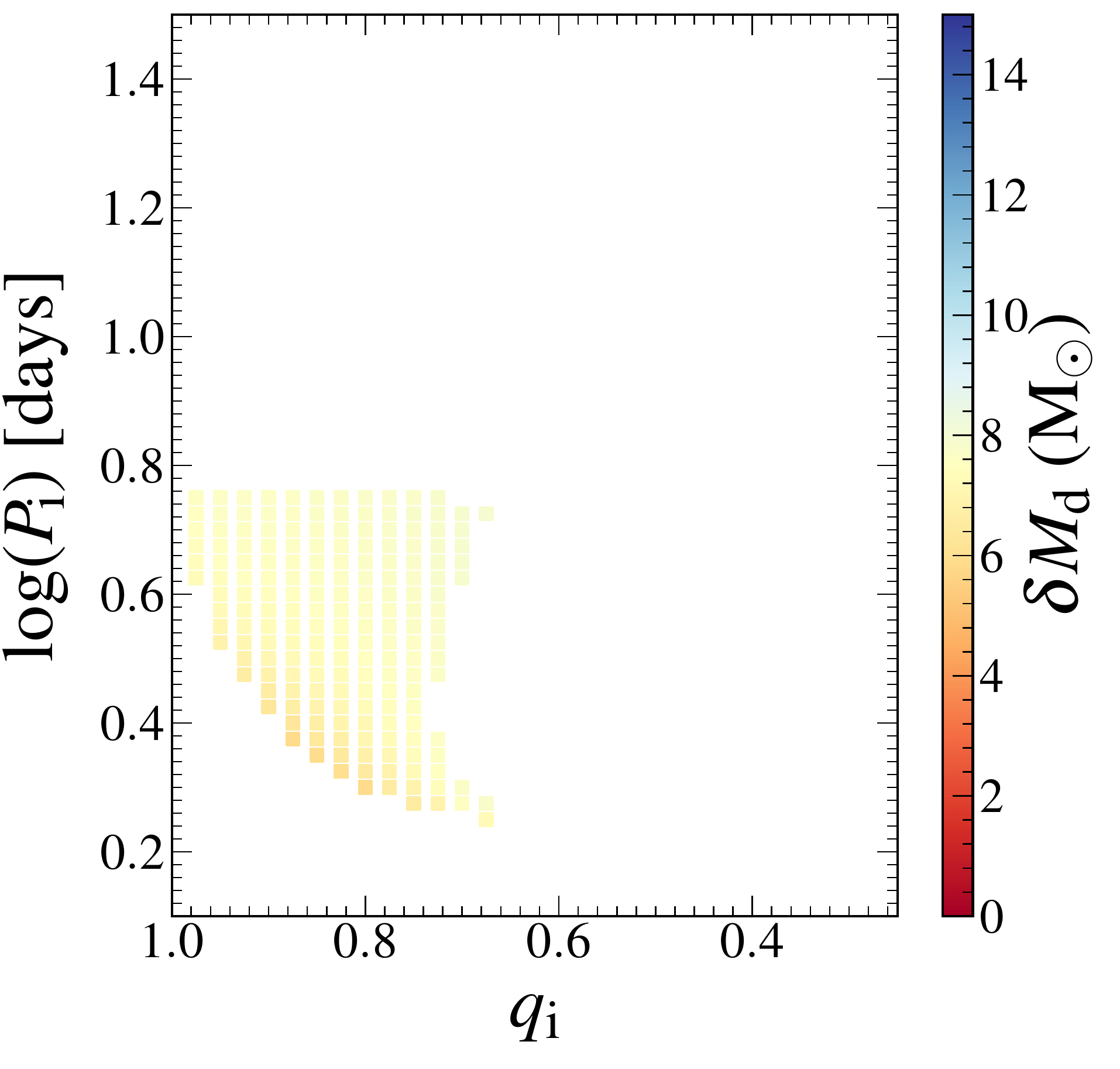}
\includegraphics[width=0.45\hsize]{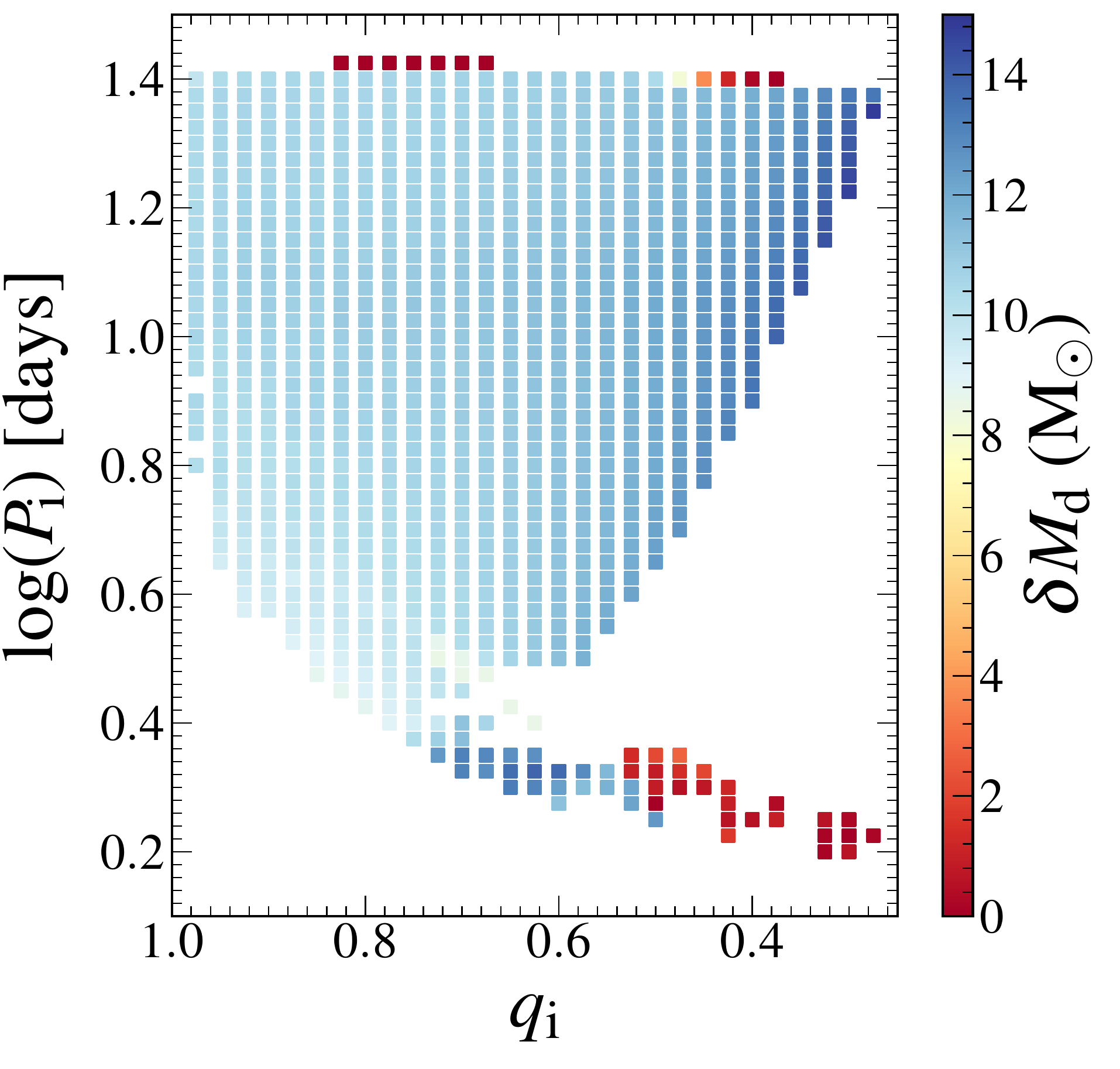}}
\centering{
\includegraphics[width=0.45\hsize]{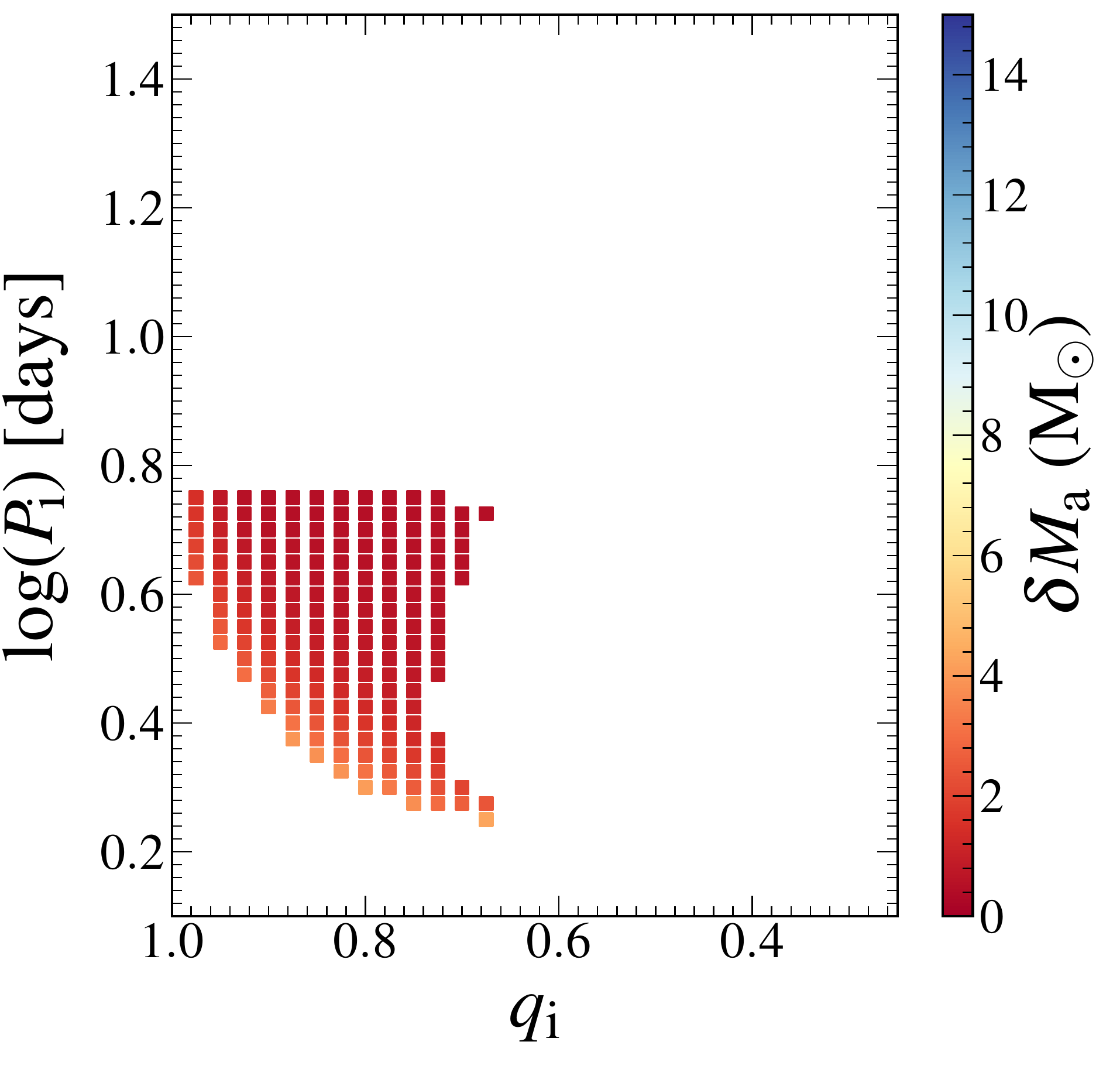}
\includegraphics[width=0.45\hsize]{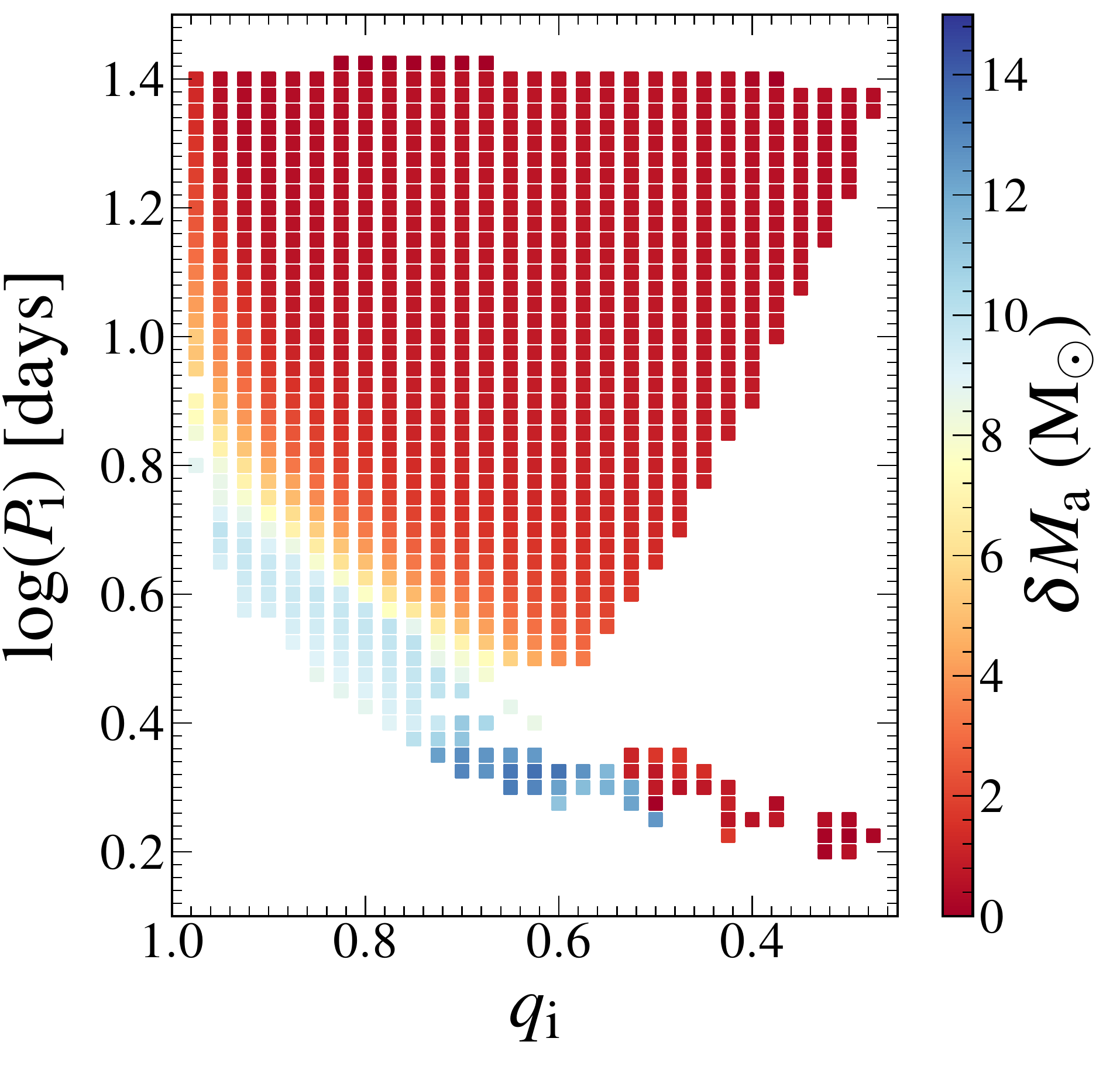}}
\caption{Amount of mass lost by the donor stars $\delta M_{d}$ (top panels)
and gained by the accretor $\delta M_{a}$ (bottom panels) during fast Case\,A mass transfer phase,
for systems with initial donor masses of $\sim 16 \mso$
and $\sim 40\mso$ in the left and right panels, respectively.
}
\label{fast_Case_A_mass_lost_gained}
\end{figure*}

\begin{figure*}
\centering{
\includegraphics[width=0.45\hsize]{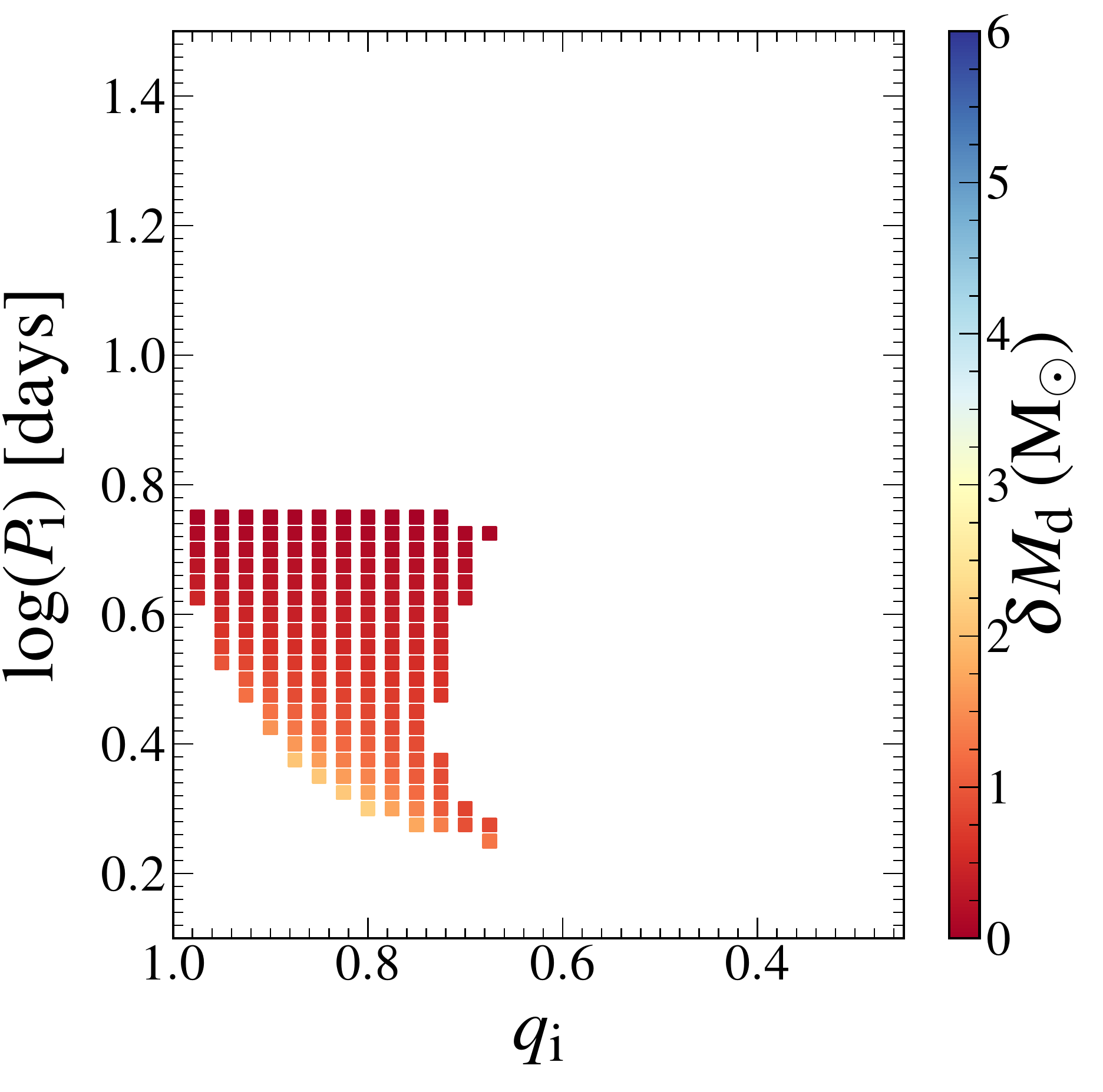}
\includegraphics[width=0.45\hsize]{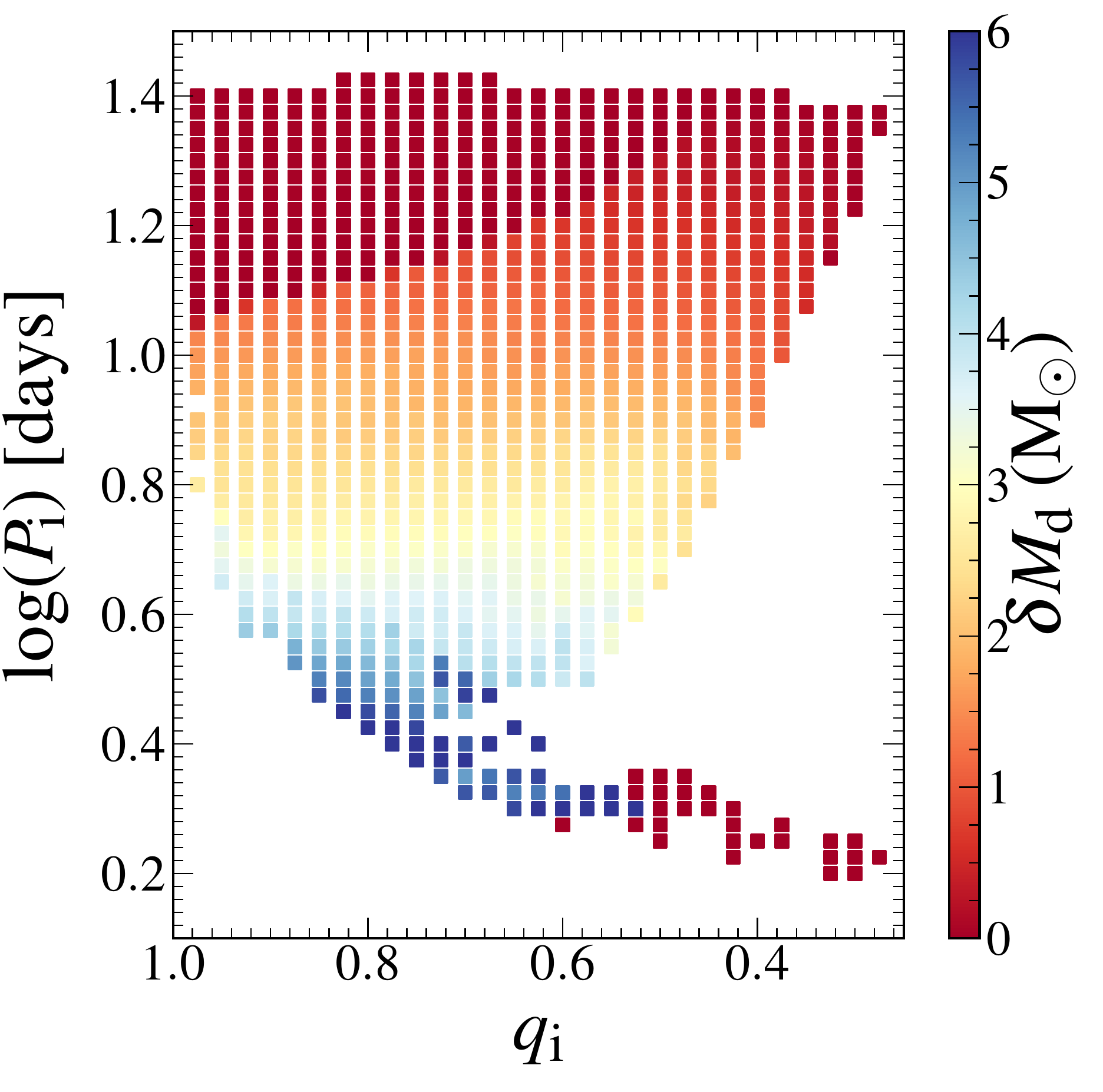}}
\centering{
\includegraphics[width=0.45\hsize]{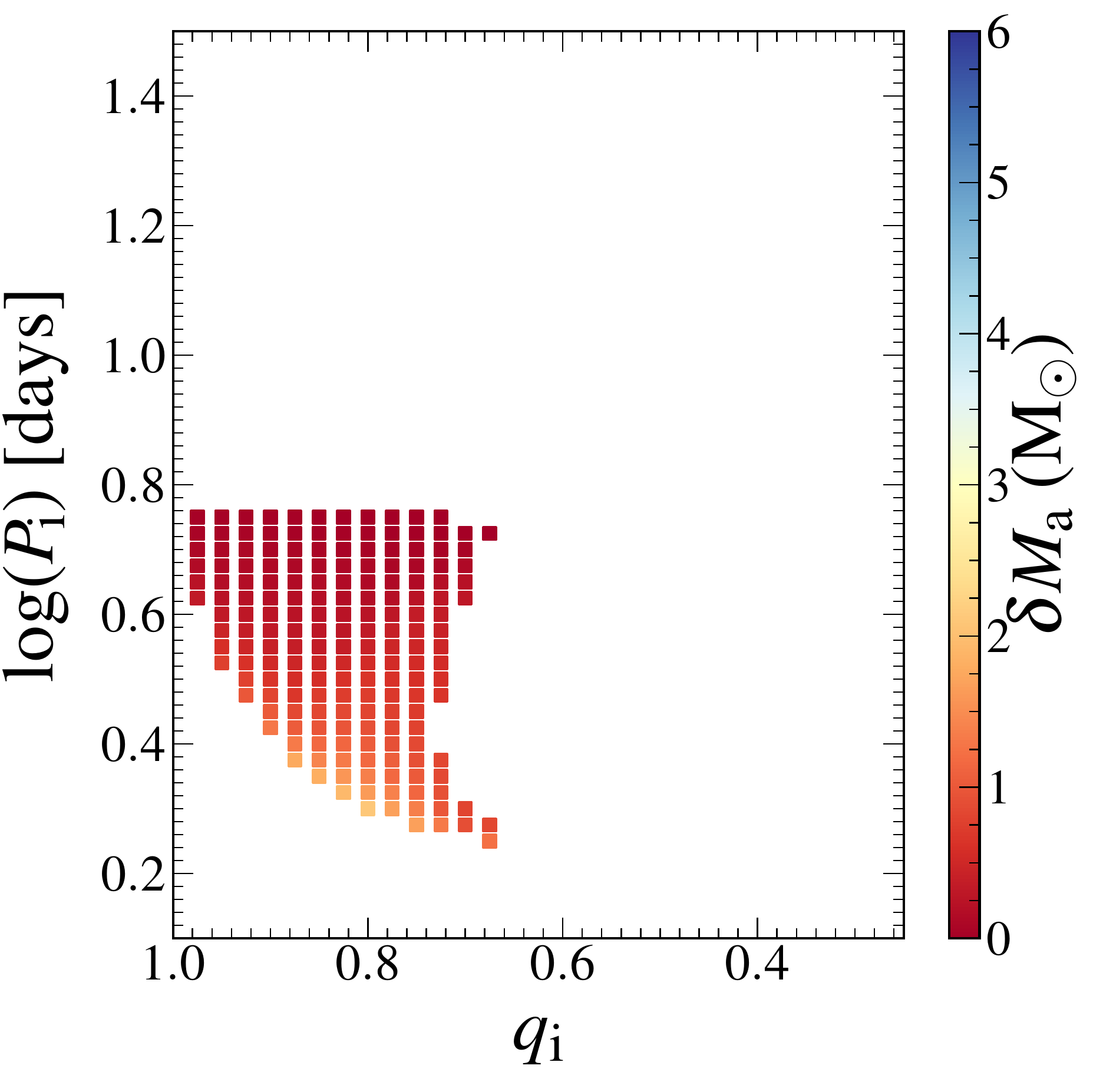}
\includegraphics[width=0.45\hsize]{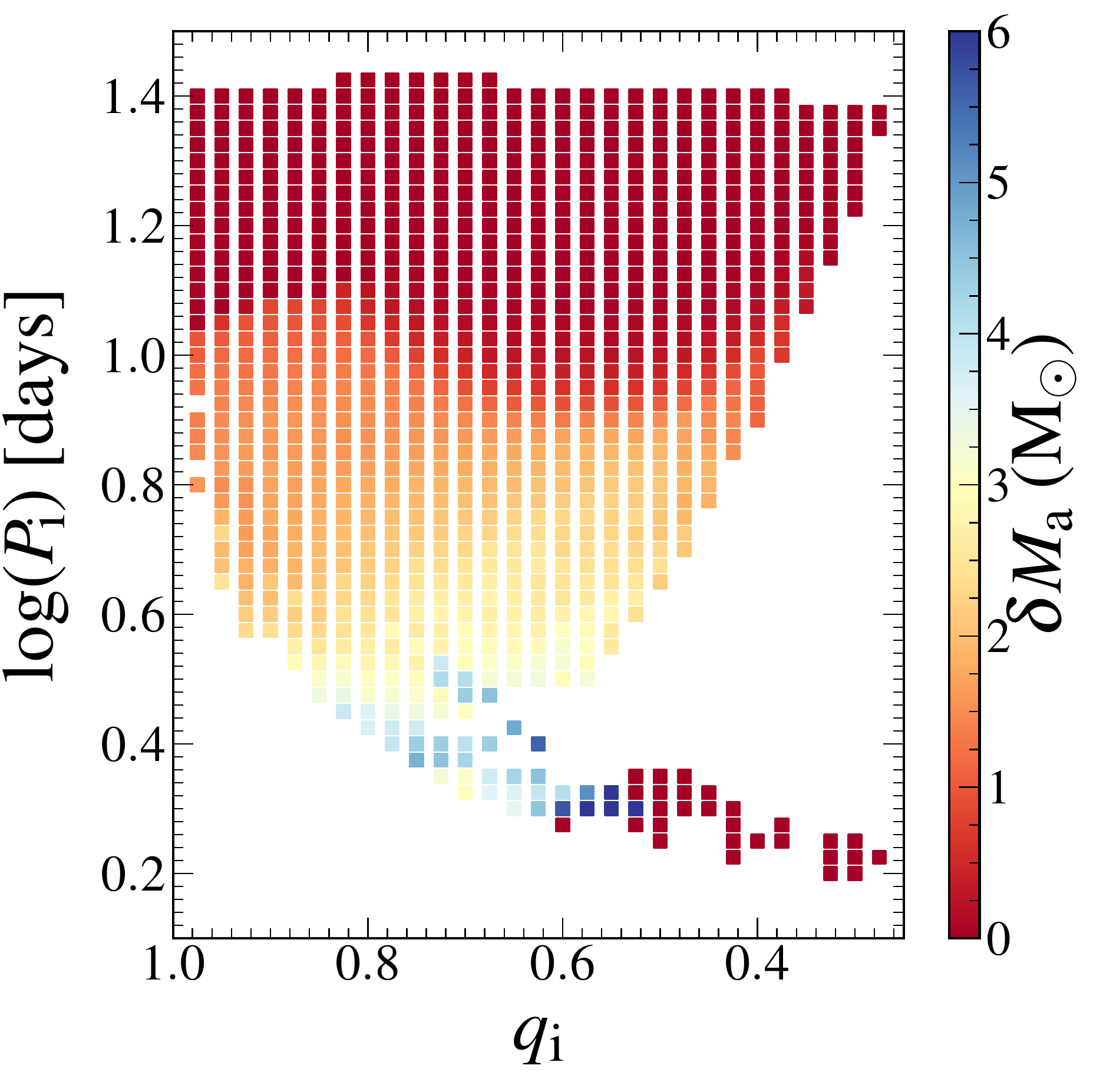}}
\caption{Amount of mass lost by the donor stars $\delta M_{d}$ (top panels)
and gained by the accretor $\delta M_{a}$ (bottom panels) during slow Case\,A mass transfer phase,
for systems with initial donor masses of $\sim 16 \mso$
and $\sim 40\mso$ in the left and right panels, respectively.
}
\label{slow_Case_A_mass_lost_gained}
\end{figure*}

\begin{figure*}
\centering{
\includegraphics[width=0.45\hsize]{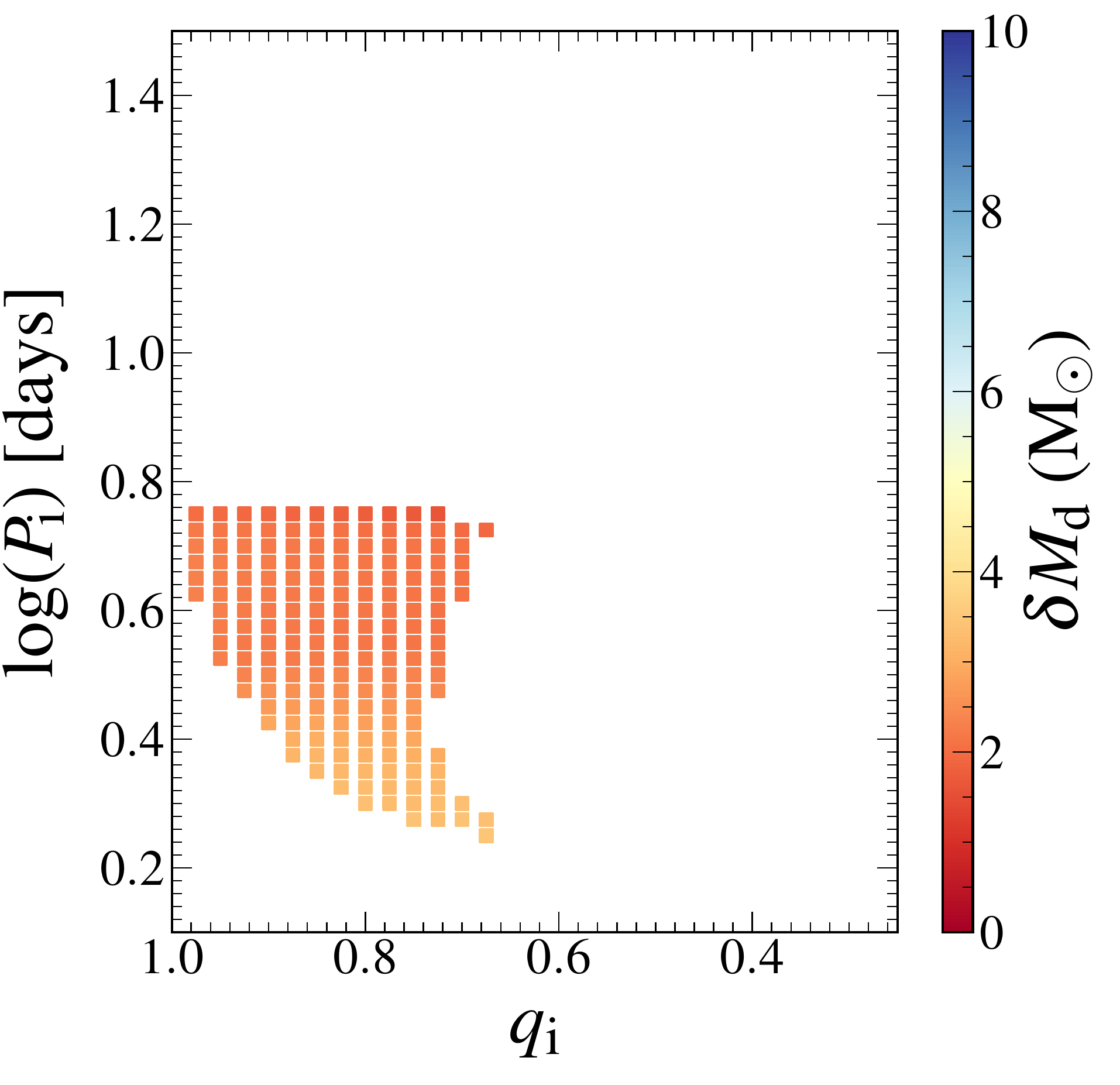}
\includegraphics[width=0.45\hsize]{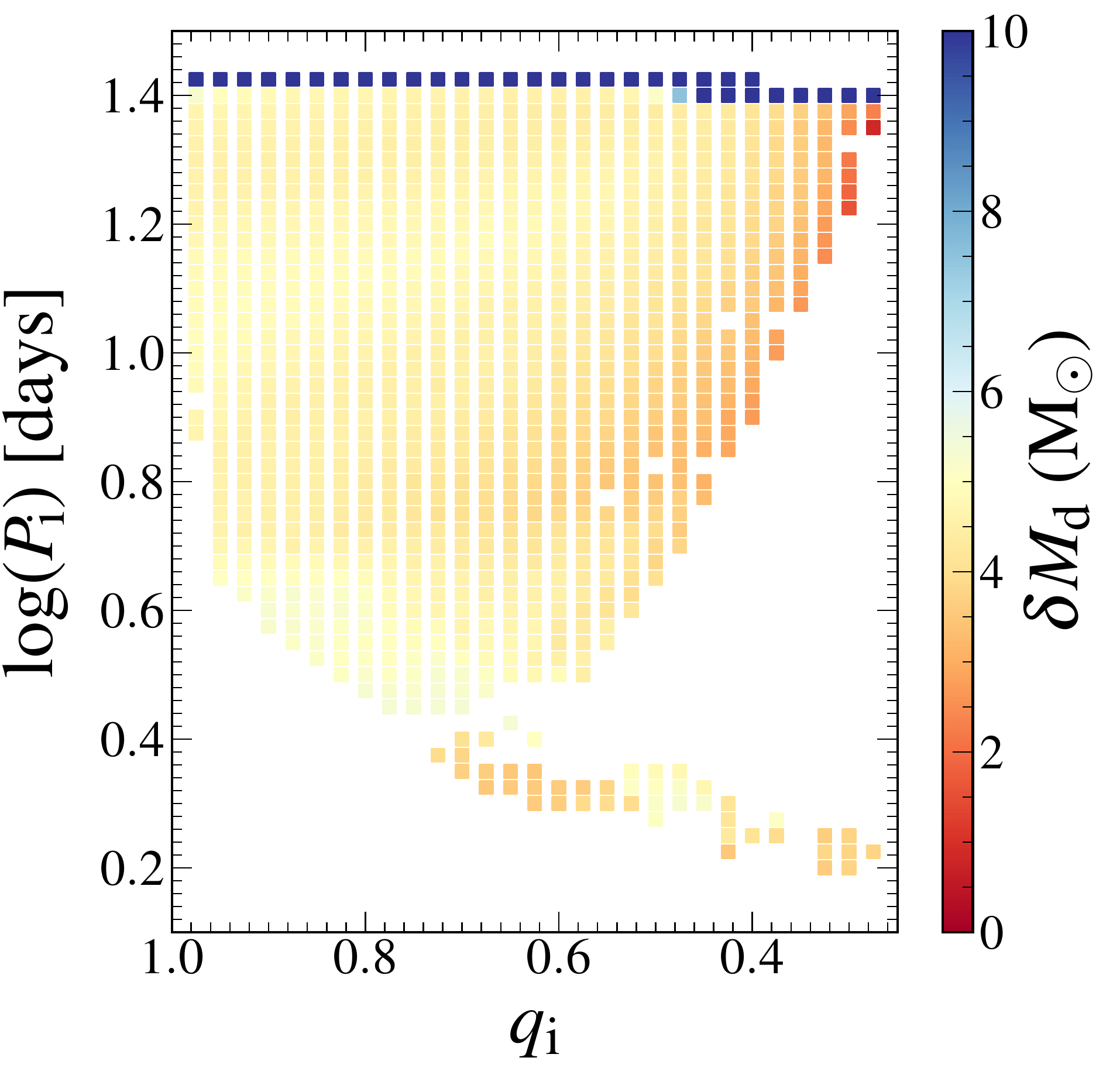}}
\centering{
\includegraphics[width=0.45\hsize]{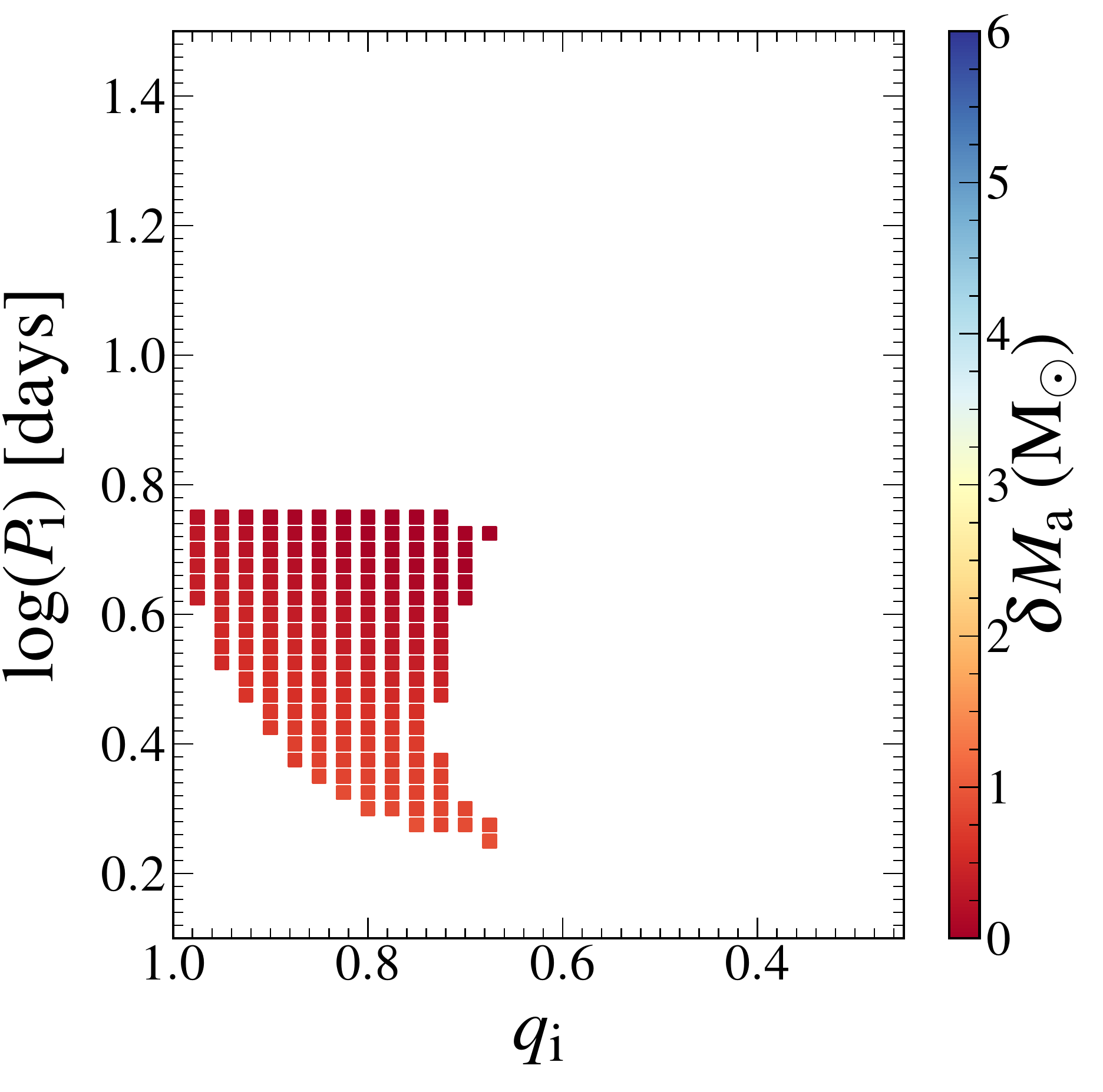}
\includegraphics[width=0.45\hsize]{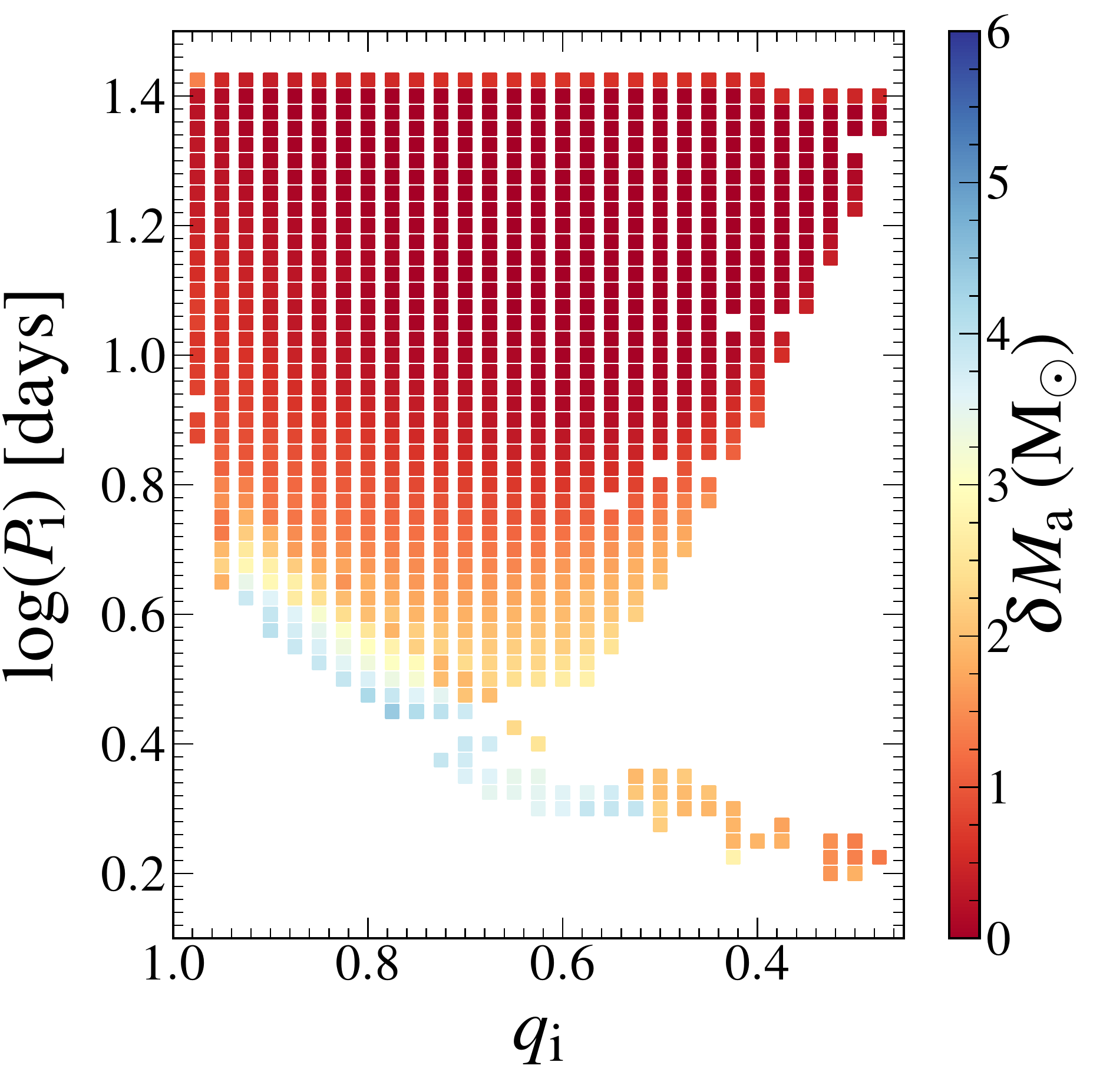}}
\caption{Amount of mass lost by the donor stars $\delta M_{d}$ (top panels)
and gained by the accretor $\delta M_{a}$ (bottom panels) during Case\,AB mass transfer phase,
for systems with initial donor masses of $\sim 16 \mso$
and $\sim 40\mso$ in the left and right panels, respectively.
}
\label{fast_Case_AB_mass_lost_gained}
\end{figure*}

\begin{figure*}
\centering{
\includegraphics[width=0.45\hsize]{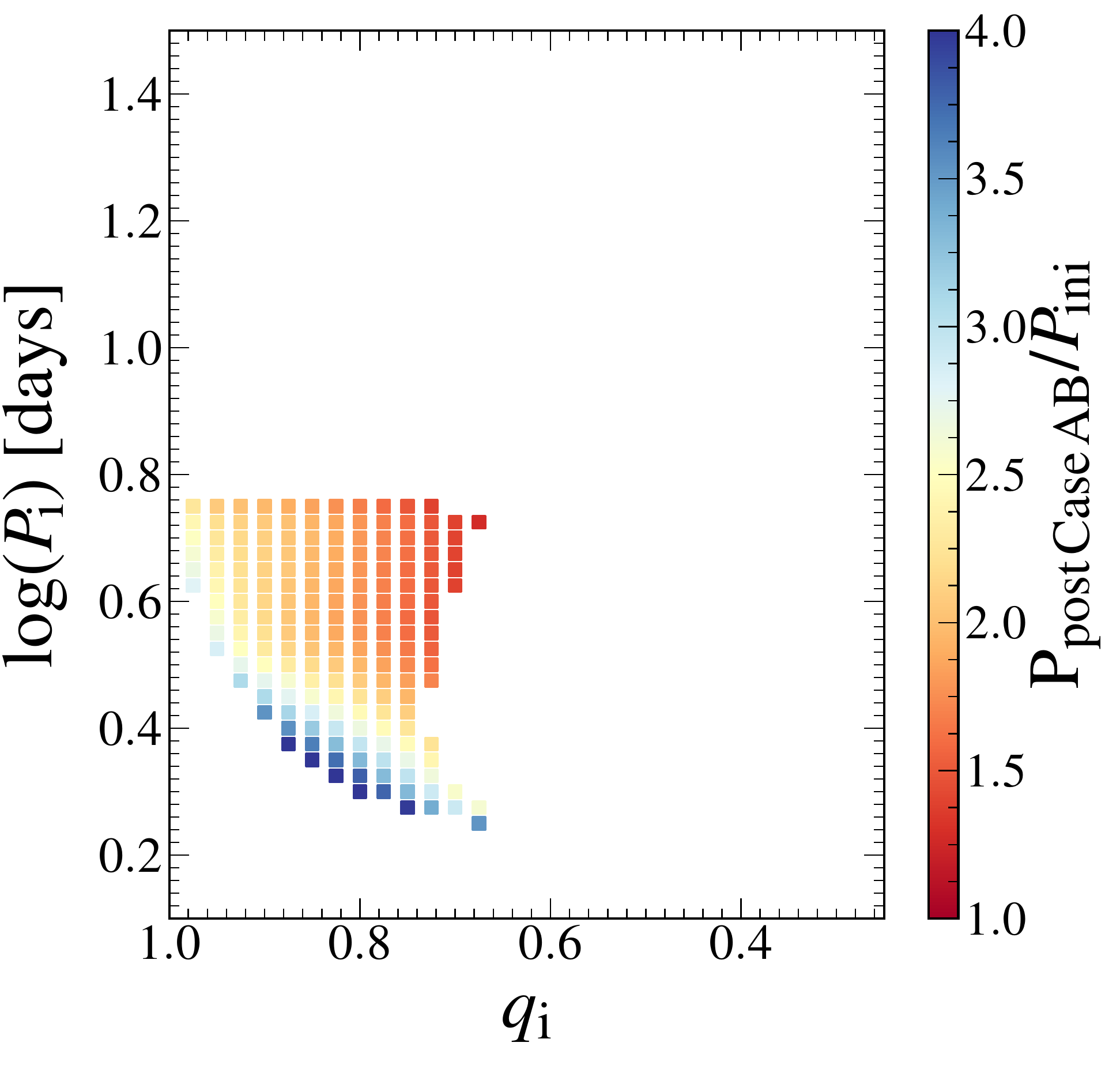}
\includegraphics[width=0.45\hsize]{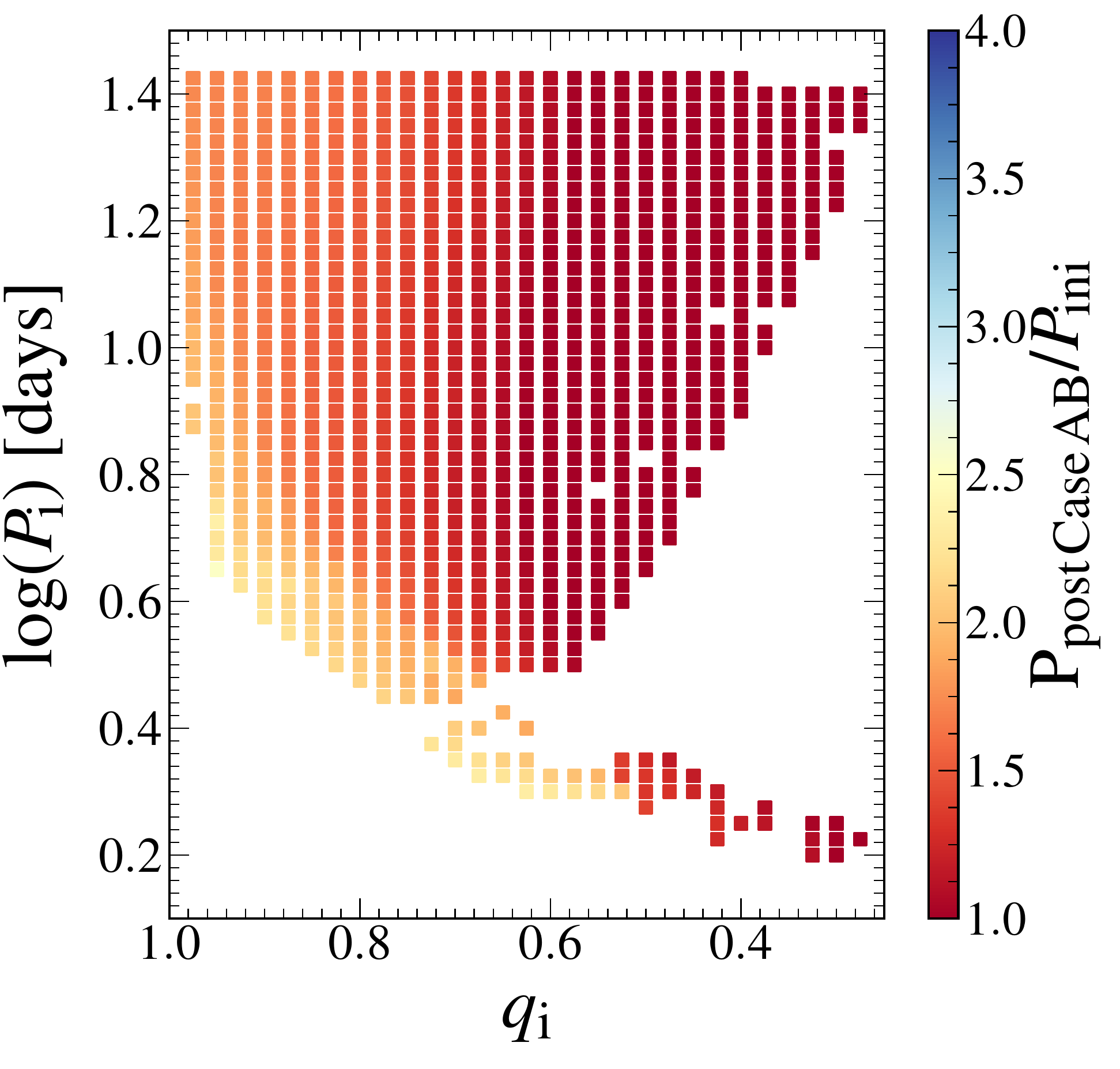}}
\caption{Ratio of the binary orbital period after Case\,AB mass transfer phase to the initial orbital period (colour coded) as function of initial orbital period and 
initial mass ratio, for systems with initial donor masses of $\sim 16 \mso$
and $\sim 40\mso$ in the left and right panels, respectively.
}
\label{periodchange}
\end{figure*}

\section{Binary properties after Case\,AB mass transfer}

Here, we show the binary properties of post Case\,AB models 
when the central helium mass fraction of the donor is 0.10.

\begin{figure*}
\centering
\includegraphics[width=0.48\hsize]{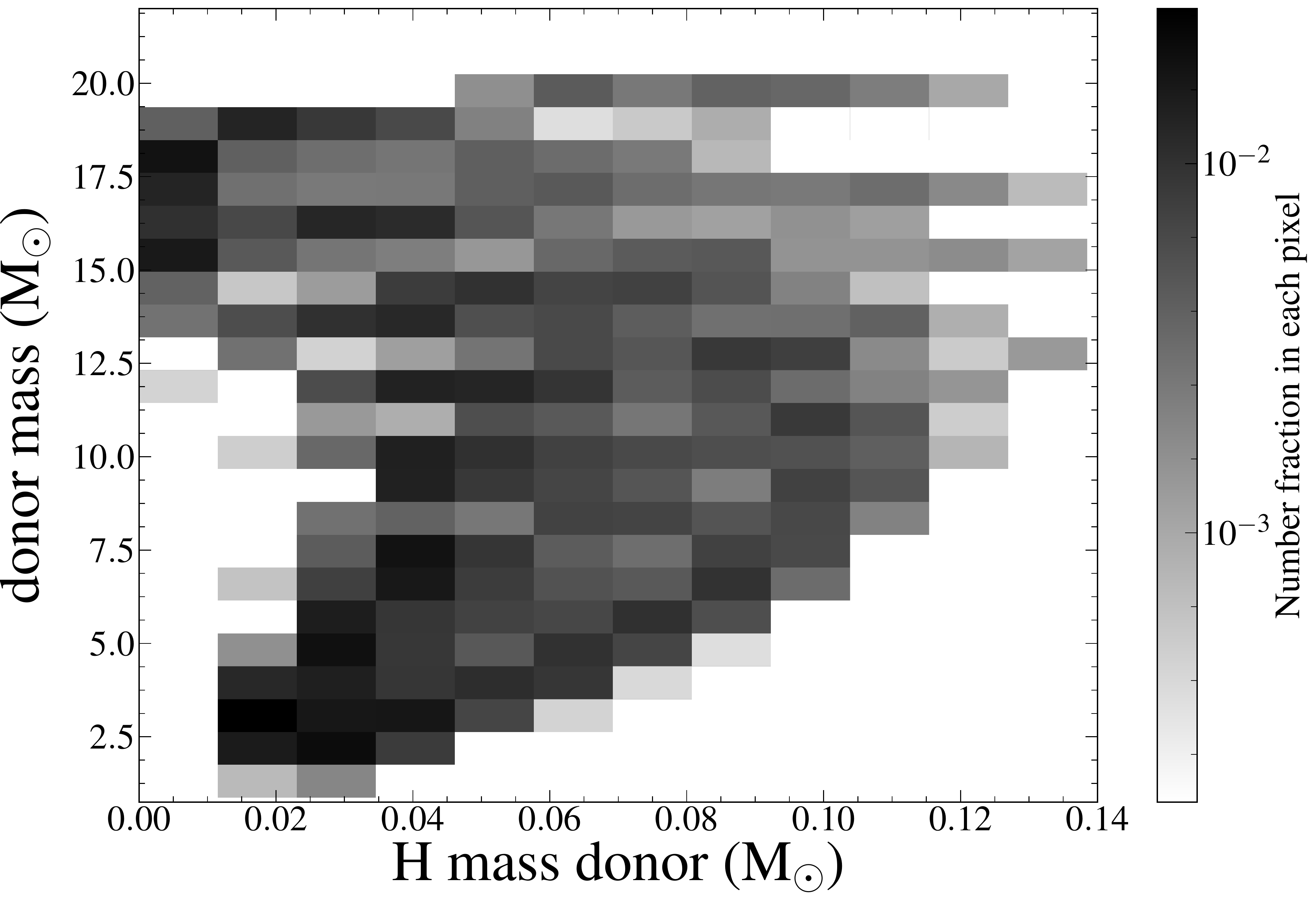}
\includegraphics[width=0.48\hsize]{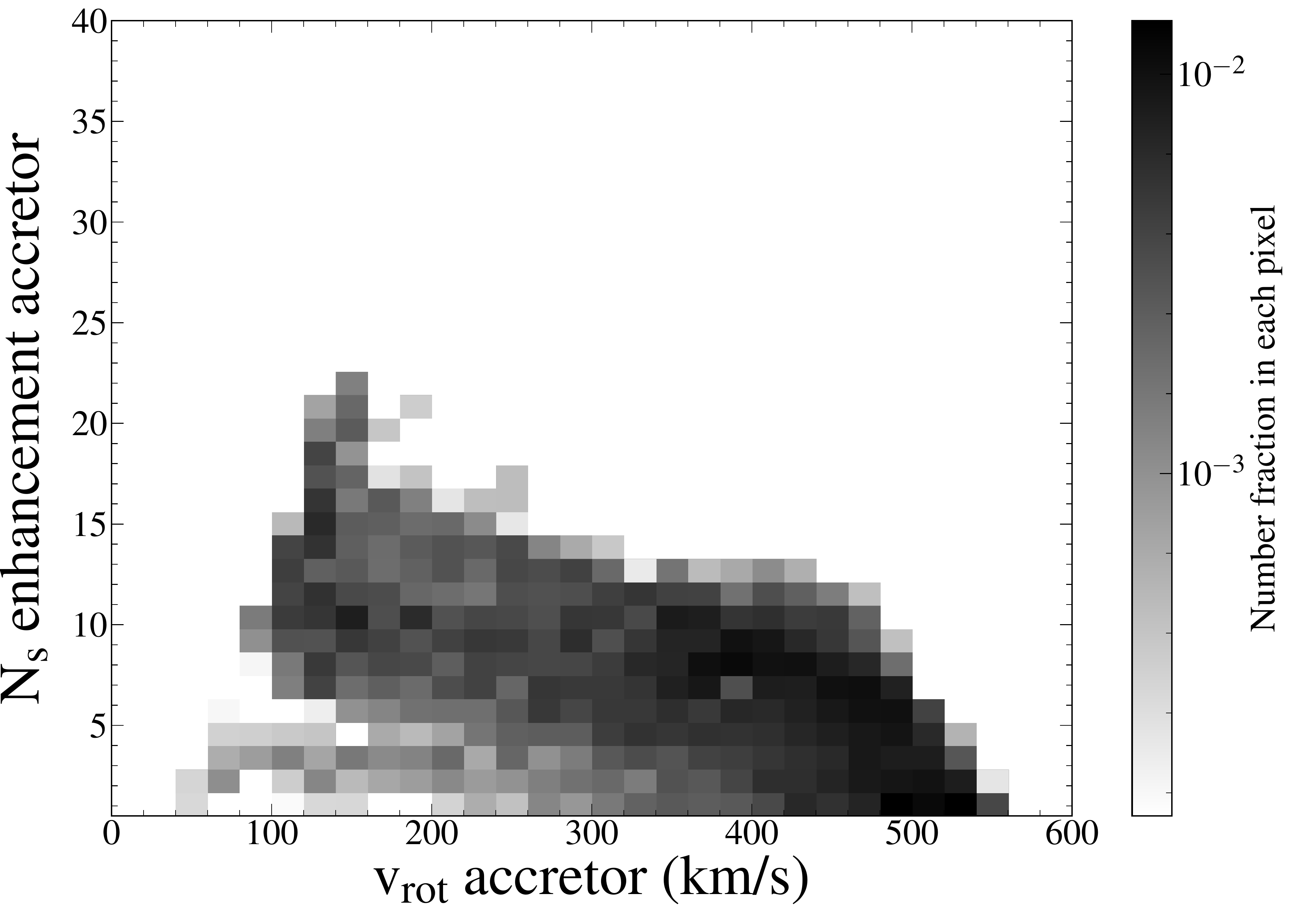}
\caption{Probability density distribution the leftover hydrogen mass vs total stellar mass of the donor (left panel)
and of the surface nitrogen mass fraction enhancement and rotational velocity for the mass gainer, for our binary models
after the Case\,AB mass transfer phase. In contrast to Fig.\,\ref{hydrogen_mass_CaseAB}, which depicts the same models at
a time where the central helium mass fraction of the donors is 0.9, here we display the information for 
a donor central helium mass fraction of 0.1.}
\label{Nenh_vrot_accretors_CaseAB1d-1}
\label{hydrogen_mass_CaseAB1d-1}
\end{figure*}

\end{appendix}

\end{document}